\pdfoutput=1
\newcommand*{\ATLASLATEXPATH}{latex/}
\documentclass[cernpreprint,texlive=2016,txfonts,UKenglish]{latex/atlasdoc}
\usepackage{latex/atlaspackage}
\usepackage{latex/atlasphysics}
\nolinenumbers

\usepackage{float}
\usepackage{lipsum}
\usepackage{\ATLASLATEXPATH atlaspackage}
\usepackage{\ATLASLATEXPATH atlasbiblatex}
\usepackage{\ATLASLATEXPATH atlascontribute}
\usepackage{\ATLASLATEXPATH atlasphysics}
\graphicspath{{logos/}{figures/}}
\usepackage{latex/atlasphysics}
\PreprintIdNumber{CERN-EP-2017-093}

\AtlasTitle{Determination of the strong coupling constant $\alpha_\mathrm{s}$ from transverse energy--energy correlations in multijet events at $\sqrt{s} = 8~\TeV$ using the ATLAS detector}
\author{The ATLAS Collaboration}
\AtlasAbstract{
Measurements of transverse energy--energy correlations and their associated asymmetries in multi-jet events using the ATLAS detector at the LHC are presented. The data used correspond to $\sqrt{s} = 8~\TeV$ proton--proton collisions with an integrated luminosity of 20.2$~\ifb$. The results are presented in bins of the scalar sum of the transverse momenta of the two leading jets, unfolded to the particle level and compared to the predictions from Monte Carlo simulations. A comparison with next-to-leading-order perturbative QCD is also performed, showing excellent agreement within the uncertainties. From this comparison, the value of the strong coupling constant is extracted for different energy regimes, thus testing the running of $\alpha_{\mathrm{s}}(\mu)$ predicted in QCD up to scales over $1~\TeV$. A global fit to the transverse energy--energy correlation distributions yields $\alpha_{\mathrm{s}}(m_\Zboson) = 0.1162 \pm 0.0011 \mbox{ (exp.)}^{+0.0084}_{-0.0070} \mbox{ (theo.)}$, while a global fit to the asymmetry distributions yields a value of $\alpha_{\mathrm{s}}(m_\Zboson) = 0.1196 \pm 0.0013 \mbox{ (exp.)}^{+0.0075}_{-0.0045} \mbox{ (theo.)}$.
}

\hypersetup{pdftitle={ATLAS document},pdfauthor={The ATLAS Collaboration}}

\newcommand\T{\rule{0pt}{2.ex}}
\newcommand\B{\rule[-1.2ex]{0pt}{0pt}}

\begin{document}

\maketitle

\tableofcontents

\clearpage

\section{Introduction}
\label{sec:intro}

Experimental studies of the energy dependence of event shape variables have proved very useful in precision tests of quantum chromodynamics (QCD). Event shape variables have been measured in $e^+ e^-$ experiments from PETRA--PEP~\cite{plutoEventShapes,tassoEventShapes,jadeEventShapes} to LEP--SLC~\cite{opalEventShapes,alephEventShapes,thrustDelphi,l3eventShapes} energies, at the $ep$ collider HERA~\cite{zeusEventShapes1,zeusEventShapes2,h1EventShapes1,h1EventShapes2,h1EventShapes3} as well as in hadron colliders from Tevatron~\cite{Aaltonen} to LHC energies~\cite{Khachatryan, Aad}.\\
Most event shape variables are based on the determination of the thrust's principal axis~\cite{brandt} or the sphericity tensor~\cite{bjo}. A notable exception is given by the energy--energy correlations (EEC), originally proposed by Basham \emph{et al.}~\cite{Basham1}, and measurements~\cite{PLUTO,CELLO,JADE,TASSO,MARKJ,MARKII,MAC,TOPAZ,ALEPH,DELPHI,OPAL,L3,SLD} of these have significantly improved the precision tests of perturbative QCD (pQCD). The EEC is defined as the energy-weighted angular distribution of hadron pairs produced in $e^+e^-$ annihilation and, by construction, the EEC as well as its associated asymmetry (AEEC) are infrared safe. The second-order corrections to these functions were found to be significantly smaller~\cite{Ali1982,Ali1984A,Richards,Glover} than for other event shape variables such as thrust.\\
The transverse energy--energy correlation (TEEC) and its associated asymmetry (ATEEC) were proposed as the appropriate generalisation to hadron colliders in Ref.~\cite{Ali1984B}, where leading-order (LO) predictions were also presented. As a jet-based quantity, it makes use of the jet transverse energy $\et=E \sin\theta$ since the energy alone is not Lorentz-invariant under longitudinal boosts along the beam direction. Here $\theta$ refers to the polar angle of the jet axis, while $E$ is the jet energy.\footnote{ATLAS uses a right-handed coordinate system with its origin at the nominal interaction point (IP) in the centre of the detector and the $z$-axis along the beam pipe. The $x$-axis points from the IP to the centre of the LHC ring, and the $y$-axis points upward. Cylindrical coordinates $(r,\varphi)$ are used in the transverse plane, $\varphi$ being the azimuthal angle around the beam pipe. The pseudorapidity is defined in terms of the polar angle $\theta$ as $\eta=-\ln\tan(\theta/2)$.} The next-to-leading-order (NLO) corrections were obtained recently~\cite{Ali2013} by using the \textsc{NLOJET++} program~\cite{Nagy1, Nagy2}. They are found to be of moderate size so that the TEEC and ATEEC functions are well suited for precision tests of QCD, including a precise determination of the strong coupling constant $\alpha_{\mathrm{s}}$. The TEEC is defined as~\cite{altarelliQCD}
\begin{equation}
\frac{1}{\sigma}\frac{\mathrm{d}\Sigma}{\mathrm{d}\cos\phi} \equiv \frac{1}{\sigma}\sum_{ij}\int\frac{\mathrm{d}\sigma}{\mathrm{d}\xti \mathrm{d}\xtj \mathrm{d}\cos\phi}\xti\xtj \mathrm{d}\xti \mathrm{d}\xtj = \frac{1}{N} \sum_{A=1}^N \sum_{ij} \frac{E^A_{\mathrm{T}i}E^A_{\mathrm{T}j}}{\left(\sum_k E^A_{\mathrm{T}k}\right)^2}\delta(\cos \phi - \cos \phi_{ij}),
\label{eq:teecDef}
\end{equation}
where the last expression is valid for a sample of $N$ hard-scattering multi-jet events, labelled by the index $A$, and the indices $i$ and $j$ run over all jets in a given event. Here, $\xti$ is the fraction of transverse energy of jet $i$ with respect to the total transverse energy, i.e. $\xti = E_{\mathrm{T}i}/\sum_k E_{\mathrm{T}k}$, $\phi_{ij}$ is the angle in the transverse plane between jet $i$ and jet $j$ and $\delta(x)$ is the Dirac delta function, which ensures $\phi = \phi_{ij}$.\\
The associated asymmetry ATEEC is then defined as the difference between the forward ($\cos\phi > 0$) and the backward ($\cos\phi < 0$) parts of the TEEC, i.e.
\[
\frac{1}{\sigma}\frac{\mathrm{d}\Sigma^{\mathrm asym}}{\mathrm{d}\cos\phi}\equiv
\left.\frac{1}{\sigma}\frac{\mathrm{d}\Sigma}{\mathrm{d}\cos\phi}\right|_{\phi} -
\left.\frac{1}{\sigma}\frac{\mathrm{d}\Sigma}{\mathrm{d}\cos\phi}\right|_{\pi -\phi}.
\label{eq:ateecDef}
\]
Recently, the ATLAS Collaboration presented a measurement of the TEEC and ATEEC~\cite{atlasEEC}, where these observables were used for a determination of the strong coupling constant $\alpha_{\mathrm{s}}(m_\Zboson)$ at an energy regime of $\langle Q \rangle = 305~\GeV$. This paper extends the previous measurement to higher energy scales up to values close to $1~\TeV$. The analysis consists in the measurement of the TEEC and ATEEC distributions in different energy regimes, determining $\alpha_{\mathrm{s}}(m_\Zboson)$ in each of them, and using these determinations to test the running of $\alpha_{\mathrm{s}}$ predicted by the QCD $\beta$-function. Precise knowledge of the running of $\alpha_{\mathrm{s}}$ is not only important as a precision test of QCD at large scales but also as a test for new physics, as the existence of new coloured fermions would imply modifications to the $\beta$-function~\cite{kaplan, newFermions}.

\section{ATLAS detector}
\label{sec:detector}
\label{sec:detector}
The ATLAS detector~\cite{detector} is a multipurpose particle physics detector with a forward-backward symmetric cylindrical geometry and a solid angle coverage of almost $4\pi$.\\
The inner tracking system covers the pseudorapidity range $|\eta|< 2.5$. It consists of a silicon pixel detector, a silicon microstrip detector and, for $|\eta|<2.0$, a transition radiation tracker. It is surrounded by a thin superconducting solenoid providing a 2 $\mathrm{T}$ magnetic field along the beam direction. A high-granularity liquid-argon sampling electromagnetic calorimeter covers the region $|\eta|<3.2$. A steel/scintillator tile hadronic calorimeter provides coverage in the range $|\eta|<1.7$. The endcap and forward regions, spanning $1.5<|\eta|<4.9$, are instrumented with liquid-argon calorimeters for electromagnetic and hadronic measurements. The muon spectrometer surrounds the calorimeters. It consists of three large air-core superconducting toroid systems and separate trigger and high-precision tracking chambers providing accurate muon tracking for $|\eta|<2.7$.\\
The trigger system~\cite{atlasTrigger} has three consecutive levels: level 1, level 2 and the event filter. The level 1 triggers are hardware-based and use coarse detector information to identify regions of interest, whereas the level 2 triggers are software-based and perform a fast online data reconstruction. Finally, the event filter uses reconstruction algorithms similar to the offline versions with the full detector granularity.

\section{Monte Carlo simulation}
\label{sec:mc}

Multi-jet production in $pp$ collisions is described by the convolution of the production cross-sections for parton--parton scattering with the parton distribution functions (PDFs).
Monte Carlo (MC) event generators differ in the approximations used to calculate the underlying short-distance QCD processes, in the way parton showers are built to take into account higher-order effects and in the fragmentation scheme responsible for long-distance effects.
\textsc{Pythia} and \textsc{Herwig++} event generators were used for the description of multi-jet production in $pp$ collisions. These event generators differ in the modelling of the parton shower, hadronisation and underlying event. \textsc{Pythia} uses $\pt$-ordered parton showers, in which the $\pt$ of the emitted parton is decreased in each step, while for the angle-ordered parton showers in \textsc{Herwig++}, the relevant scale is related to the angle between the emitted and the incoming parton. The generated events were processed with the ATLAS full detector simulation~\cite{atlasim} based on \textsc{Geant4}~\cite{geant}.\\
The baseline MC samples were generated using \textsc{Pythia 8.160}~\cite{pythia} with the matrix elements for the underlying $2 \to 2$ processes calculated at LO using the \textsc{CT10 LO} PDFs~\cite{ct10} and matched to $\pt$-ordered parton showers. A set of tuned parameters called the \textsc{AU2CT10} tune~\cite{atlasPythia8Tunes} was used to model the underlying event (UE). The hadronisation follows the Lund string model~\cite{lund}.\\
A different set of samples were generated with \textsc{Herwig++ 2.5.2}~\cite{Herwigpp}, using the \textsc{LO CTEQ6L1} PDFs~\cite{pdf1} and the \textsc{CTEQ6L1-UE-EE-3} tune for the underlying event~\cite{herwigTunes}. \textsc{Herwig++} uses angle-ordered parton showers, a cluster hadronisation scheme and the underlying-event parameterisation is given by \textsc{Jimmy}~\cite{Jimmy}.\\
Additional samples are generated using \textsc{Sherpa 1.4.5}~\cite{sherpa}, which calculates matrix elements for $2\to N$ processes at LO, which are then convolved with the \textsc{CT10 LO} PDFs, and uses the \textsc{CKKW}~\cite{CKKW} method for the parton shower matching. These samples were generated with up to three hard-scattering partons in the final state.\\
In order to compensate for the steeply falling $\pt$ spectrum, MC samples are generated in seven intervals of the leading-jet transverse momentum. Each of these samples contain of the order of $6\times 10^6$ events for \textsc{Pythia8} and $1.4\times 10^6$ events for \textsc{Herwig++} and \textsc{Sherpa}.\\
All MC simulated samples described above are subject to a reweighting algorithm in order to match the average number of $pp$ interactions per bunch-crossing observed in the data. The average number of interactions per bunch-crossing amounts to $\langle\mu \rangle = 20.4$ in data, and to $\langle\mu\rangle = 22.0$ in the MC simulation.

\section{Data sample and jet calibration}
\label{sec:dataSample}

The data used were recorded in 2012 at $\sqrt{s}=8~\TeV$ and collected using a single-jet trigger. It requires at least one jet, reconstructed with the anti-$k_t$ algorithm~\cite{jets} with radius parameter $R=0.4$ as implemented in \textsc{FastJet}~\cite{fastjet}. The jet transverse energy measured by the trigger system is required to be greater than $360~\GeV$ at the trigger level. This trigger is fully efficient for values of the scalar sum of the calibrated transverse momenta of the two leading jets, $p_{\mathrm{T}1}+p_{\mathrm{T}2}$, denoted hereafter by $H_{\mathrm{T}2}$, above $730~\GeV$. This is the lowest unprescaled trigger for the 2012 data-taking period, and the integrated luminosity of the full data sample is $20.2~\ifb$.\\
Events are required to have at least one vertex, with two or more associated tracks with transverse momentum $\pt > 400~\MeV$. The vertex maximising $\sum \pt^2$, where the sum is performed over tracks, is chosen as the primary vertex.\\
In the analysis, jets are reconstructed with the same algorithm as used in the trigger, the anti-$k_t$ algorithm with radius parameter $R=0.4$. The input objects to the jet algorithm are topological clusters of energy deposits in the calorimeters~\cite{lampl}. The baseline calibration for these clusters corrects their energy using local hadronic calibration~\cite{lc1,lc2}. The four-momentum of an uncalibrated jet is defined as the sum of the four-momenta of its constituent clusters, which are considered massless. Thus, the resulting jets are massive. However, the effect of this mass is marginal for jets in the kinematic range considered in this paper, as the difference between transverse energy and transverse momentum is at the per-mille level for these jets.\\
The jet calibration procedure includes energy corrections for multiple $pp$ interactions in the same or neighbouring bunch crossings, known as ``pile-up'', as well as angular corrections to ensure that the jet originates from the primary vertex. Effects due to energy losses in inactive material, shower leakage, the magnetic field, as well as inefficiencies in energy clustering and jet reconstruction, are taken into account. This is done using an MC-based correction, in bins of $\eta$ and $\pt$, derived from the relation of the reconstructed jet energy to the energy of the corresponding particle-level jet, not including muons or non-interacting particles. In a final step, an in situ calibration corrects for residual differences in the jet response between the MC simulation and the data using $\pt$-balance techniques for dijet, $\gamma+$jet, $\Zboson$+jet and multi-jet final states. The total jet energy scale (JES) uncertainty is given by a set of independent sources, correlated in $\pt$. The uncertainty in the $\pt$ value of individual jets due to the JES increases from (1--4)$\%$ for $|\eta| < 1.8$ to $5\%$ for $1.8 <|\eta|<4.5$~\cite{JES}.\\
The selected jets must fulfill $\pt > 100~\GeV$ and $|\eta| < 2.5$. The two leading jets are further required to fulfil $H_{\mathrm{T}2} > 800~\GeV$. In addition, jets are required to satisfy quality criteria that reject beam-induced backgrounds (jet cleaning)~\cite{cleaning}.\\
The number of selected events in data is $6.2\times 10^6$, with an average jet multiplicity $\langle N_{\mathrm{jet}} \rangle = 2.3$. In order to study the dependence of the TEEC and ATEEC on the energy scale, and thus the running of the strong coupling, the data are further binned in $H_{\mathrm{T}2}$. The binning is chosen as a compromise between reaching the highest available energy scales while keeping a sufficient statistical precision in the TEEC distributions, and thus in the determination of $\alpha_{\mathrm{s}}$. Table~\ref{tab:htBins} summarises this choice, as well as the number of events in each energy bin and the average value of the chosen scale $Q = H_{\mathrm{T}2}/2$, obtained from detector-level data.
\begin{table}[H]
\begin{center}
\begin{tabular}{crc}
\hline
\T\B $H_{\mathrm{T}2}$ range [GeV] & Number of events & $\langle Q\rangle = \langle H_{\mathrm{T}2} \rangle / 2$ [GeV]\\
\hline
$[800 , 850]$  &  1\,809\,497  &  $412$ \\
$[850 , 900]$  &  1\,240\,059  &  $437$ \\
$[900 , 1000]$  &  1\,465\,814  &  $472$ \\
$[1000 , 1100]$  &  745\,898  &  $522$ \\
$[1100 , 1400]$  &  740\,563  &  $604$ \\
$[1400, 5000]$  &  192\,204  &  $810$\\
\hline
\end{tabular}
\end{center}
\caption{Summary of the $H_{\mathrm{T}2}$ bins used in the analysis. The table shows the number of events falling into each energy bin together with the value of the scale $Q$ at which the coupling constant $\alpha_{\mathrm{s}}$ is measured.}
\label{tab:htBins}
\end{table}

\section{Results at the detector level}
\label{sec:detResults}

The data sample described in Section~\ref{sec:dataSample} is used to measure the TEEC and ATEEC functions. In order to study the kinematical dependence of such observables, and thus the running of the strong coupling with the energy scale involved in the hard process, the binning introduced in Table~\ref{tab:htBins} is used. Figure~\ref{fig:teecDet} compares the TEEC and ATEEC distributions, measured in two of these bins, with the MC predictions from \textsc{Pythia8}, \textsc{Herwig++} and \textsc{Sherpa}.
\begin{figure}[h]
\centering
\includegraphics[width=7.cm,height=7.3cm]{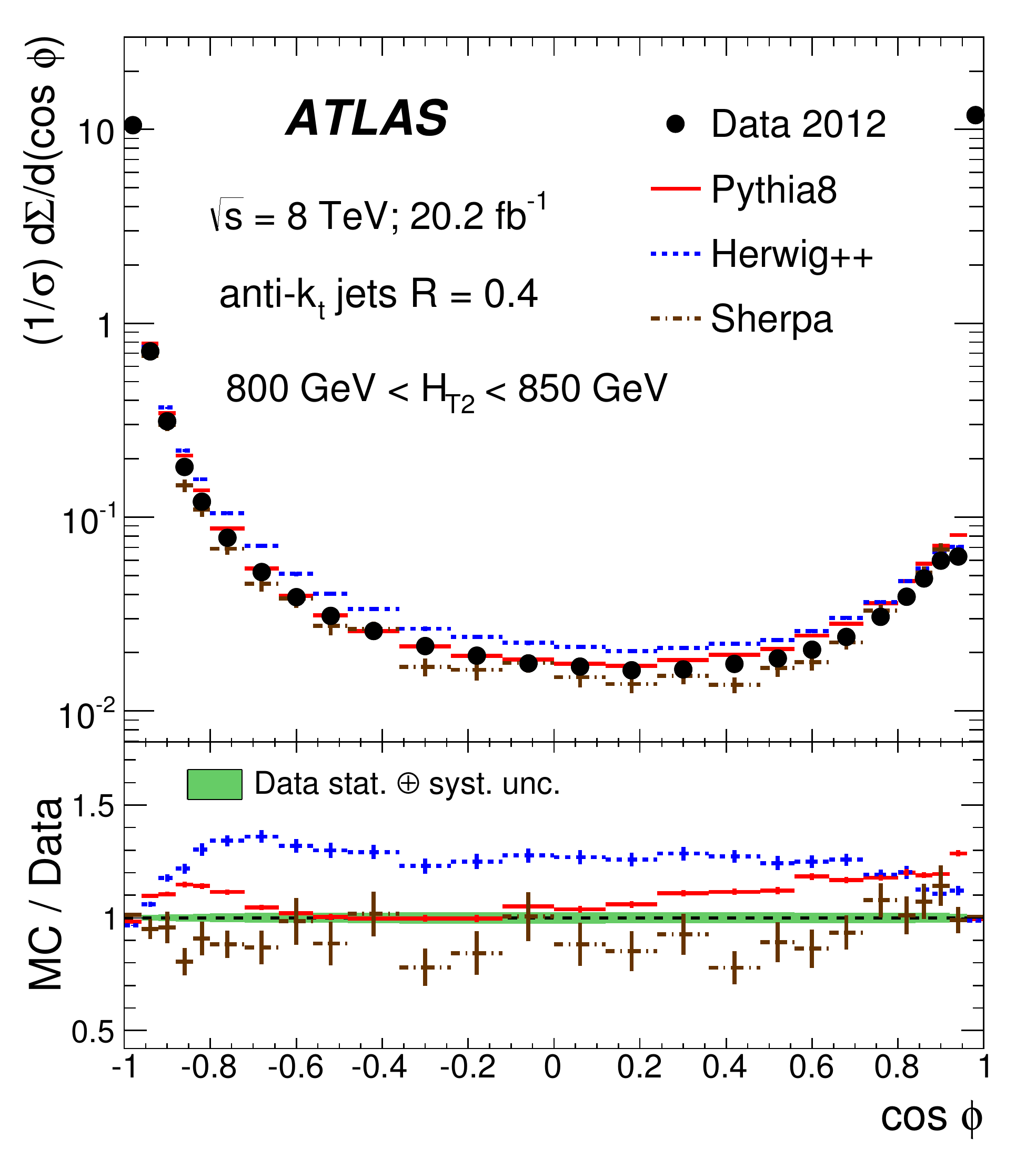}
\hspace{0.4cm}
\includegraphics[width=7.cm,height=7.3cm]{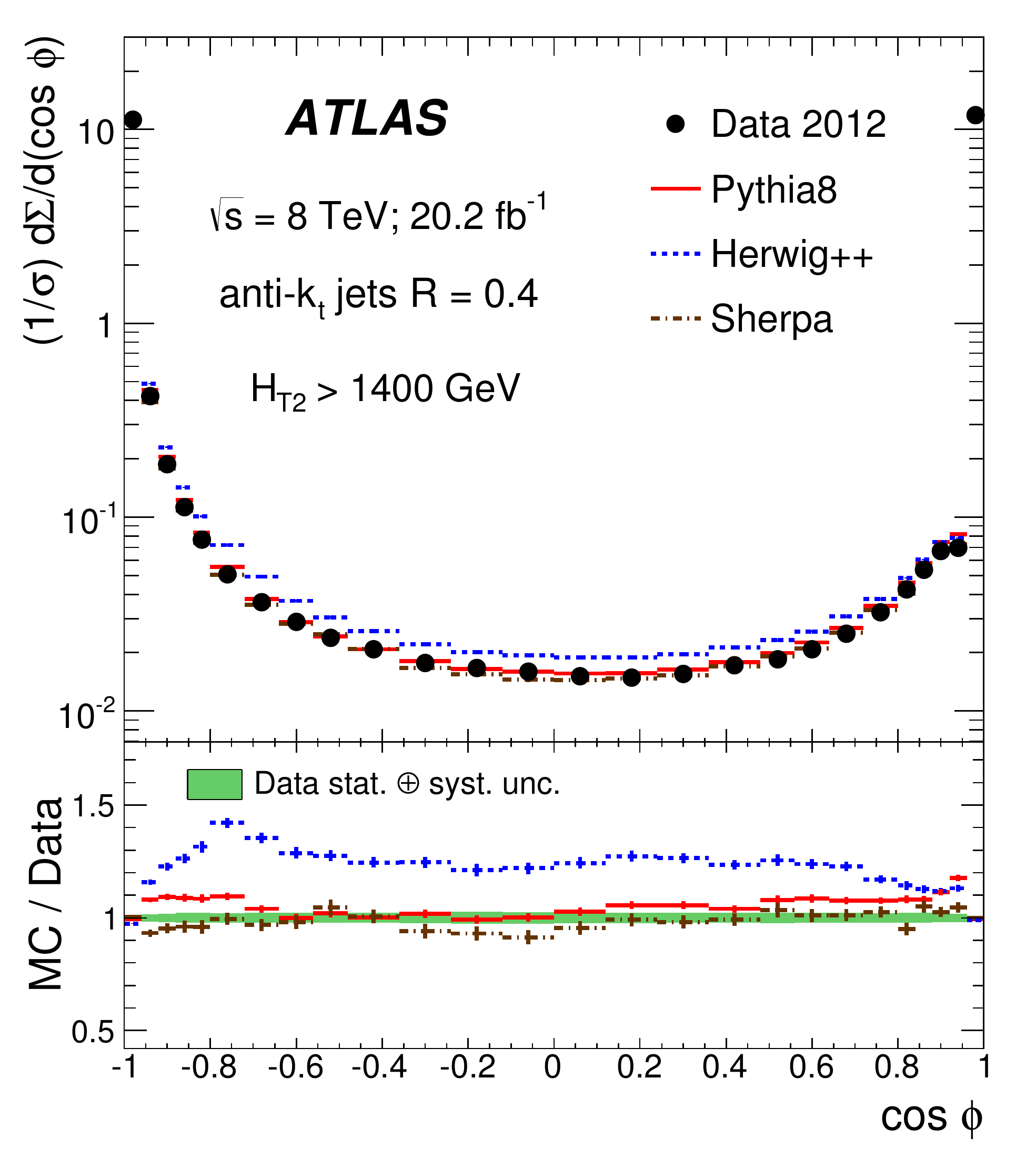}
\includegraphics[width=7.cm,height=7.3cm]{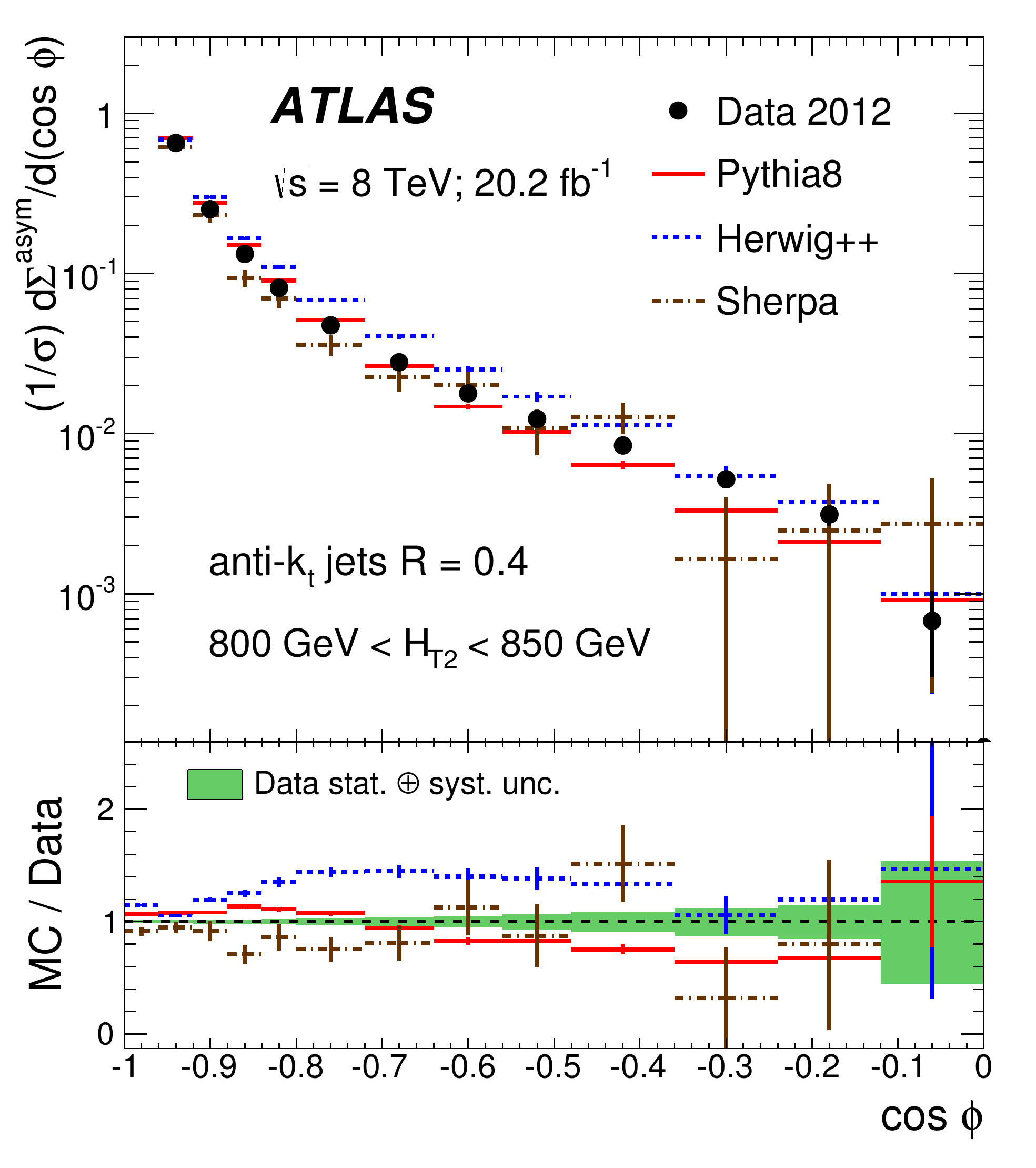}
\hspace{0.4cm}
\includegraphics[width=7.cm,height=7.3cm]{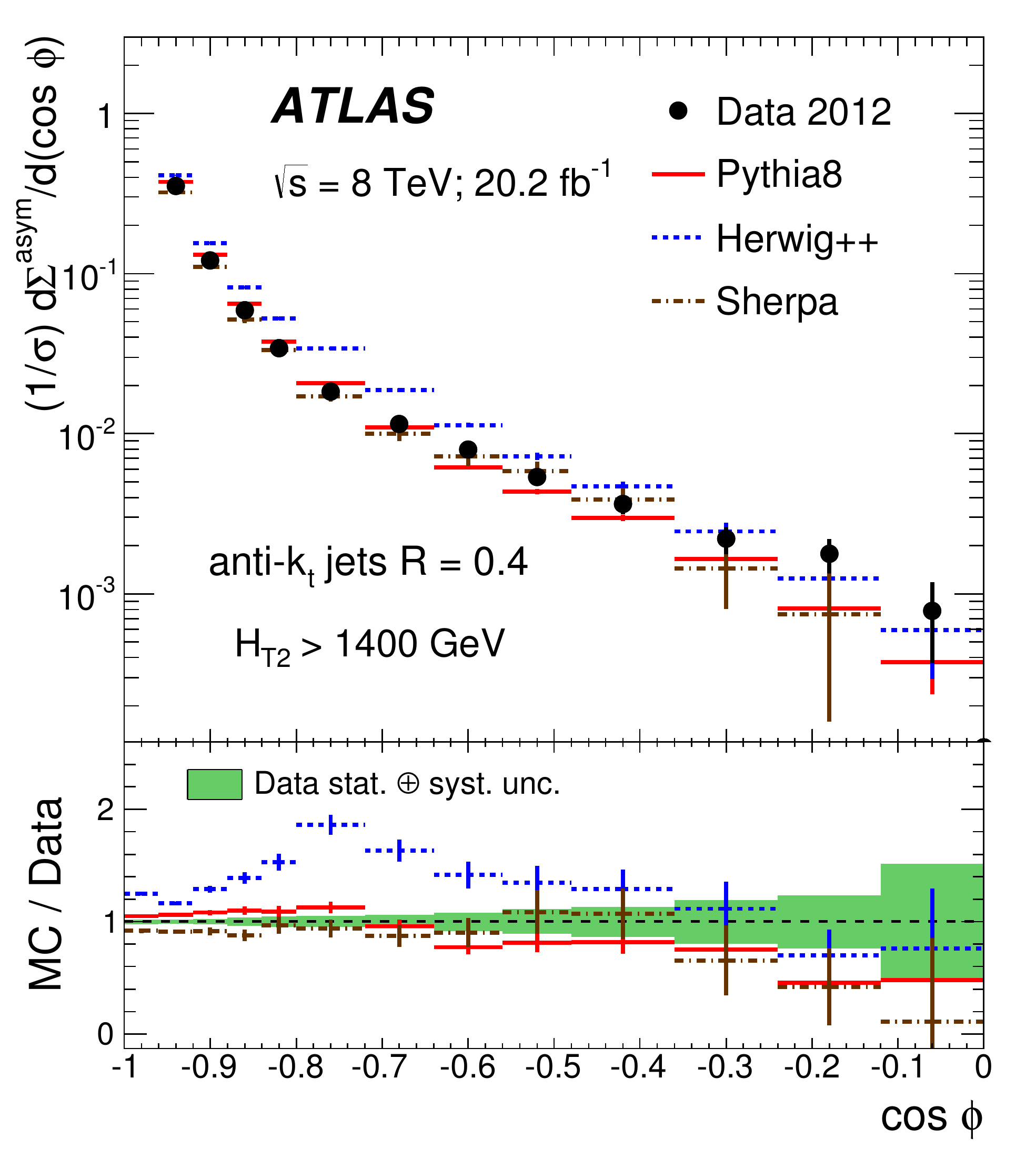}
\caption{Detector-level distributions for the TEEC (top) and ATEEC functions (bottom) for the first and the last $H_{\mathrm{T}2}$ intervals chosen in this analysis, together with MC predictions from \textsc{Pythia8}, \textsc{Herwig++} and \textsc{Sherpa}. The total uncertainty, including statistical and detector experimental sources, i.e. those not related to unfolding corrections, is also indicated using an error bar for the distributions and a green-shaded band for the ratios. The systematic uncertainties are discussed in Section~\ref{sec:systematics}.}
\label{fig:teecDet}
\end{figure}

The TEEC distributions show two peaks in the regions close to the kinematical endpoints $\cos\phi = \pm 1$. The first one, at $\cos\phi = -1$ is due to the back-to-back configuration in two-jet events, which dominate the sample, while the second peak at $\cos\phi = +1$ is due to the self-correlations of one jet with itself. These self-correlations are included in Equation~(\ref{eq:teecDef}) and are necessary for the correct normalisation of the TEEC functions. The central regions of the TEEC distributions shown in Figure~\ref{fig:teecDet} are dominated by gluon radiation, which is decorrelated from the main event axis as predicted by QCD and measured in Refs.~\cite{azimDec1, azimDec2}.\\
Among the MC predictions considered here, \textsc{Pythia8} and \textsc{Sherpa} are the ones which fit the data best, while \textsc{Herwig++} shows significant discrepancies with the data.

\section{Correction to particle level}
\label{sec:unfolding}

In order to allow comparison with particle-level MC predictions, as well as NLO theoretical predictions, the detector-level distributions presented in Section~\ref{sec:detResults} need to be corrected for detector effects. Particle-level jets are reconstructed in the MC samples using the anti-$k_t$ algorithm with $R = 0.4$, applied to final-state particles with an average lifetime $\tau > 10^{-11}~\mathrm{s}$, including muons and neutrinos. The kinematical requirements for particle-level jets are the same as for the definition of TEEC / ATEEC at the detector level.\\
In the data, an unfolding procedure is used which relies on an iterative Bayesian unfolding method~\cite{bayes} as implemented in the \textsc{RooUnfold} program~\cite{roounfold}. The method makes use of a transfer matrix for each distribution, which takes into account any inefficiencies in the detector, as well as its finite resolution. The \textsc{Pythia8} MC sample is used to determine the transfer matrices from the particle-level to detector-level TEEC distributions. Pairs of jets not entering the transfer matrices are accounted for using inefficiency correction factors.\\
The excellent azimuthal resolution of the ATLAS detector, together with the reduction of the energy scale and resolution effects by the weighting procedure involved in the definition of the TEEC function, are reflected in the fact that the transfer matrices have very small off-diagonal terms (smaller than 10\%), leading to very small migrations between bins.\\
The statistical uncertainty is propagated through the unfolding procedure by using pseudo-experiments. A set of $10^3$ replicas is considered for each measured distribution by applying a Poisson-distributed fluctuation around the nominal measured distribution. Each of these replicas is unfolded using a fluctuated version of the transfer matrix, which produces the corresponding set of $10^3$ replicas of the unfolded spectra. The statistical uncertainty is defined as the standard deviation of all replicas.

\section{Systematic uncertainties}
\label{sec:systematics}

The dominant sources are those associated with the MC model used in the unfolding procedure and the JES uncertainty in the jet calibration procedure.
\begin{itemize}
\item{} \textbf{Jet Energy Scale:} The uncertainty in the jet calibration procedure~\cite{JES} is propagated to the TEEC by varying each jet energy and transverse momentum by one standard deviation of each of the 67 nuisance parameters of the JES uncertainty, which depend on both the jet transverse momentum and pseudorapidity. The total JES uncertainty is evaluated as the sum in quadrature of all nuisance parameters, and amounts to 2\%.
\end{itemize}

\begin{figure}[H]
\centering
\includegraphics[width=8.cm,height=5.8cm]{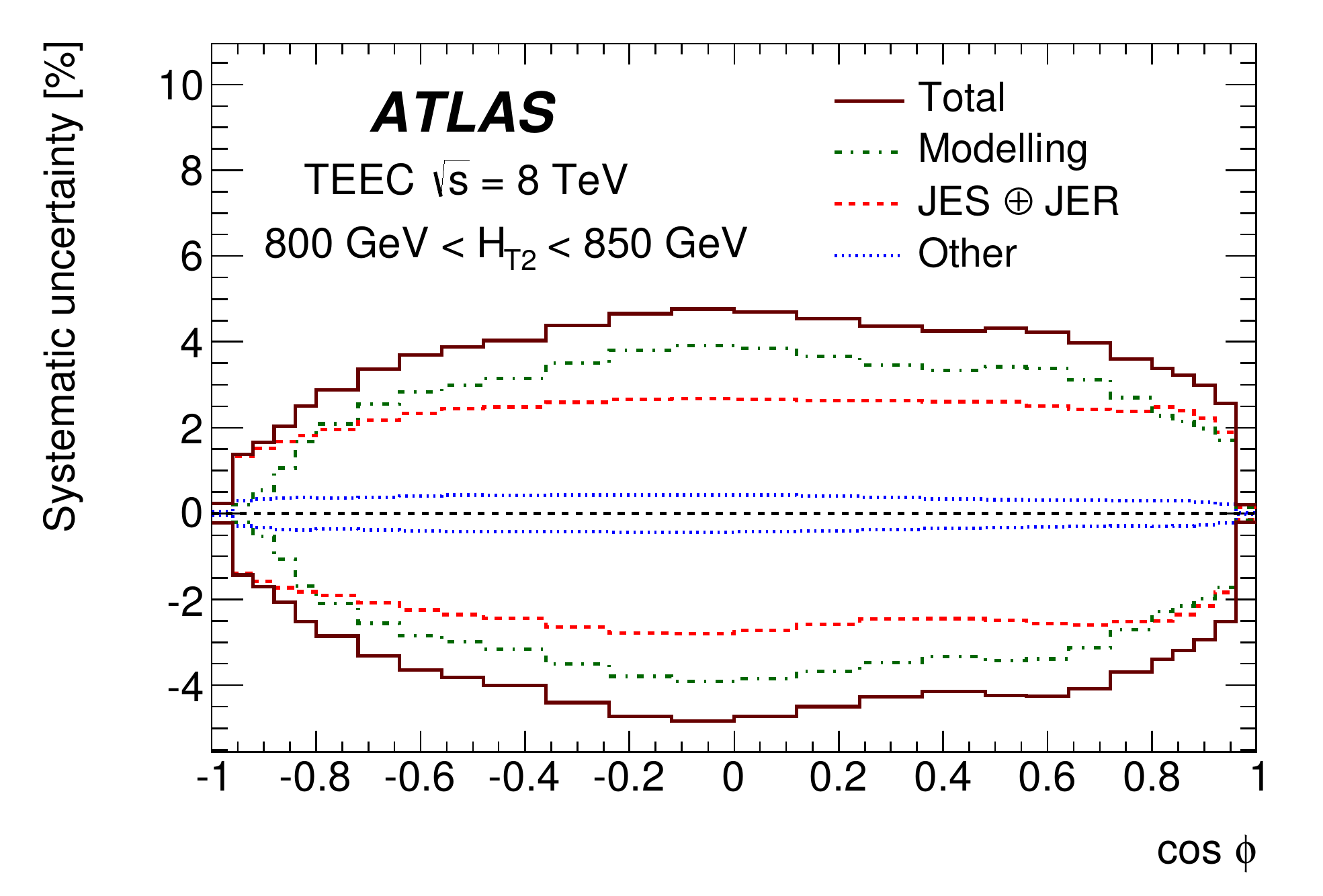}
\includegraphics[width=8.cm,height=5.8cm]{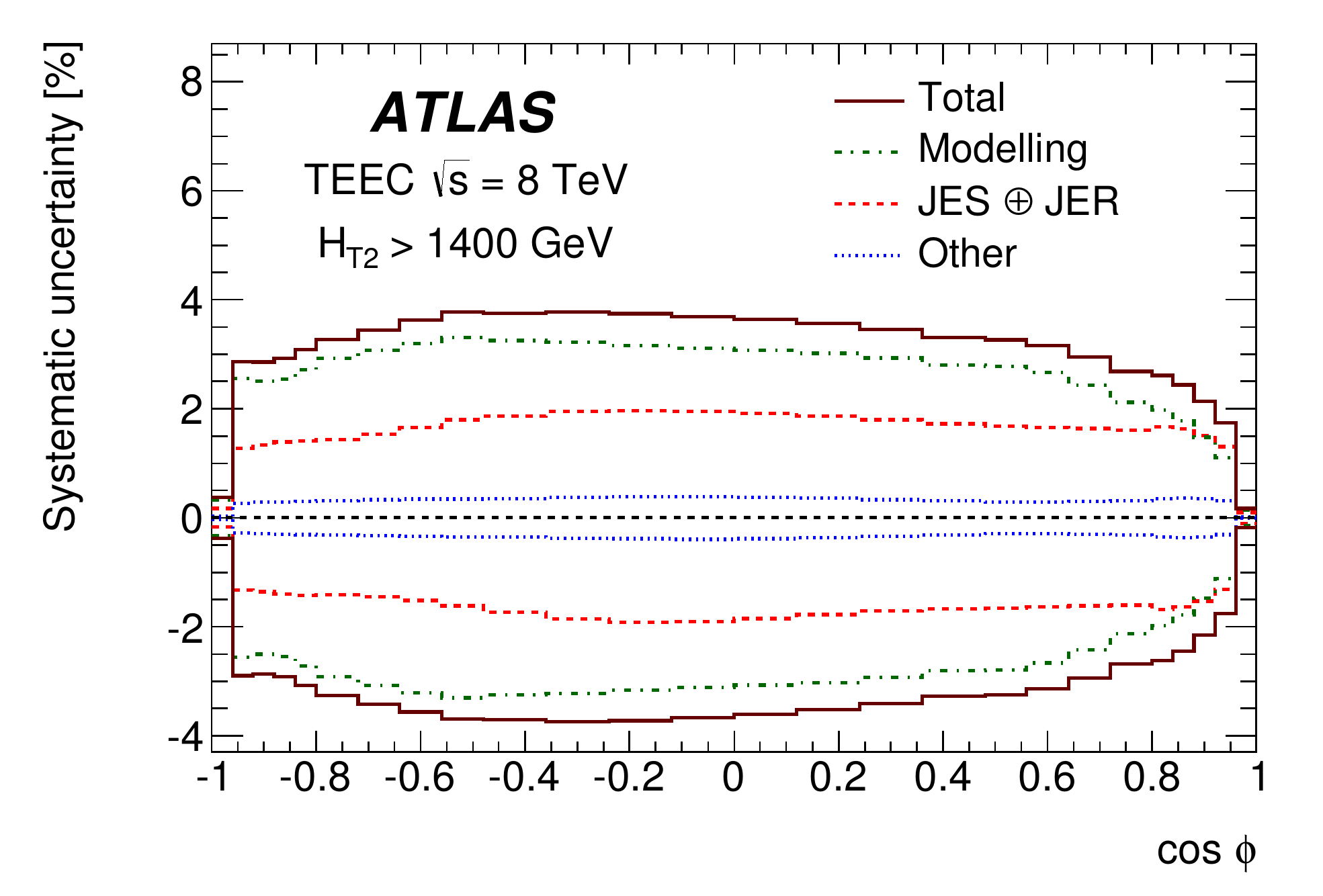}
\includegraphics[width=8.cm,height=5.8cm]{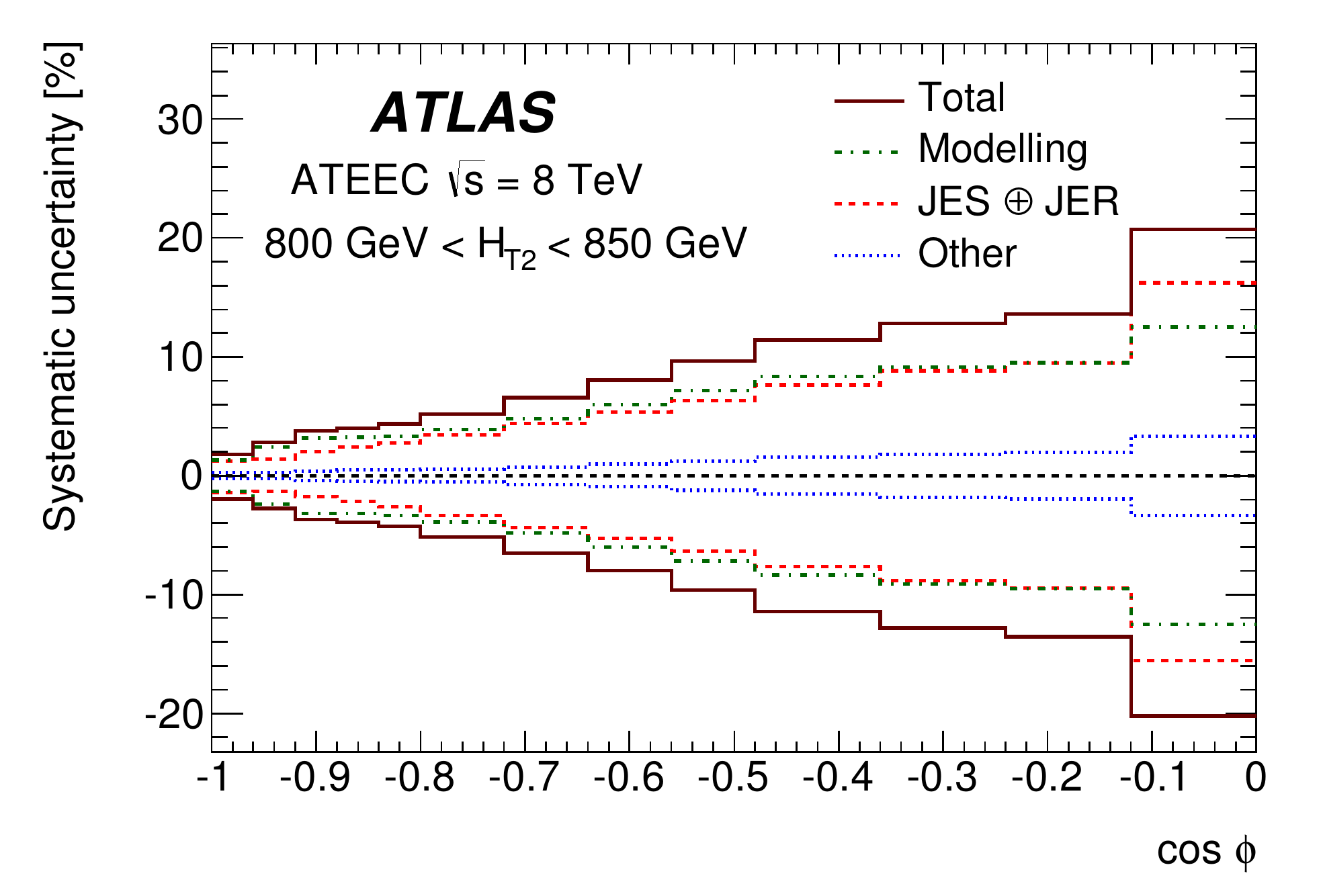}
\includegraphics[width=8.cm,height=5.8cm]{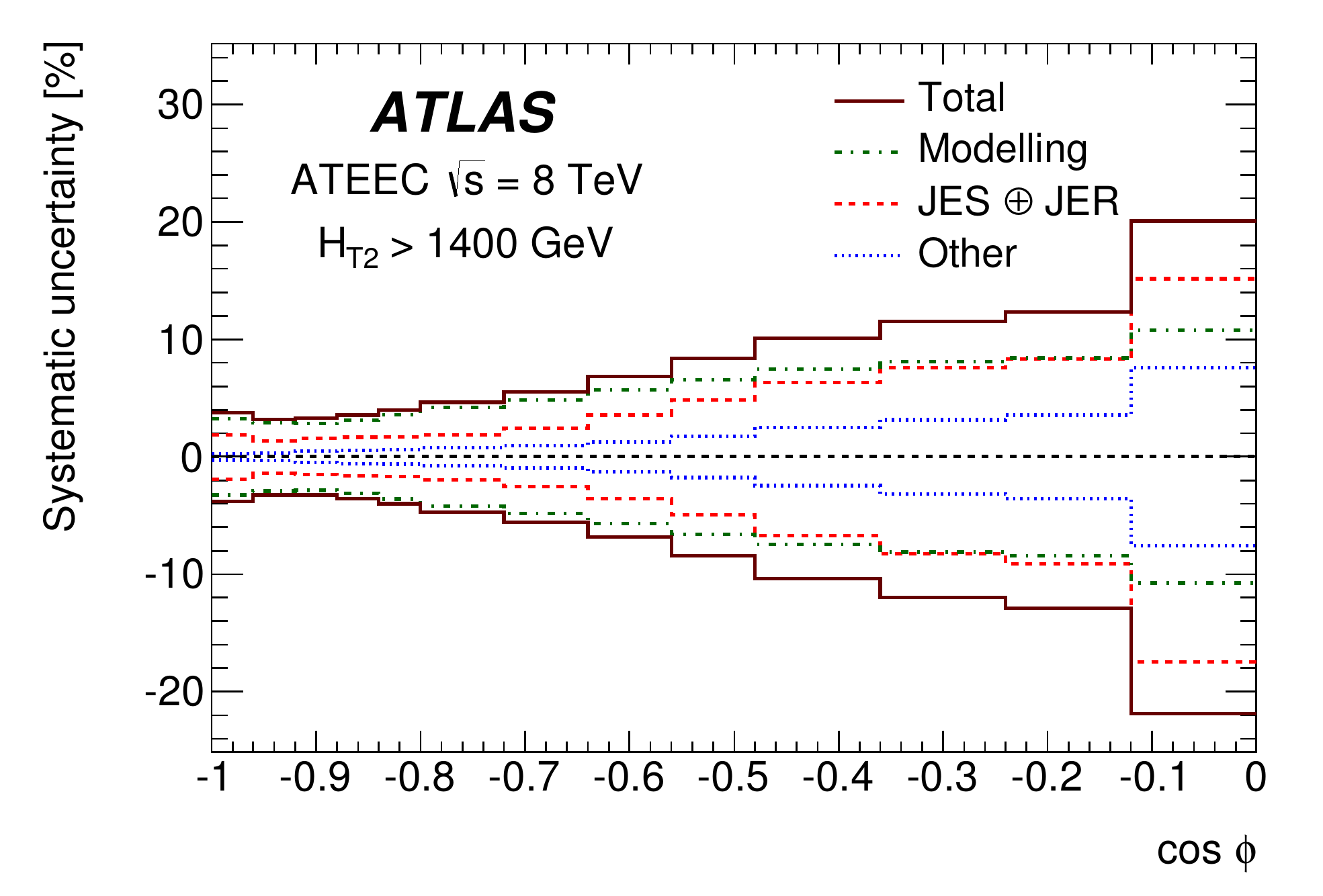}
\caption{Systematic uncertainties in the measured TEEC (top) and ATEEC distributions (bottom) for the first and the last bins in $H_{\mathrm{T}2}$. The total uncertainty is below 5\% in all bins of the TEEC distributions.}
\label{fig:sysErrorsTEEC}
\end{figure}

\newpage

\begin{itemize}
\item{} \textbf{Jet Energy Resolution:} The effect on the TEEC function of the jet energy resolution uncertainty~\cite{JER} is estimated by smearing the energy and transverse momentum by a smearing factor depending on both $\pt$ and $\eta$. This amounts to approximately 1\% in the TEEC distributions.

\item{} \textbf{Monte Carlo modelling:} The modelling uncertainty is estimated by performing the unfolding procedure described in Section~\ref{sec:unfolding} with different MC approaches. The difference between the unfolded distributions using \textsc{Pythia} and \textsc{Herwig++} defines the envelope of the uncertainty. This was cross-checked using the difference between \textsc{Pythia} and \textsc{Sherpa}, leading to similar results. This is the dominant experimental uncertainty for this measurement, being always below 5\% for the TEEC distributions, and being larger for low $H_{\mathrm{T}2}$.

\item{} \textbf{Unfolding:} The mismodelling of the data made by the MC simulation is accounted for as an additional source of uncertainty. This is assessed by reweighting the transfer matrices so that the level of agreement between the detector-level projection and the data is enhanced. The modified detector-level distributions are then unfolded using the method described in Section~\ref{sec:unfolding}. The difference between the modified particle-level distribution and the nominal one is then taken as the uncertainty. This uncertainty is smaller than 0.5\% for the full $\cos\phi$ range for all bins in $H_{\mathrm{T}2}$. The impact of this uncertainty on the TEEC function is below 1\%.

\item{} \textbf{Jet Angular Resolution:} The uncertainty in the jet angular resolution is propagated to the TEEC measurements by smearing the azimuthal coordinate $\varphi$ of each jet by 10\% of the resolution in the MC simulation. This is motivated by the track-to-cluster matching studies done in Ref.~\cite{azimDec1}. This impacts the TEEC measurement at the level of 0.5\%.

\item{} \textbf{Jet cleaning:} The modelling of the efficiency of the jet-cleaning cuts is considered as an additional source of experimental uncertainty. This is studied by tightening the jet cleaning-requirements in both data and MC simulation, and considering the double ratio between them. The differences are below 0.5\%.

\end{itemize}
In order to mitigate statistical fluctuations, the resulting systematic uncertainties are smoothed using a Gaussian kernel algorithm. The impact of these systematic uncertainties is summarised in Figure~\ref{fig:sysErrorsTEEC}, where the relative errors are shown for the TEEC and ATEEC distributions for each $H_{\mathrm{T}2}$ bin considered.

\section{Experimental results}
\label{sec:unfResults}

The results of the unfolding are compared with particle-level MC predictions, including the estimated systematic uncertainties. Figure~\ref{fig:teecPart} shows this comparison for the TEEC, while the ATEEC results are shown in Figure~\ref{fig:ateecPart}. The level of agreement seen here between data and MC simulation is similar to that at detector level. \textsc{Pythia} and \textsc{Sherpa} broadly describe the data, while the \textsc{Herwig++} description is disfavoured. 

\begin{figure}[H]
\centering
\includegraphics[width=6.8cm,height=7.cm]{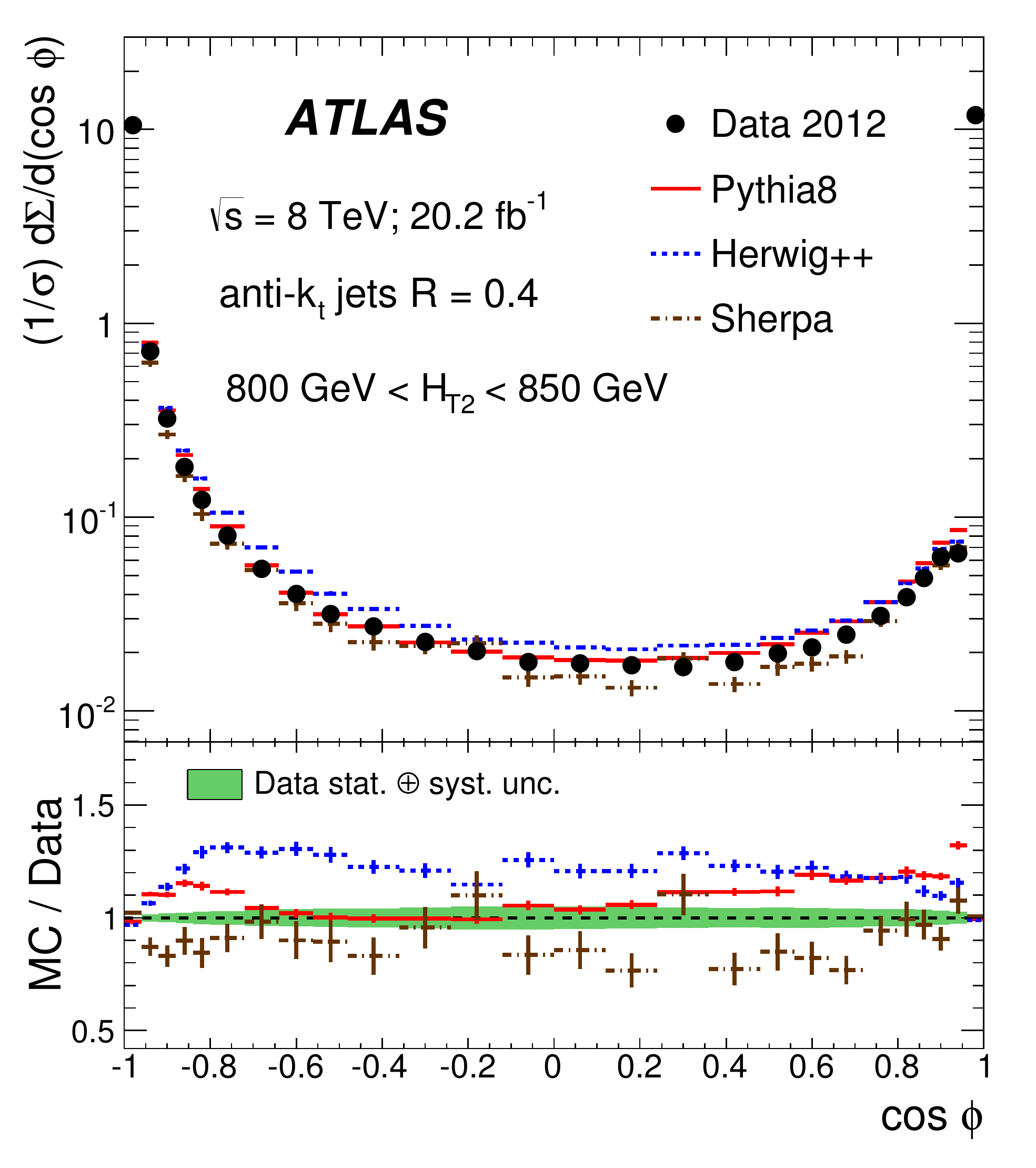}
\hspace{0.4cm}
\includegraphics[width=6.8cm,height=7.cm]{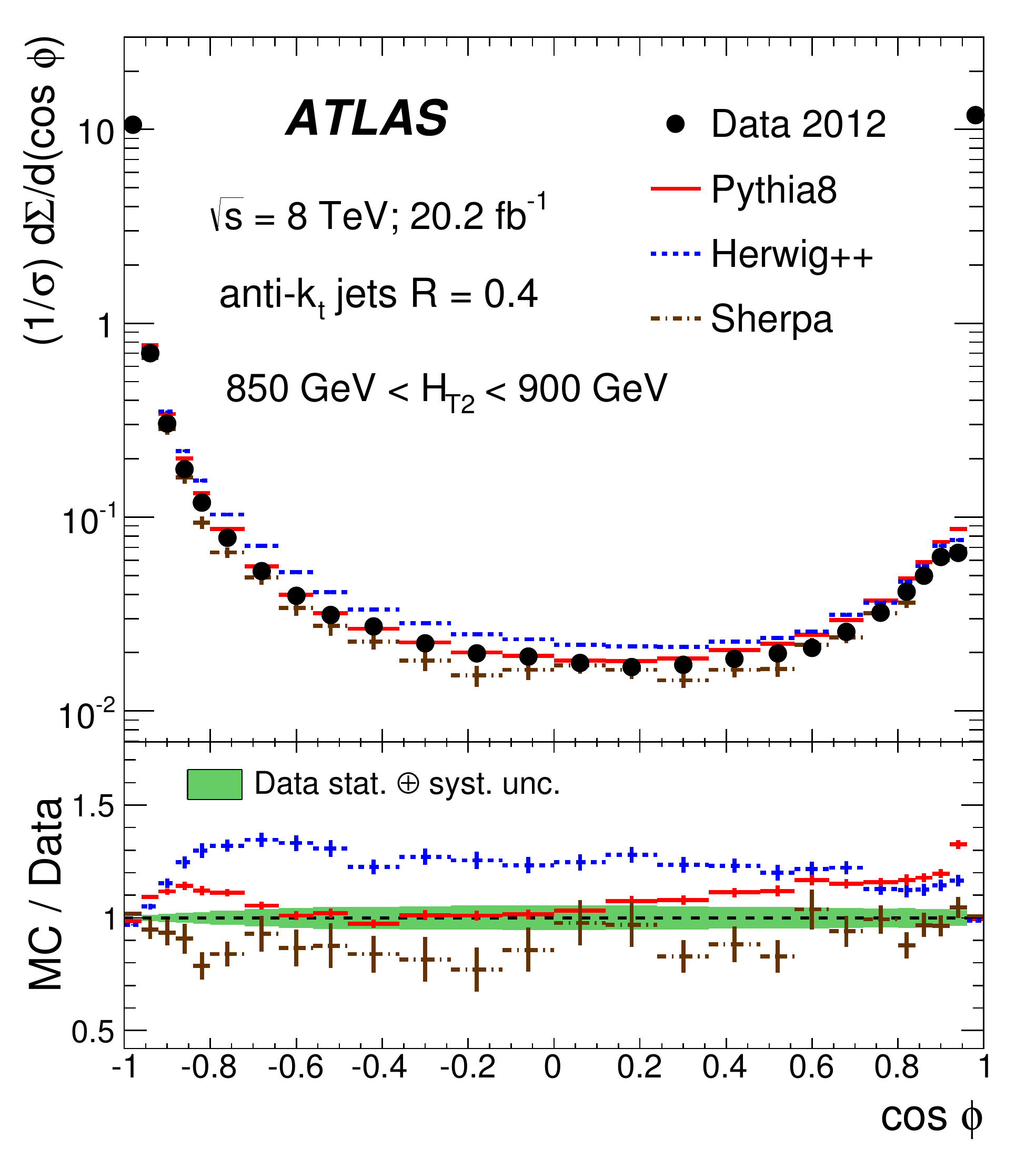}
\includegraphics[width=6.8cm,height=7.cm]{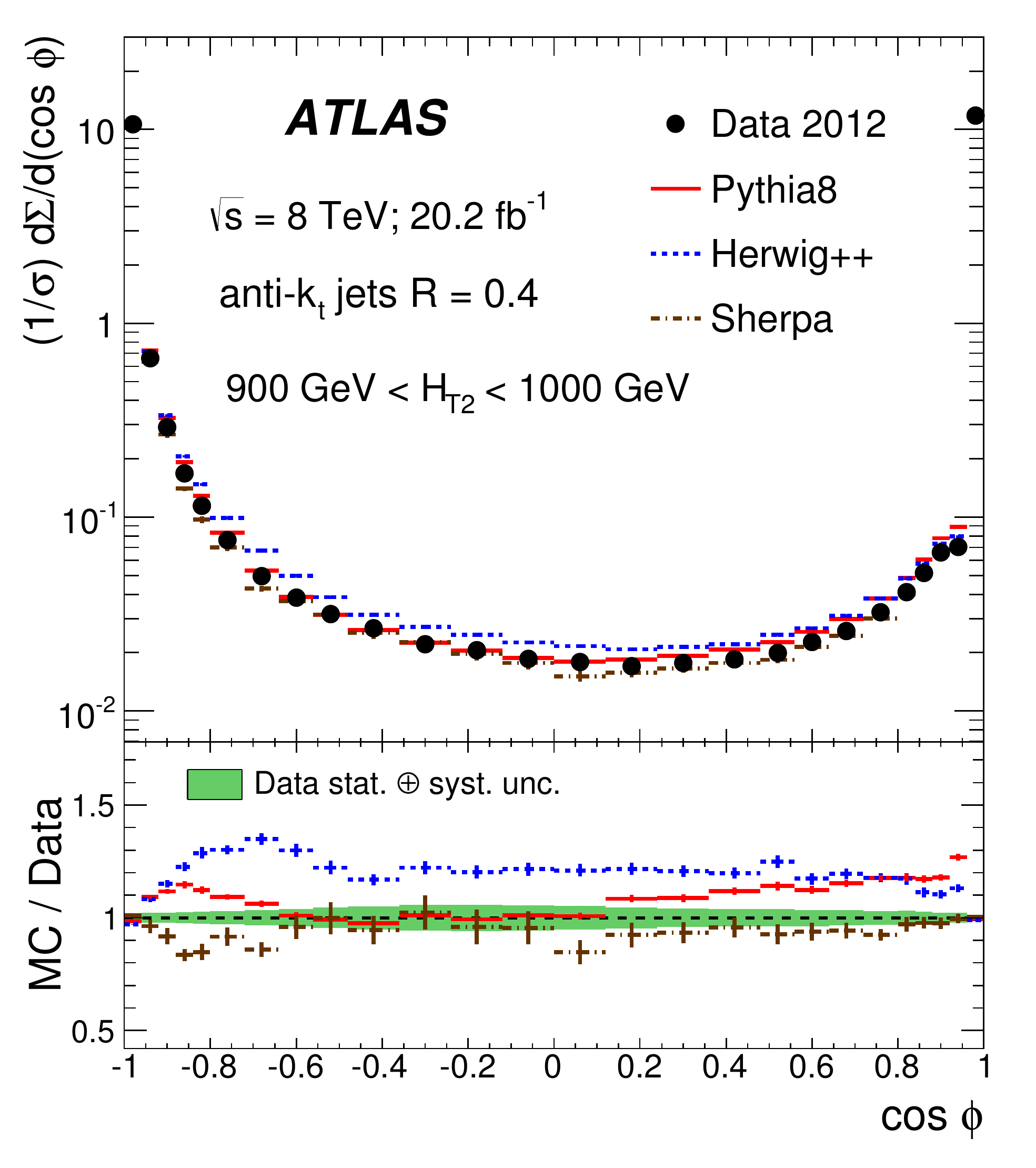}
\hspace{0.4cm}
\includegraphics[width=6.8cm,height=7.cm]{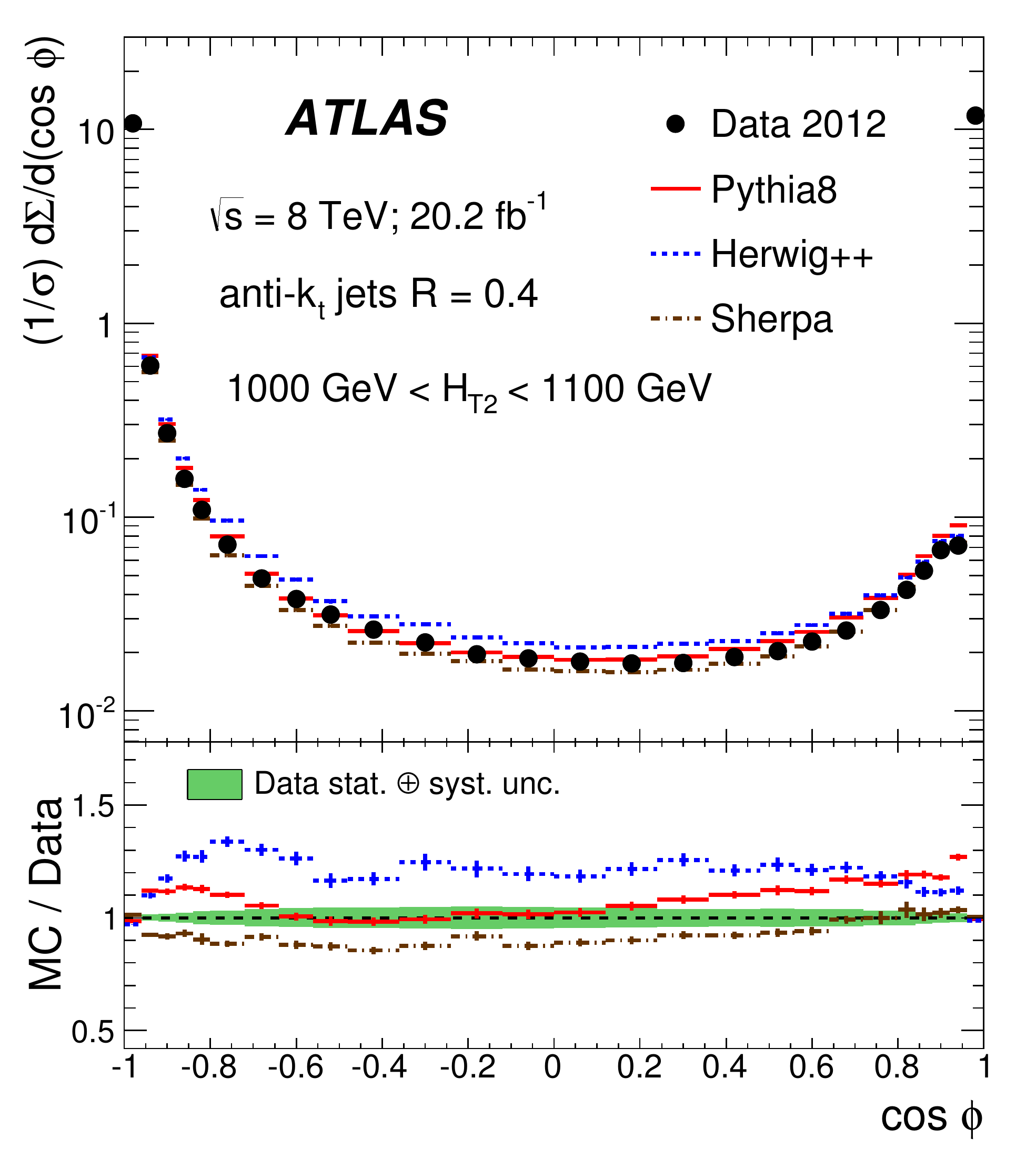}
\includegraphics[width=6.8cm,height=7.cm]{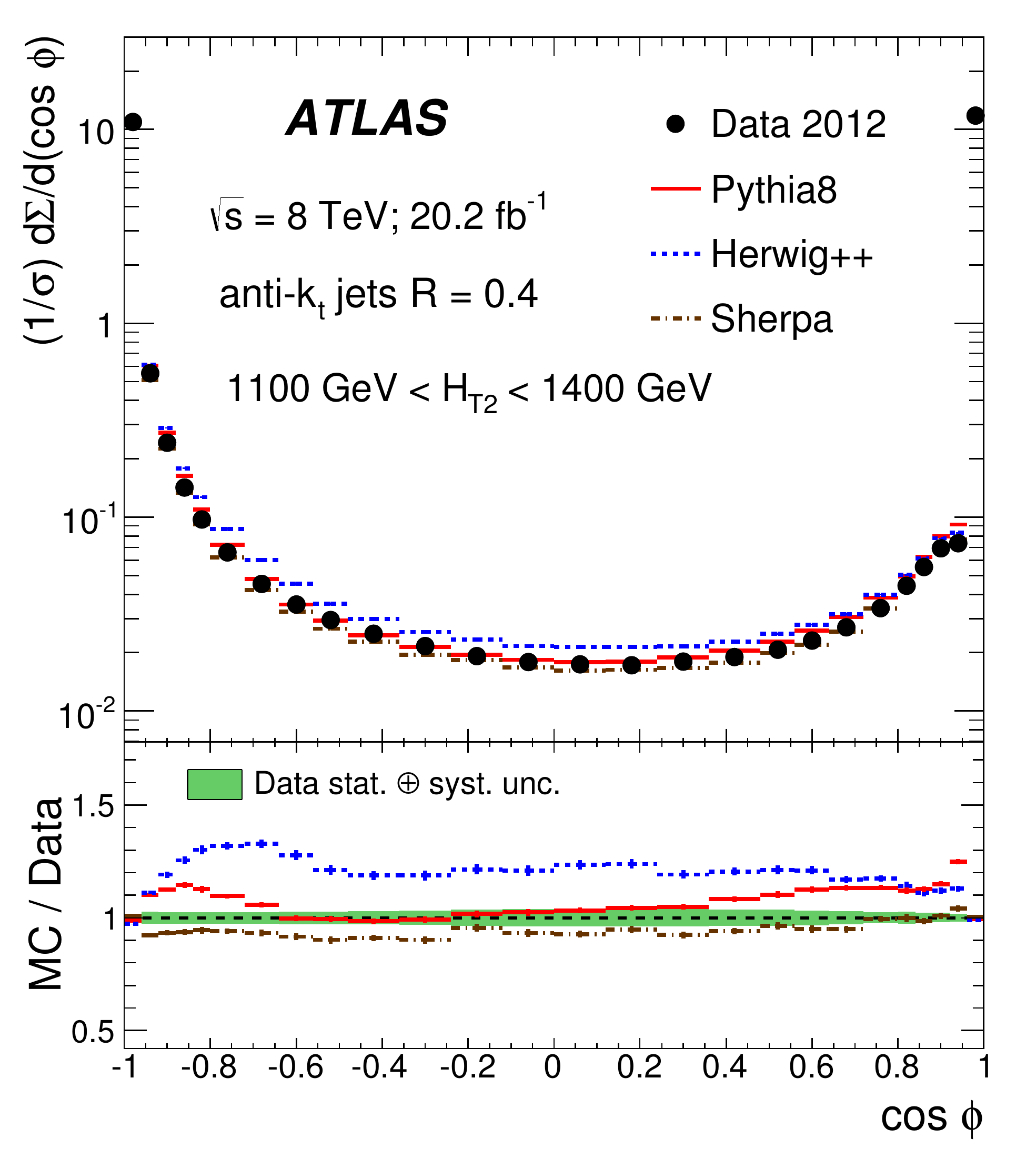}
\hspace{0.4cm}
\includegraphics[width=6.8cm,height=7.cm]{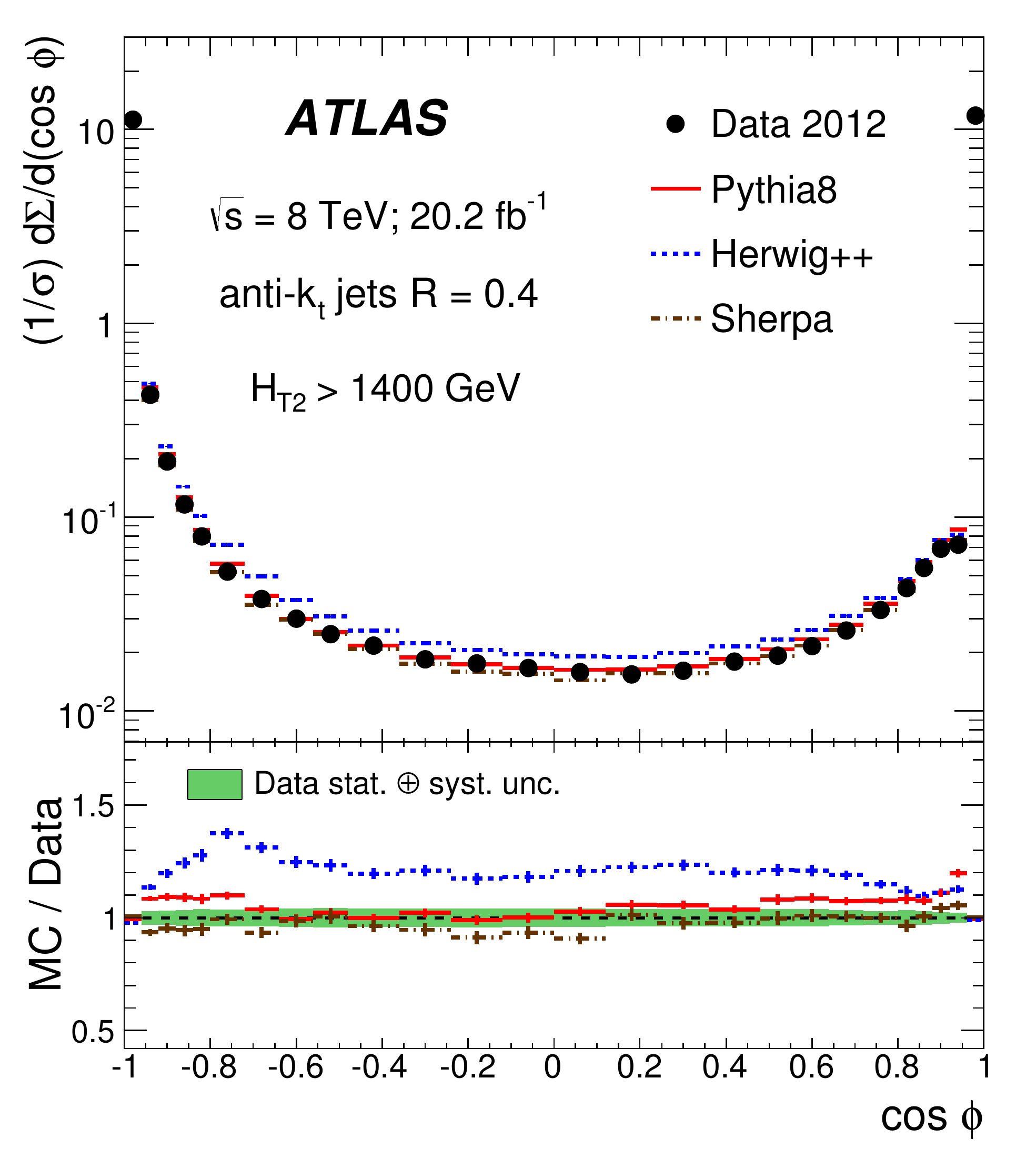}
\caption{Particle-level distributions for the TEEC functions in each of the $H_{\mathrm{T}2}$ intervals chosen in this analysis, together with MC predictions from \textsc{Pythia8}, \textsc{Herwig++} and \textsc{Sherpa}. The total uncertainty, including statistical and other experimental sources is also indicated using an error bar for the distributions and a green-shaded band for the ratios.}
\label{fig:teecPart}
\end{figure}

\begin{figure}[H]
\centering
\includegraphics[width=6.8cm,height=7.cm]{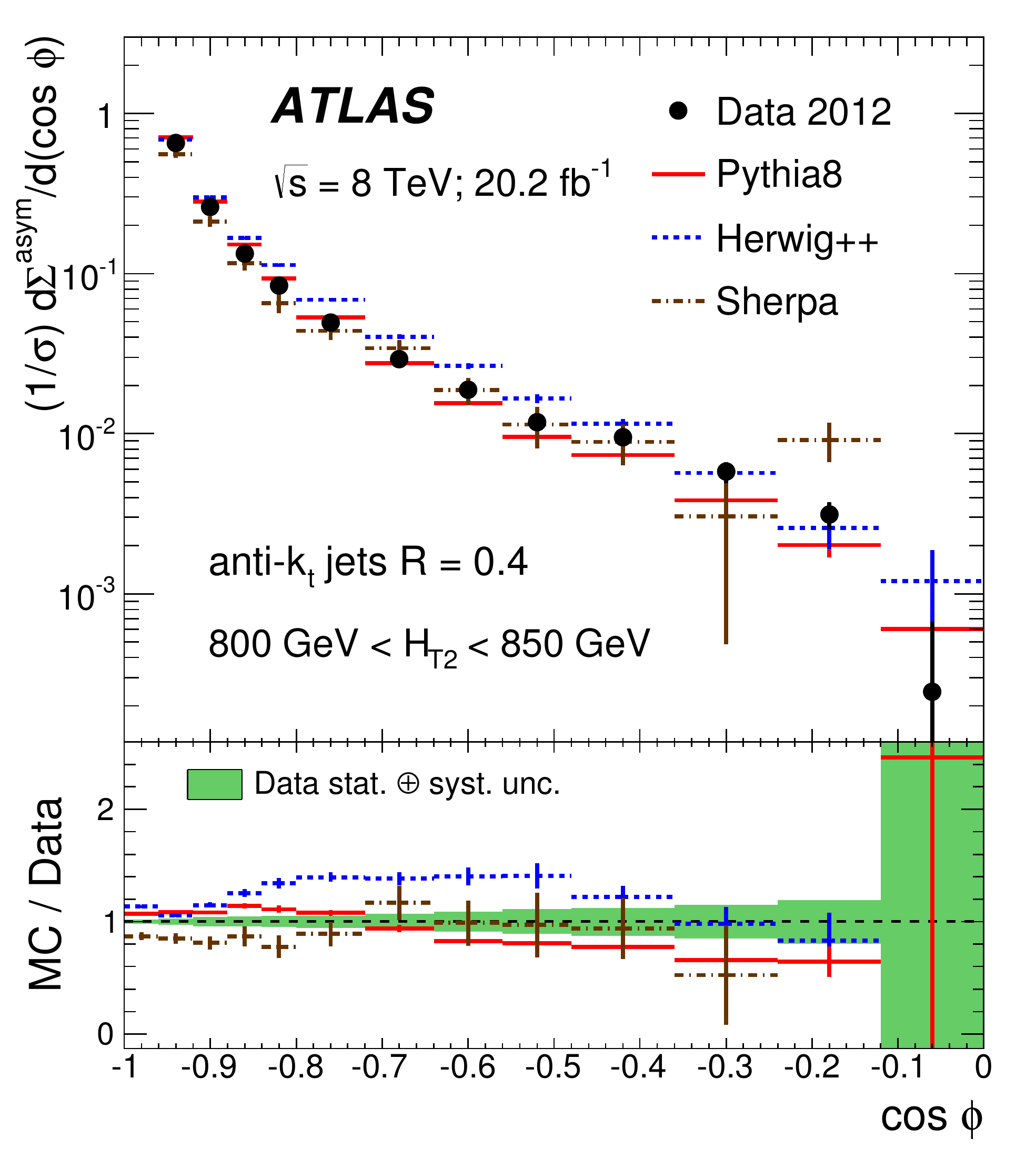}
\hspace{0.4cm}
\includegraphics[width=6.8cm,height=7.cm]{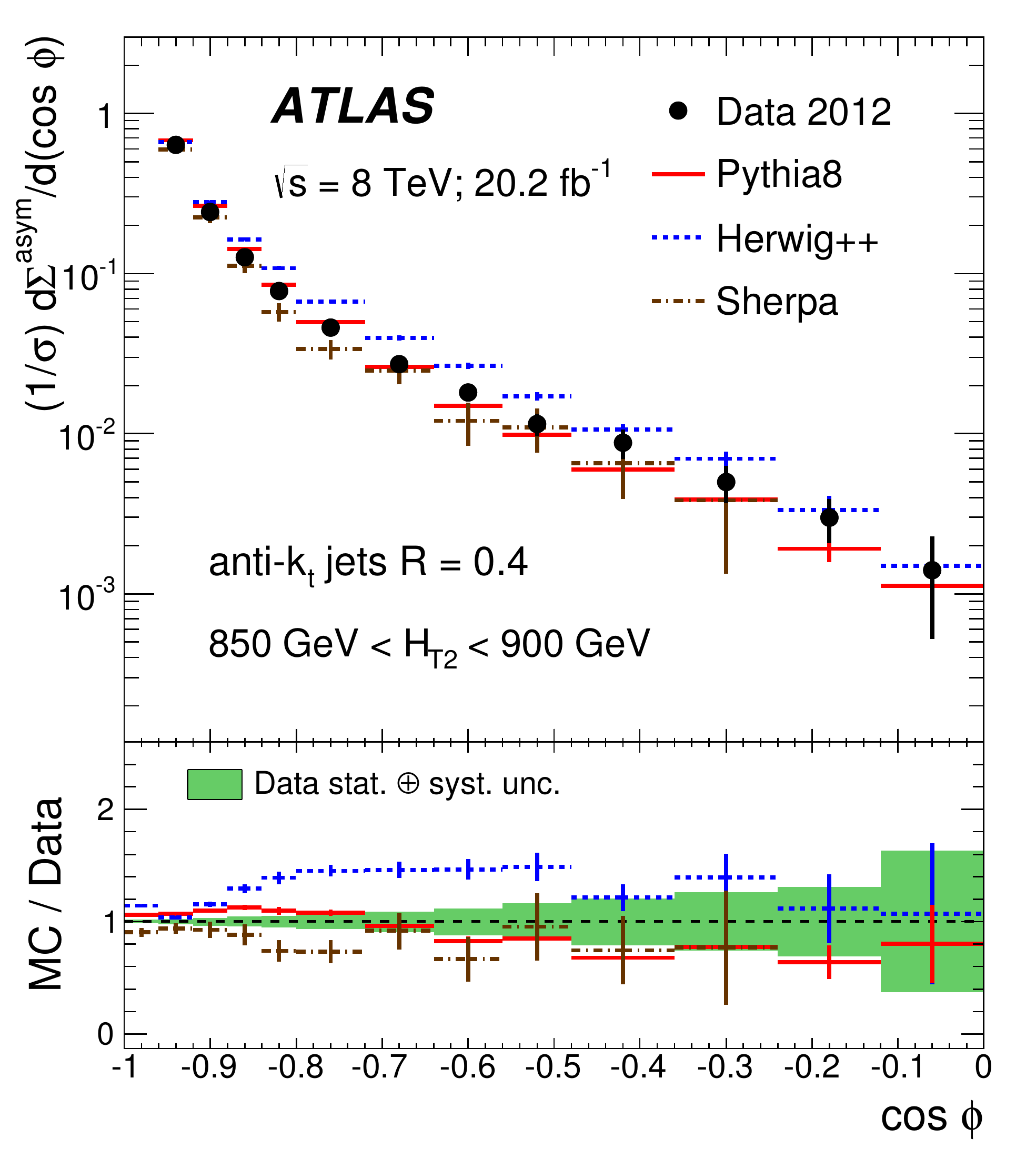}
\includegraphics[width=6.8cm,height=7.cm]{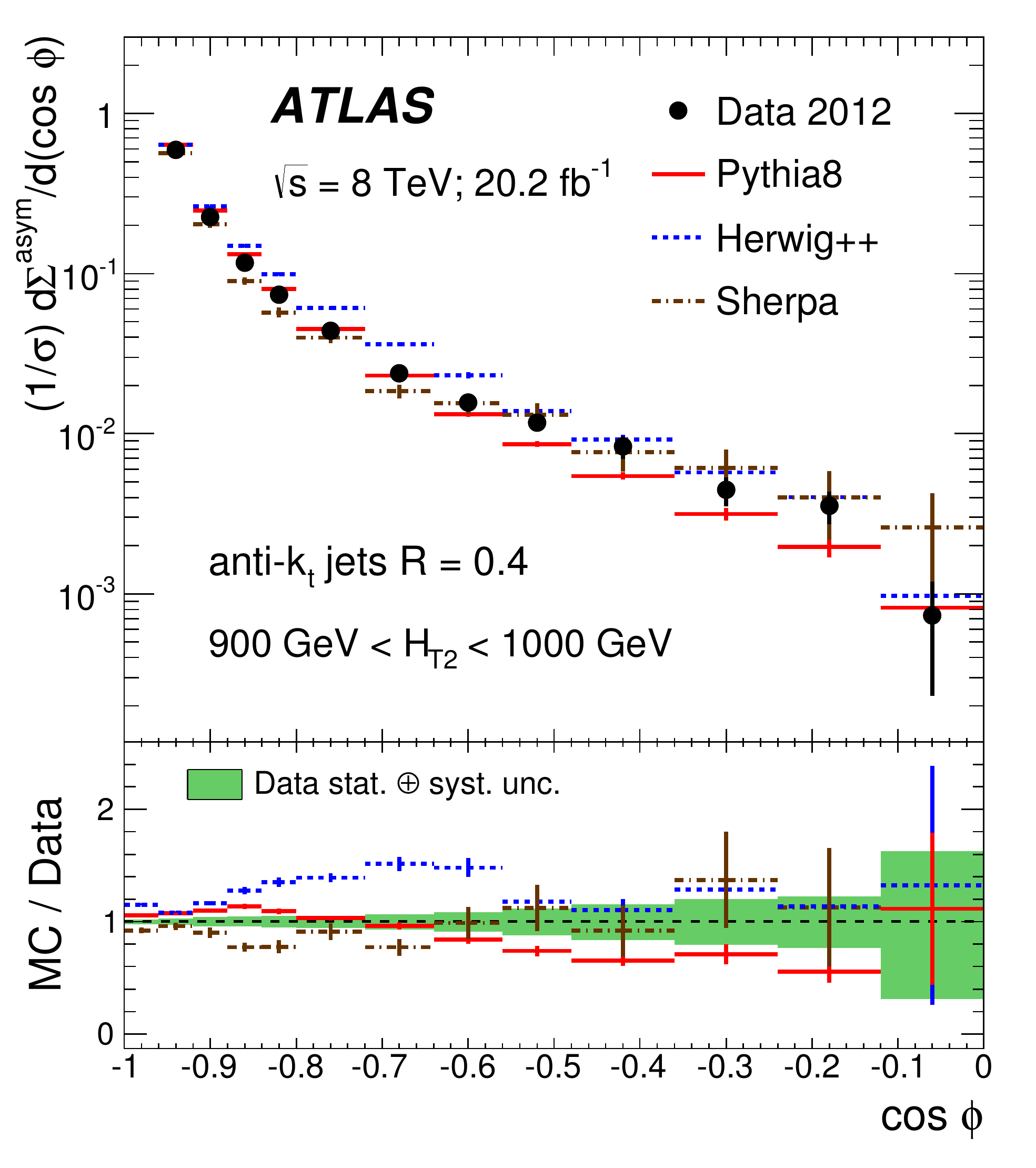}
\hspace{0.4cm}
\includegraphics[width=6.8cm,height=7.cm]{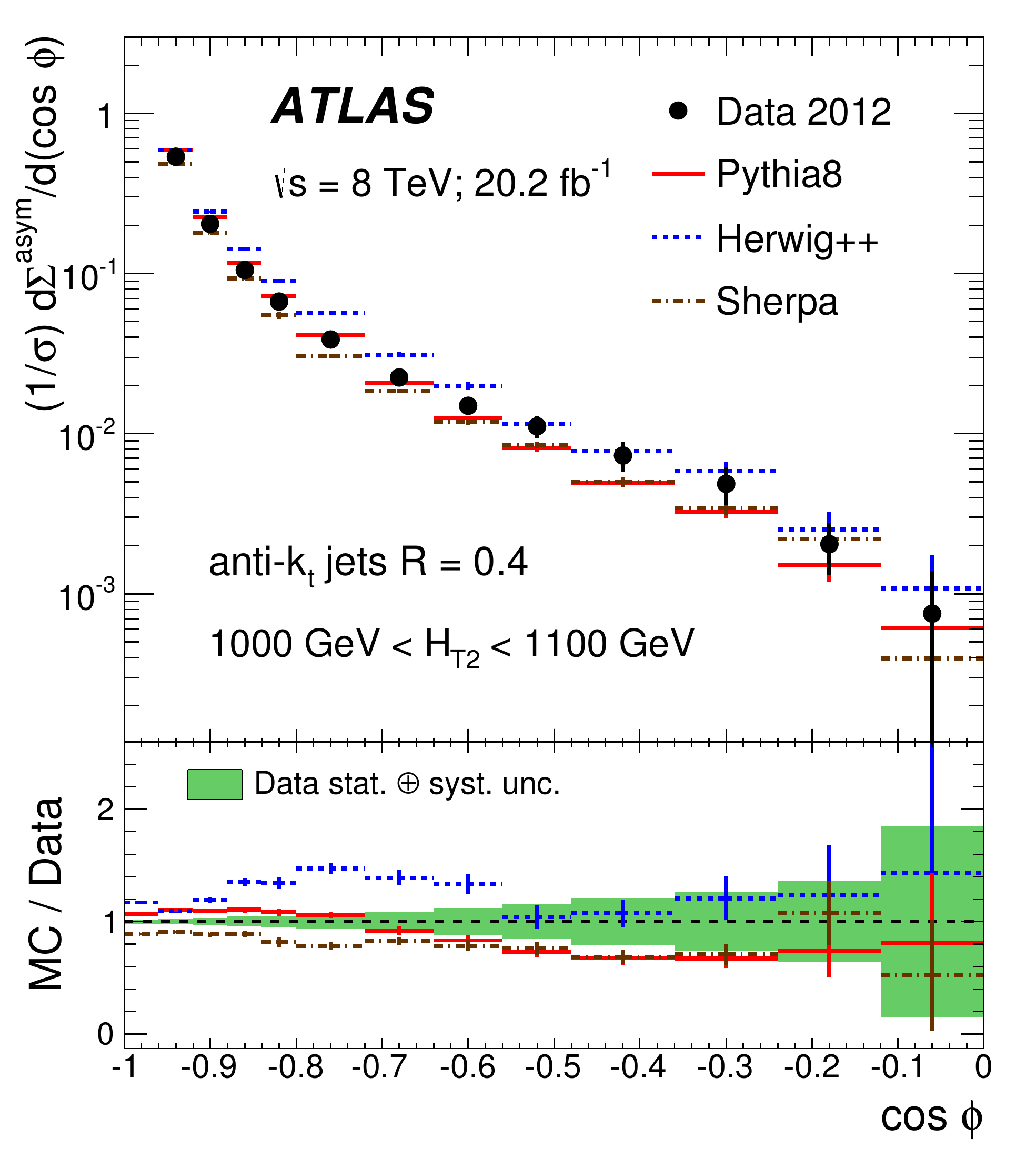}
\includegraphics[width=6.8cm,height=7.cm]{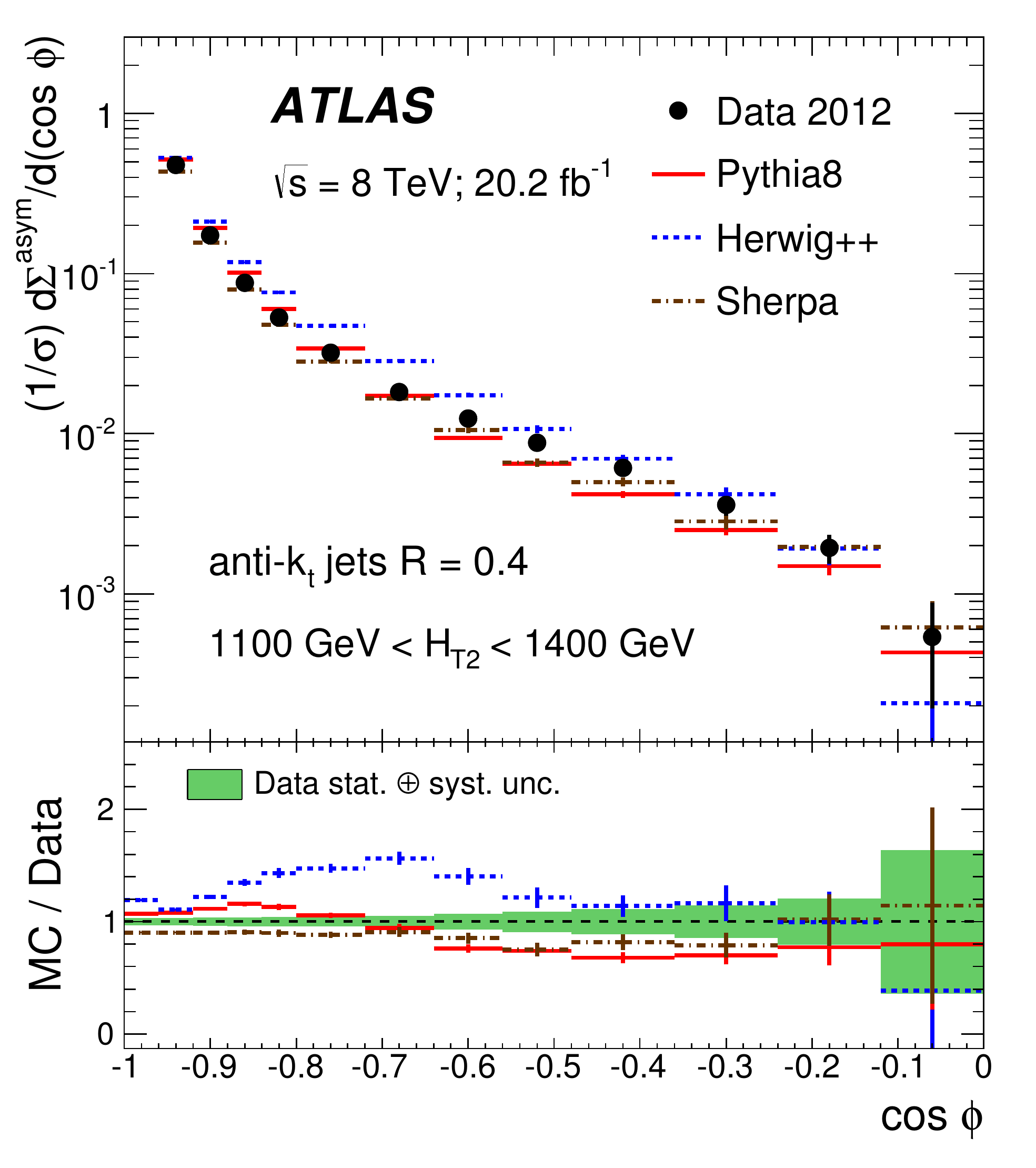}
\hspace{0.4cm}
\includegraphics[width=6.8cm,height=7.cm]{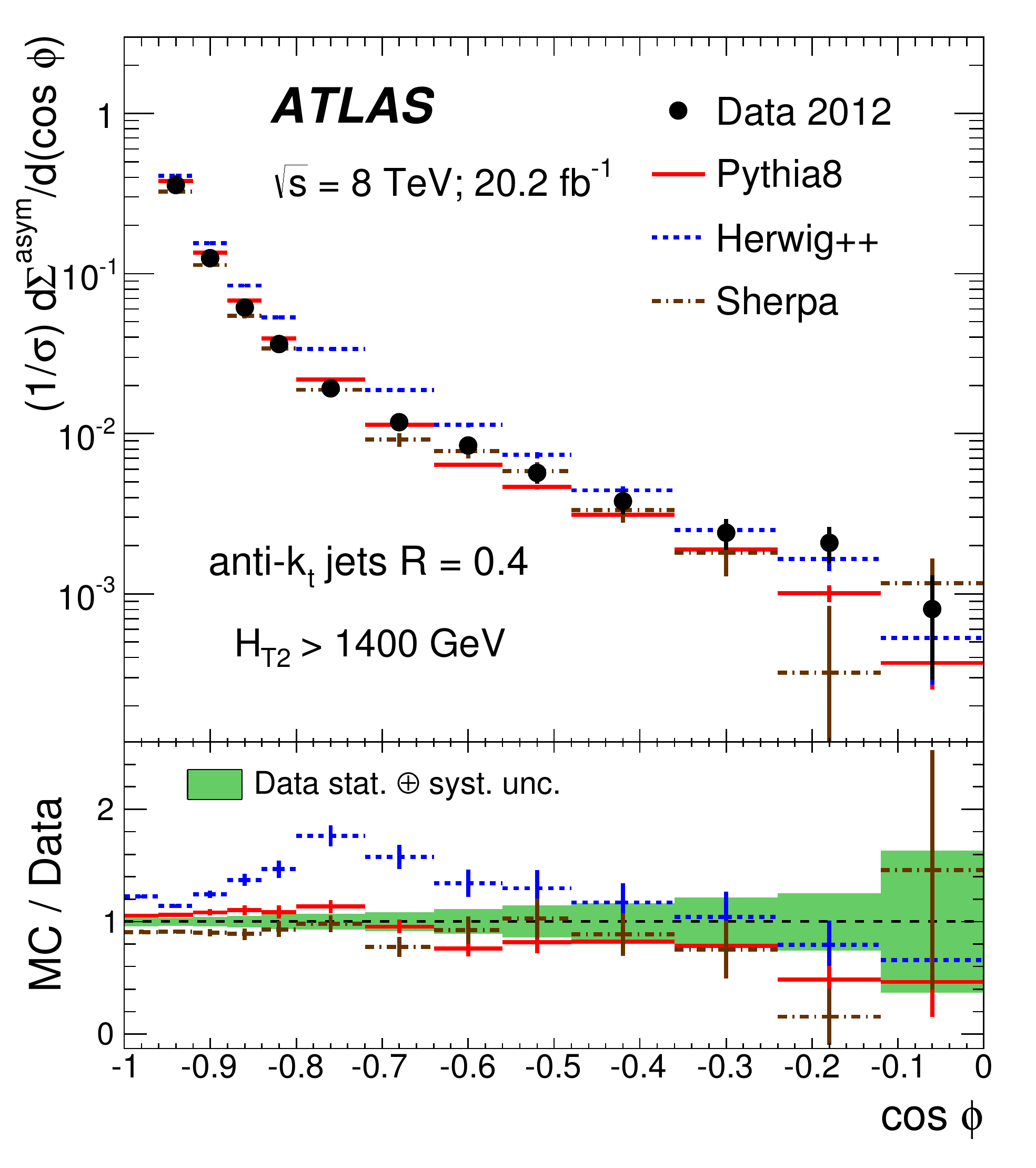}
\caption{Particle-level distributions for the ATEEC functions in each of the $H_{\mathrm{T}2}$ intervals chosen in this analysis, together with MC predictions from \textsc{Pythia8}, \textsc{Herwig++} and \textsc{Sherpa}. The total uncertainty, including statistical and other experimental sources is also indicated using an error bar for the distributions and a green-shaded band for the ratios.}
\label{fig:ateecPart}
\end{figure}

\section{Theoretical predictions}
\label{sec:theory}

The theoretical predictions for the TEEC and ATEEC functions are calculated using perturbative QCD at NLO as implemented in \textsc{NLOJET++}~\cite{Nagy1,Nagy2}. Typically $\mathcal{O}(10^{10})$ events are generated for the calculation. The partonic cross-sections, $\hat{\sigma}$, are convolved with the NNLO PDF sets from \textsc{MMHT 2014}~\cite{mmht}, \textsc{CT14}~\cite{ct14}, \textsc{NNPDF 3.0}~\cite{nnpdf3} and \textsc{HERAPDF 2.0}~\cite{herapdf} using the \textsc{LHAPDF6} package~\cite{lhapdf}. The value of $\alpha_{\mathrm{s}}(m_\Zboson)$ used in the partonic matrix-element calculation is chosen to be the same as that of the PDF. At leading order in $\alpha_{\mathrm{s}}$, the TEEC function defined in Equation~(\ref{eq:teecDef}) can be expressed as
\begin{equation}
\frac{1}{\sigma}\frac{\mathrm{d}\Sigma}{\mathrm{d}\phi} =
\frac{\Sigma_{a_i,b_i} f_{a_1/p}(x_1) f_{a_2/p}(x_2) \otimes \hat{\Sigma}^{a_1 a_2 \to b_1 b_2 b_3}}
     {\Sigma_{a_i,b_i} f_{a_1/p}(x_1) f_{a_2/p}(x_2) \otimes \hat{\sigma}^{a_1 a_2 \to b_1 b_2 } } ,
\label{eq:TEEC-QCD}
\end{equation}
where $\hat{\Sigma}^{a_1 a_2 \to b_1 b_2 b_3}$ is the partonic cross-section weighted by the fractions of transverse energy of the outgoing partons, $\xti\xtj$ as in Equation~(\ref{eq:teecDef}); $x_i$ ($ i=1,2$) are the fractional longitudinal momenta carried by the initial-state partons, $f_{a_1/p}(x_1)$ and $f_{a_2/p}(x_2)$ are the PDFs and $\otimes$ denotes a convolution over $x_1$,$x_2$.\\
At $\mathcal{O}(\alpha_{\mathrm{s}}^4)$, the numerator in Equation~(\ref{eq:TEEC-QCD}) entails calculations of the $2\to 3$ partonic subprocesses at NLO accuracy, and the $2\to 4$ partonic subprocesses at LO. In order to avoid the double collinear singularities appearing in the latter, the angular range is restricted to $|\cos \phi| \leq 0.92$. This avoids calculating the two-loop virtual corrections to the $2\to 2$ subprocesses. Thus, with the azimuthal angle cut, the denominator in Equation~(\ref{eq:TEEC-QCD}) includes the $2\to 2$ and $2\to 3$ subprocesses up to and including the $\mathcal{O}(\alpha_{\mathrm{s}}^3)$ corrections.\\
The nominal renormalisation and factorisation scales are defined as a function of the transverse momenta of the two leading jets as follows~\cite{renScale}
\[
\mu_\mathrm{R} = \frac{p_{\mathrm{T}1}+p_{\mathrm{T}2}}{2}; \ \ \ \mu_\mathrm{F} = \frac{p_{\mathrm{T}1}+p_{\mathrm{T}2}}{4}.
\label{eq:renScale}
\]
This choice eases the comparison with the previous measurement at $\sqrt{s} = 7~\TeV$~\cite{atlasEEC}, where the renormalisation scale was the same. The relevant scale for the perturbative calculation is the renormalisation scale, as variations of the factorisation scale lead to small variations of the physical observable. The scale choice for the NLO pQCD templates used to extract $\alpha_{\mathrm{s}}$ as well as for the presentation of the measurement is not uniquely defined. The nominal scale choice, $H_{\mathrm{T}2}/2$, used in  this paper is based on previous publications \cite{cmsR32,atlasEEC}. However, it should be noted that other scale choices, which explicitly take into account the kinematics of the third jet, are also viable options and can be considered in future measurements.\\
The following comments are in order. The \textsc{NLOJet++} calculations are performed in the limit of massless quarks. PDFs are based on the $n_{\mathrm{f}}=5$ scheme. There is therefore a residual uncertainty due to the mass of the top quark. This is expected to be small since at LHC energies $\sigma_{t\bar{t}} \ll \sigma_{\mathrm{QCD}}$. The correct treatment of top quark mass effects in the initial as well as in final state is not yet available.
\subsection{Non-perturbative corrections}
\label{subsec:nonPert}
The pQCD predictions obtained using \textsc{NLOJET++} are generated at the parton level only. In order to compare these predictions with the data, one needs to correct for non-perturbative (NP) effects, namely hadronisation and the underlying event. Here, doing this relies on bin-by-bin correction factors calculated as the ratio of the MC predictions for TEEC distributions with hadronisation and UE turned on to those with hadronisation and UE turned off. These factors, which are calculated using several MC models, are used to correct the pQCD prediction to the particle level by multiplying each bin of the theoretical distributions. Figure~\ref{fig:nonPert} shows the distributions of the factors for the TEEC as a function of $\cos\phi$ and for two bins in the energy scale $H_{\mathrm{T}2}$. They were calculated using several models, namely \textsc{Pythia8} with the \textsc{AU2}~\cite{atlasTunes2} and \textsc{4C} tunes~\cite{4cTune} and \textsc{Herwig++} with the \textsc{LHC-UE-EE-3-CTEQ6L1} and \textsc{LHC-UE-EE-3-LOMOD} tunes~\cite{herwigTunes}. From these four posibilities, \textsc{Pythia8} with the \textsc{AU2} tune is used for the nominal corrections.

\subsection{Theoretical uncertainties}
\label{subsec:theoUncert}
The theoretical uncertainties are divided into three classes: those corresponding to the renormalisation and factorisation scale variations, the ones corresponding to the PDF eigenvectors, and the ones for the non-perturbative corrections.

\begin{figure}[H]
\centering
\includegraphics[width=8.cm,height=6.0cm]{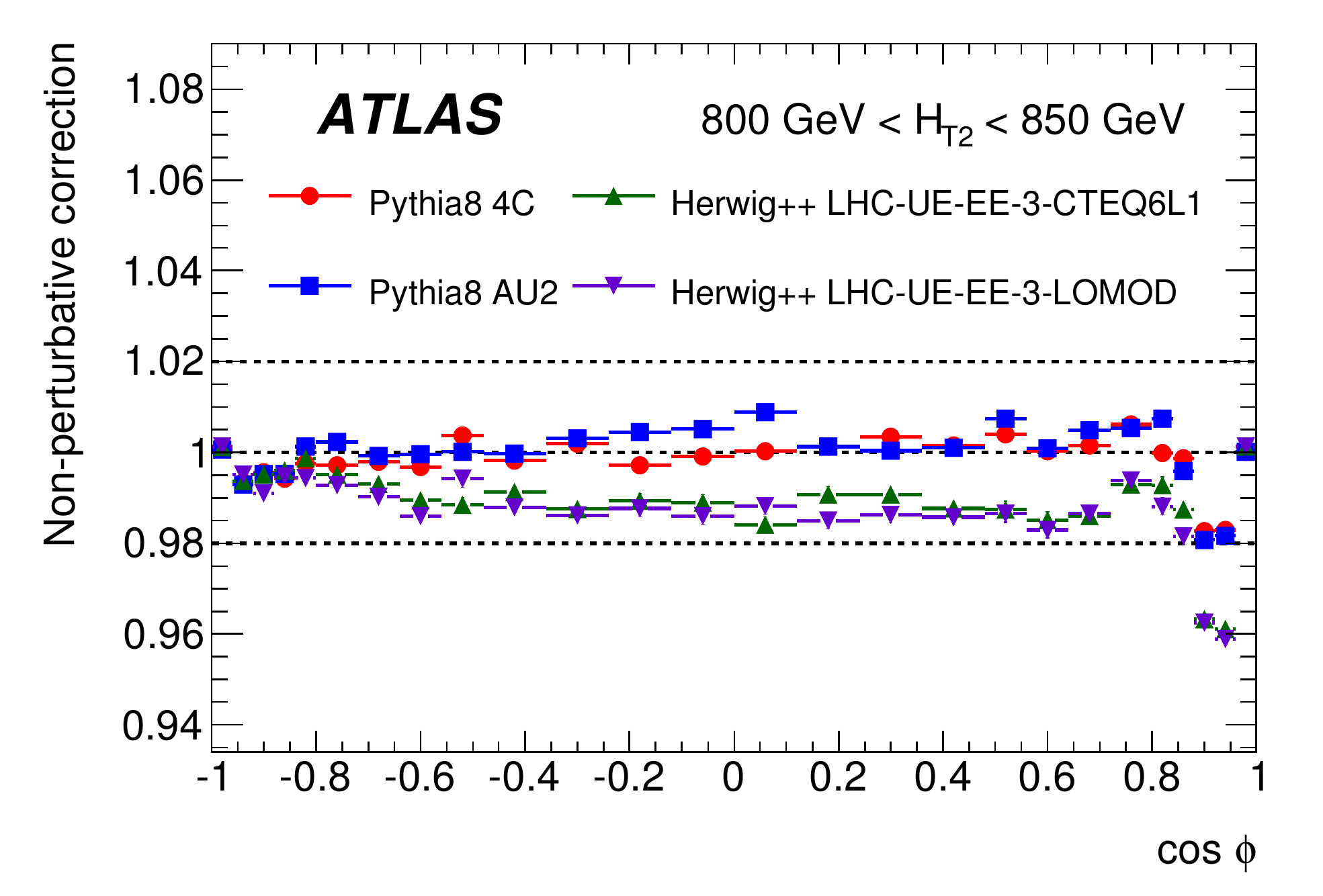}
\includegraphics[width=8.cm,height=6.0cm]{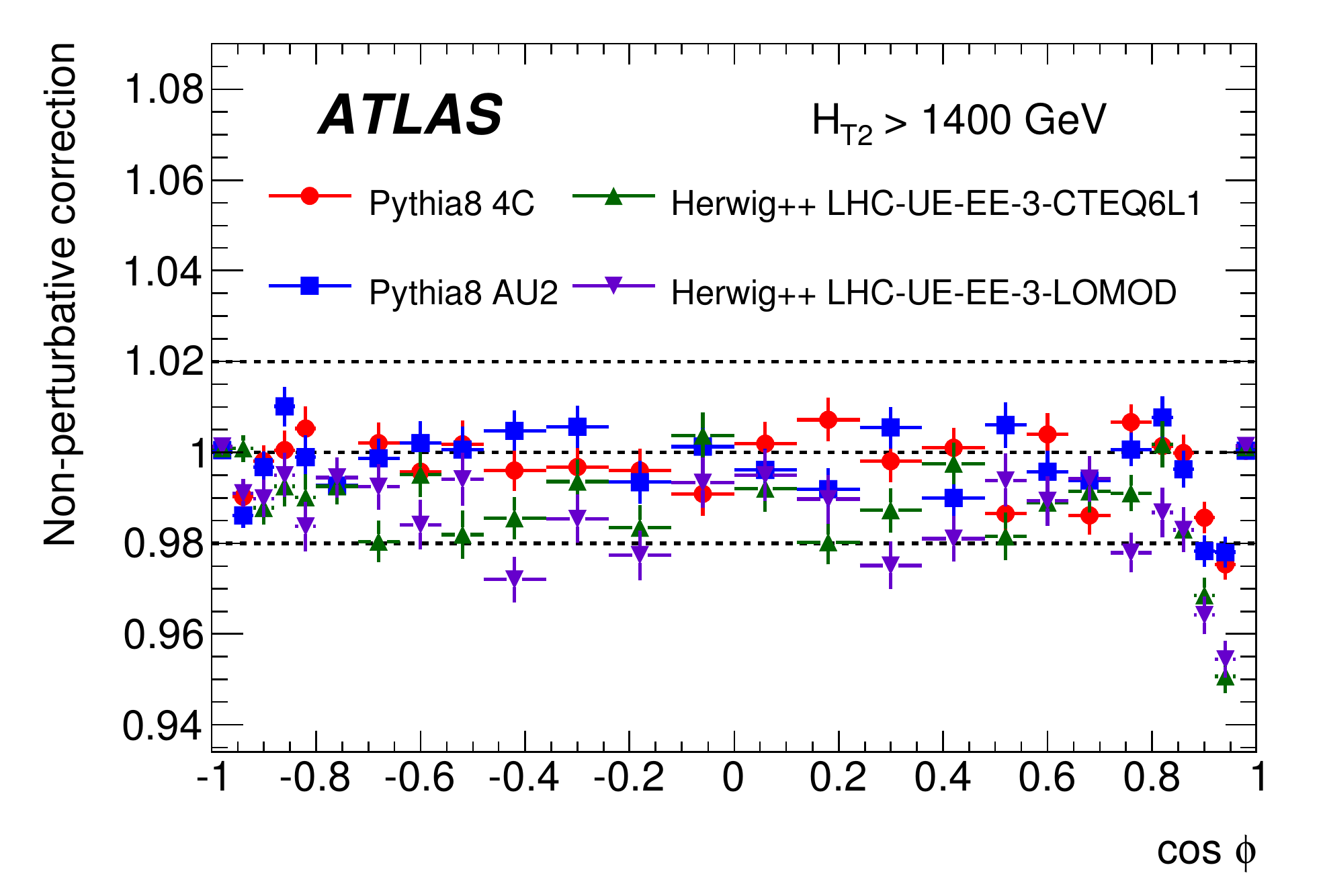}
\caption{Non-perturbative correction factors for TEEC in the first and last bins of $H_{\mathrm{T}2}$ as a function of $\cos\phi$.}
\label{fig:nonPert}
\end{figure}

\begin{itemize}
\item{} The theoretical uncertainty due to the choice of renormalisation and factorisation scales is defined as the envelope of all the variations of the TEEC and ATEEC distributions obtained by varying up and down the scales $\mu_\mathrm{R}, \mu_\mathrm{F}$ by a factor of two, excluding those configurations in which both scales are varied in opposite directions. This is the dominant theoretical uncertainty in this measurement, which can reach 20\% in the central region of the TEEC distributions.
\item{} The parton distribution functions are varied following the set of eigenvectors / replicas provided by each PDF group~\cite{mmht, ct14, nnpdf3, herapdf}. The propagation of the corresponding uncertainty to the TEEC and ATEEC is done following the recommendations for each particular set of distribution functions. The size of this uncertainty is around 1\% for each TEEC bin.
\item{} The uncertainty in the non-perturbative corrections is estimated as the envelope of all models used for the calculation of the correction factors in Figure~\ref{fig:nonPert}. This uncertainty is around 1\% for each of the TEEC bins considered in the NLO predictions, i.e. those with $|\cos\phi|\leq 0.92$.
\item{} The uncertainty due to $\alpha_{\mathrm{s}}$ is also considered for the comparison of the data with the theoretical predictions. This is estimated by varying $\alpha_{\mathrm{s}}$ by the uncertainty in its value for each PDF set, as indicated in Refs.~~\cite{mmht, ct14, nnpdf3, herapdf}.
\end{itemize}
The total theoretical uncertainty is obtained by adding these four theoretical uncertainties in quadrature. The total uncertainty can reach 20\% for the central part of the TEEC, due to the large value of the scale uncertainty in this region.
\section{Comparison of theoretical predictions and experimental results}
\label{subsec:dataTheory}
The unfolded data obtained in Section~\ref{sec:unfResults} are compared to the pQCD predictions, once corrected for non-perturbative effects. Figures~\ref{fig:dataTheoryTEEC} and~\ref{fig:dataTheoryATEEC} show the ratios of the data to the theoretical predictions for the TEEC and ATEEC functions, respectively. The theoretical predictions were calculated, as a function of $\cos\phi$ and for each of the $H_{\mathrm{T}2}$ bins considered, using the \textsc{NNPDF 3.0} PDFs with $\alpha_{\mathrm{s}}(m_\Zboson) = 0.1180$.
\begin{figure}[H]
\centering
\includegraphics[width=13.5cm, height=7.9cm]{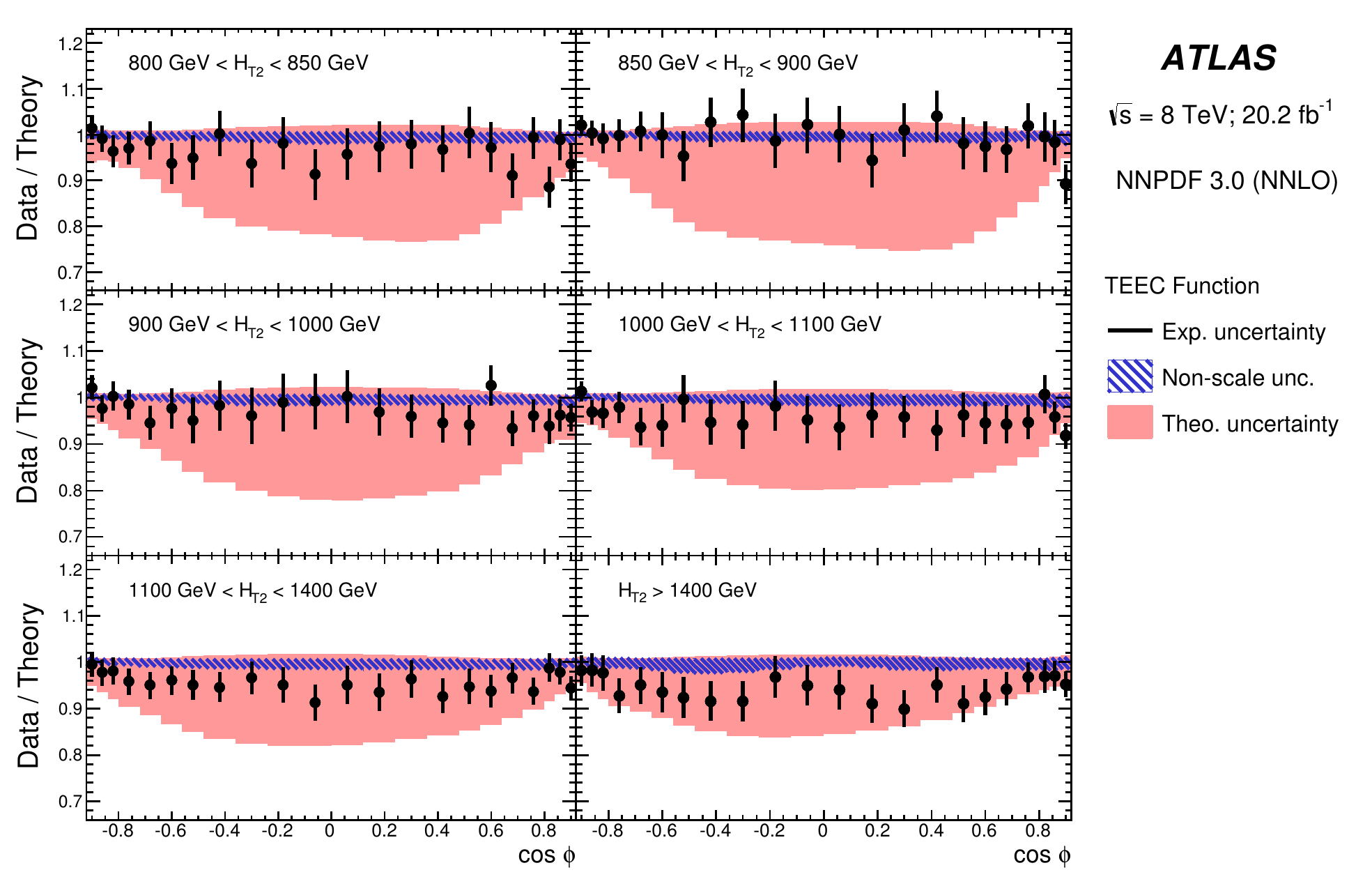}
\caption{Ratios of the TEEC data in each $H_{\mathrm{T}2}$ bin to the NLO pQCD predictions obtained using the \textsc{NNPDF 3.0} parton distribution functions, and corrected for non-perturbative effects.}
\label{fig:dataTheoryTEEC}
\end{figure}
\begin{figure}[H]
\centering
\includegraphics[width=13.5cm, height=7.9cm]{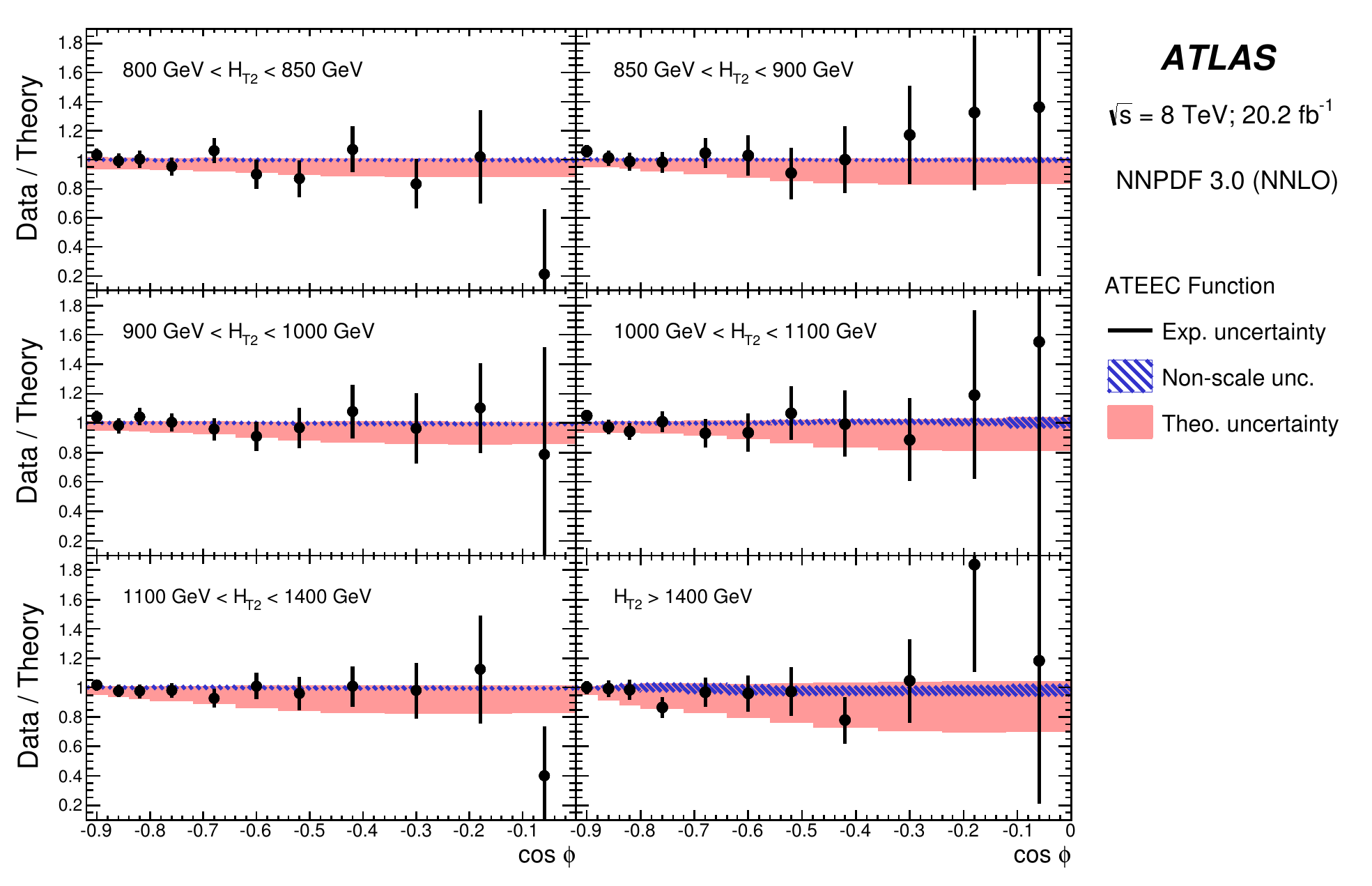}
\caption{Ratios of the ATEEC data in each $H_{\mathrm{T}2}$ bin to the NLO pQCD predictions obtained using the \textsc{NNPDF 3.0} parton distribution functions, and corrected for non-perturbative effects.}
\label{fig:dataTheoryATEEC}
\end{figure}
\newpage
From the comparisons in Figures~\ref{fig:dataTheoryTEEC} and~\ref{fig:dataTheoryATEEC}, one can conclude that perturbative QCD correctly describes the data within the experimental and theoretical uncertainties. 
\section{Determination of $\alpha_{\mathrm{s}}$ and test of asymptotic freedom}
\label{sec:alphaSDet}
From the comparisons made in the previous section, one can determine the strong coupling constant at the scale given by the pole mass of the $\Zboson$ boson, $\alpha_{\mathrm{s}}(m_\Zboson)$, by considering the following $\chi^2$ function
\begin{equation}
\chi^2(\alpha_{\mathrm{s}},\vec{\lambda}) = \sum_{\mathrm{bins}}\frac{(x_i-F_i(\alpha_{\mathrm{s}},\vec{\lambda}))^2}{\Delta x_i^2+\Delta\xi_i ^2} + \sum_{k}\lambda_k^2,
\label{eq:chi2a}
\end{equation}
where the theoretical predictions are varied according to
\begin{equation}
F_{i}(\alpha_{\mathrm{s}},\vec{\lambda}) = \psi_{i}(\alpha_{\mathrm{s}})\left(1+\sum_k\lambda_k\sigma_k^{(i)}\right).
\label{eq:chi2b}
\end{equation}
In Equations~(\ref{eq:chi2a}) and~(\ref{eq:chi2b}), $\alpha_{\mathrm{s}}$ stands for $\alpha_{\mathrm{s}}(m_\Zboson)$; $x_i$ is the value of the $i$-th point of the distribution as measured in data, while $\Delta x_i$ is its statistical uncertainty. The statistical uncertainty in the theoretical predictions is also included as $\Delta \xi_i$, while $\sigma_k^{(i)}$ is the relative value of the $k$-th source of systematic uncertainty in bin $i$.\\
This technique takes into account the correlations between the different sources of systematic uncertainty discussed in Section~\ref{sec:systematics} by introducing the nuisance parameters $\{\lambda_k\}$, one for each source of experimental uncertainty. Thus, the minimum of the $\chi^2$ function defined in Equation~(\ref{eq:chi2a}) is found in a 74-dimensional space, in which 73 correspond to nuisance parameters $\left\{\lambda_i\right\}$ and one to $\alpha_{\mathrm{s}}(m_\Zboson)$.\\
The method also requires an analytical expression for the dependence of the fitted observable on the strong coupling constant, which is given by $\psi_{i}(\alpha_{\mathrm{s}})$ for bin $i$. For each PDF set, the corresponding $\alpha_{\mathrm{s}}(m_\Zboson)$ variation range is considered and the theoretical prediction is obtained for each value of $\alpha_{\mathrm{s}}(m_\Zboson)$. The functions $\psi_{i}(\alpha_{\mathrm{s}})$ are then obtained by fitting the values of the TEEC (ATEEC) in each $(H_{\mathrm{T}2}, \cos\phi)$ bin to a second-order polynomial. For both the TEEC and ATEEC functions, the fits to extract $\alpha_{\mathrm{s}}(m_\Zboson)$ are repeated separately for each $H_{\mathrm{T}2}$ interval, thus determining a value of $\alpha_{\mathrm{s}}(m_\Zboson)$ for each energy bin. The theoretical uncertainties are determined by shifting the theory distributions by each of the uncertainties separately, recalculating the functions $\psi_i(\alpha_{\mathrm{s}})$ and determining a new value of $\alpha_{\mathrm{s}}(m_\Zboson)$. The uncertainty is determined by taking the difference between this value and the nominal one.\\
 Each of the obtained values of $\alpha_{\mathrm{s}}(m_\Zboson)$ is then evolved to the corresponding measured scale using the NLO solution to the renormalisation group equation (RGE), given by
\begin{equation}
\alpha_{\mathrm{s}}(Q^2) = \frac{1}{\beta_0\log x} \left[1- \frac{\beta_1}{\beta_0^2} \frac{\log\left(\log x \right)}{\log x}\right]; \ \ \ x = \frac{Q^2}{\Lambda^2},
\label{eq:asRunNLO}
\end{equation}
where the coefficients $\beta_0$ and $\beta_1$ are given by
\[
\beta_0 = \frac{1}{4\pi}\left(11-\frac{2}{3}n_{\mathrm{f}}\right); \ \ \ \ \beta_1 = \frac{1}{(4\pi)^2}\left(102-\frac{38}{3}n_{\mathrm{f}}\right),
\label{eq:asRunNLO2}
\]
and $\Lambda$ is the QCD scale, determined in each case from the fitted value of $\alpha_s(m_\Zboson)$. Here, $n_{\mathrm{f}}$ is the number of active flavours at the scale $Q$, i.e. the number of quarks with mass $m < Q$. Therefore, $n_{\mathrm{f}} = 6$ in the six bins considered in Table \ref{tab:htBins}. When evolving $\alpha_{\mathrm{s}}(m_Z)$ to $\alpha_{\mathrm{s}}(Q)$, the proper transition rules for $n_{\mathrm{f}}=5$ to $n_{\mathrm{f}}=6$ are applied so that $\alpha_{\mathrm{s}}(Q)$ is a continuous function across quark thresholds. Finally, the results are combined by performing a global fit, where all bins are merged together.

\newpage

\subsection{Fits to individual TEEC functions}
\label{sec:teecFits}
The values of $\alpha_{\mathrm{s}}(m_\Zboson)$ obtained from fits to the TEEC function in each $H_{\mathrm{T}2}$ bin are summarised in Table~\ref{tab:asTEEC1}. The theoretical predictions used for this extraction use \textsc{NNPDF 3.0} as the nominal PDF set. 
\begin{table}[H]
\small
\begin{center}
\begin{tabular}{ccc}
\hline
\T\B $\langle Q \rangle$ (GeV) & $\alpha_{\mathrm{s}}(m_Z)$ value (\textsc{NNPDF 3.0}) & $\chi^2/N_{\mathrm{dof}}$\\
\hline
\T \B 412 & 0.1171 $\pm$ 0.0021 (exp.) $^{+ 0.0081 }_{- 0.0022 }$ (scale) $\pm$ 0.0013 (PDF) $\pm$ 0.0001 (NP) & 24.3 / 21 \\
\T \B 437 & 0.1178 $\pm$ 0.0017 (exp.) $^{+ 0.0073 }_{- 0.0017 }$ (scale) $\pm$ 0.0014 (PDF) $\pm$ 0.0002 (NP) & 28.3 / 21 \\
\T \B 472 & 0.1177 $\pm$ 0.0017 (exp.) $^{+ 0.0079 }_{- 0.0023 }$ (scale) $\pm$ 0.0015 (PDF) $\pm$ 0.0001 (NP) & 27.7 / 21 \\
\T \B 522 & 0.1163 $\pm$ 0.0017 (exp.) $^{+ 0.0067 }_{- 0.0016 }$ (scale) $\pm$ 0.0016 (PDF) $\pm$ 0.0001 (NP) & 22.8 / 21 \\
\T \B 604 & 0.1181 $\pm$ 0.0017 (exp.) $^{+ 0.0082 }_{- 0.0022 }$ (scale) $\pm$ 0.0017 (PDF) $\pm$ 0.0005 (NP) & 24.3 / 21 \\
\T \B 810 & 0.1186 $\pm$ 0.0023 (exp.) $^{+ 0.0085 }_{- 0.0035 }$ (scale) $\pm$ 0.0020 (PDF) $\pm$ 0.0004 (NP) & 23.7 / 21 \\
\hline
\end{tabular}
\end{center}
\caption{Values of the strong coupling constant at the $\Zboson$ boson mass scale, $\alpha_{\mathrm{s}}(m_\Zboson)$ obtained from fits to the TEEC function for each $H_{\mathrm{T}2}$ interval using the \textsc{NNPDF 3.0} parton distribution functions. The values of the average scale $\langle Q \rangle$ for each energy bin are shown in the first column, while the values of the $\chi^2$ function at the minimum are shown in the third column. The uncertainty referred to as NP is the one related to the non-perturbative corrections.}
\label{tab:asTEEC1}
\end{table}

The values summarised in Table~\ref{tab:asTEEC1} are in good agreement with the 2016 world average value~\cite{pdg}, as well as with previous measurements, in particular with previous extractions using LHC data~\cite{atlasEEC, cmsR32, cms3jet, cmsIncJet, cmsIncJet2, cmsTTbar, bogdan}. The values of the $\chi^2$ indicate that agreement between the data and the theoretical predictions is good. The nuisance parameters for the TEEC fits are generally compatible with zero. One remarkable exception is the nuisance parameter associated to the modelling uncertainty, which deviates by half standard deviation with a very small error bar. This is an indication that these data can be used to further tune MC event generators which model multi-jet production.\\
Figure~\ref{fig:teecAfterFit} compares the data with the theoretical predictions after the fit, i.e. where the fitted values of $\alpha_{\mathrm{s}}(m_\Zboson)$ and the nuisance parameters are already constrained. Table~\ref{tab:asTEEC2} shows the values of $\alpha_{\mathrm{s}}$ evolved from $m_\Zboson$ to the corresponding scale $Q$ using Equation~(\ref{eq:asRunNLO}). The appendix includes tables in which the values of $\alpha_{\mathrm{s}}(m_\Zboson)$ obtained from the TEEC fits are extrapolated to different values of $Q$, given by the averages of kinematical quantities other than $H_{\mathrm{T}2}/2$.
\begin{table}[H]
\begin{center}
\begin{tabular}{cc}
\hline
\T\B $\langle Q \rangle$ (GeV) & $\alpha_{\mathrm{s}}(Q^2)$ value (\textsc{NNPDF 3.0})\\
\hline
\T \B 412 & 0.0966 $\pm$ 0.0014 (exp.) $^{+ 0.0054 }_{- 0.0015 }$ (scale) $\pm$ 0.0009 (PDF) $\pm$ 0.0001 (NP) \\
\T \B 437 & 0.0964 $\pm$ 0.0012 (exp.) $^{+ 0.0048 }_{- 0.0011 }$ (scale) $\pm$ 0.0009 (PDF) $\pm$ 0.0002 (NP) \\
\T \B 472 & 0.0955 $\pm$ 0.0011 (exp.) $^{+ 0.0051 }_{- 0.0015 }$ (scale) $\pm$ 0.0009 (PDF) $\pm$ 0.0001 (NP) \\
\T \B 522 & 0.0936 $\pm$ 0.0011 (exp.) $^{+ 0.0043 }_{- 0.0010 }$ (scale) $\pm$ 0.0010 (PDF) $\pm$ 0.0001 (NP) \\
\T \B 604 & 0.0933 $\pm$ 0.0011 (exp.) $^{+ 0.0050 }_{- 0.0014 }$ (scale) $\pm$ 0.0011 (PDF) $\pm$ 0.0003 (NP) \\
\T \B 810 & 0.0907 $\pm$ 0.0013 (exp.) $^{+ 0.0049 }_{- 0.0020 }$ (scale) $\pm$ 0.0011 (PDF) $\pm$ 0.0002 (NP) \\
\hline
\end{tabular}
\end{center}
\caption{Values of the strong coupling constant at the measurement scales, $\alpha_{\mathrm{s}}(Q^2)$ obtained from fits to the TEEC function for each $H_{\mathrm{T}2}$ interval using the \textsc{NNPDF 3.0} parton distribution functions. The uncertainty referred to as NP is the one related to the non-perturbative corrections.}
\label{tab:asTEEC2}
\end{table}

\begin{figure}[H]
\centering
\includegraphics[width=7.5cm,height=7.3cm]{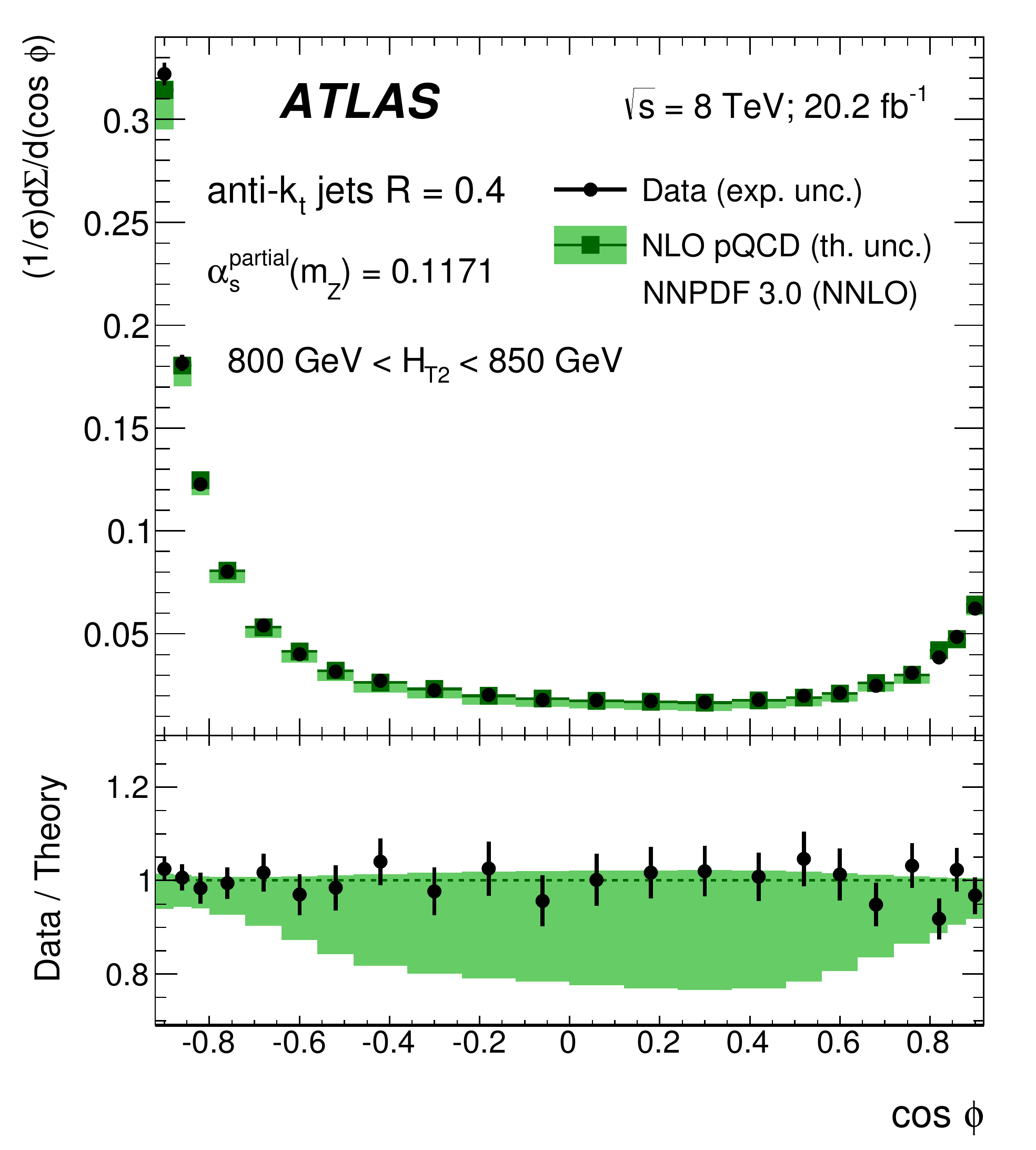}
\includegraphics[width=7.5cm,height=7.3cm]{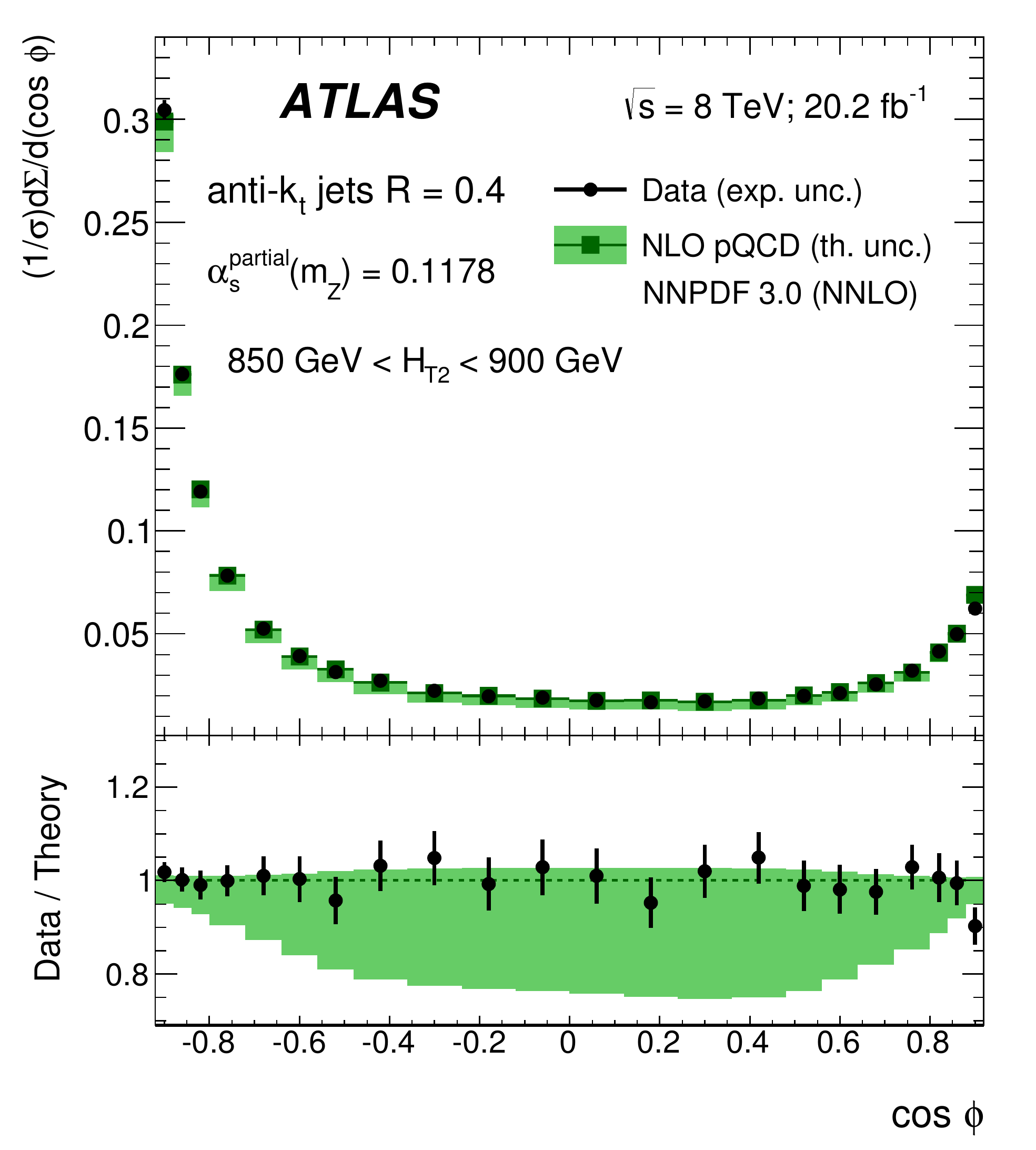}

\includegraphics[width=7.5cm,height=7.3cm]{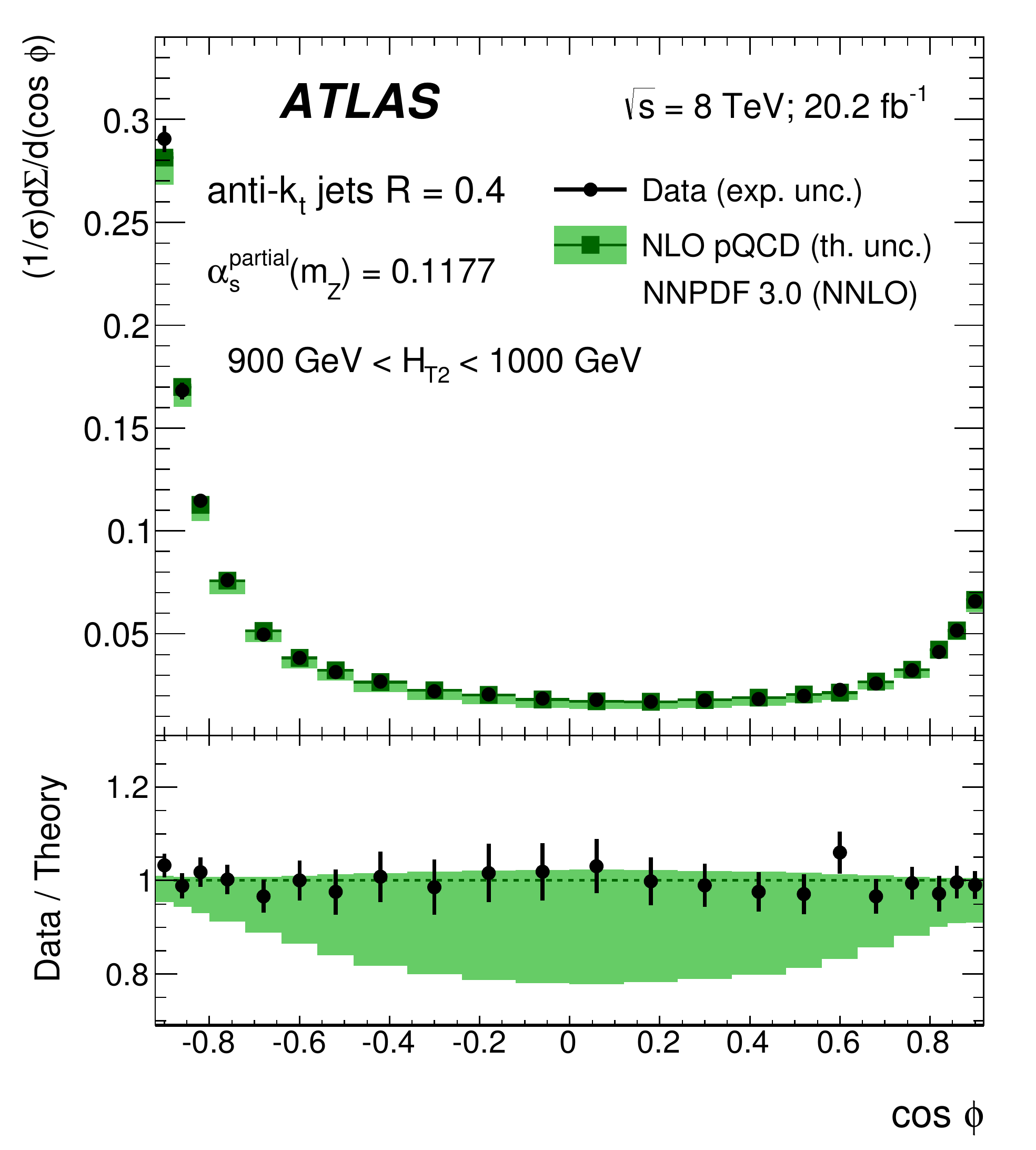}
\includegraphics[width=7.5cm,height=7.3cm]{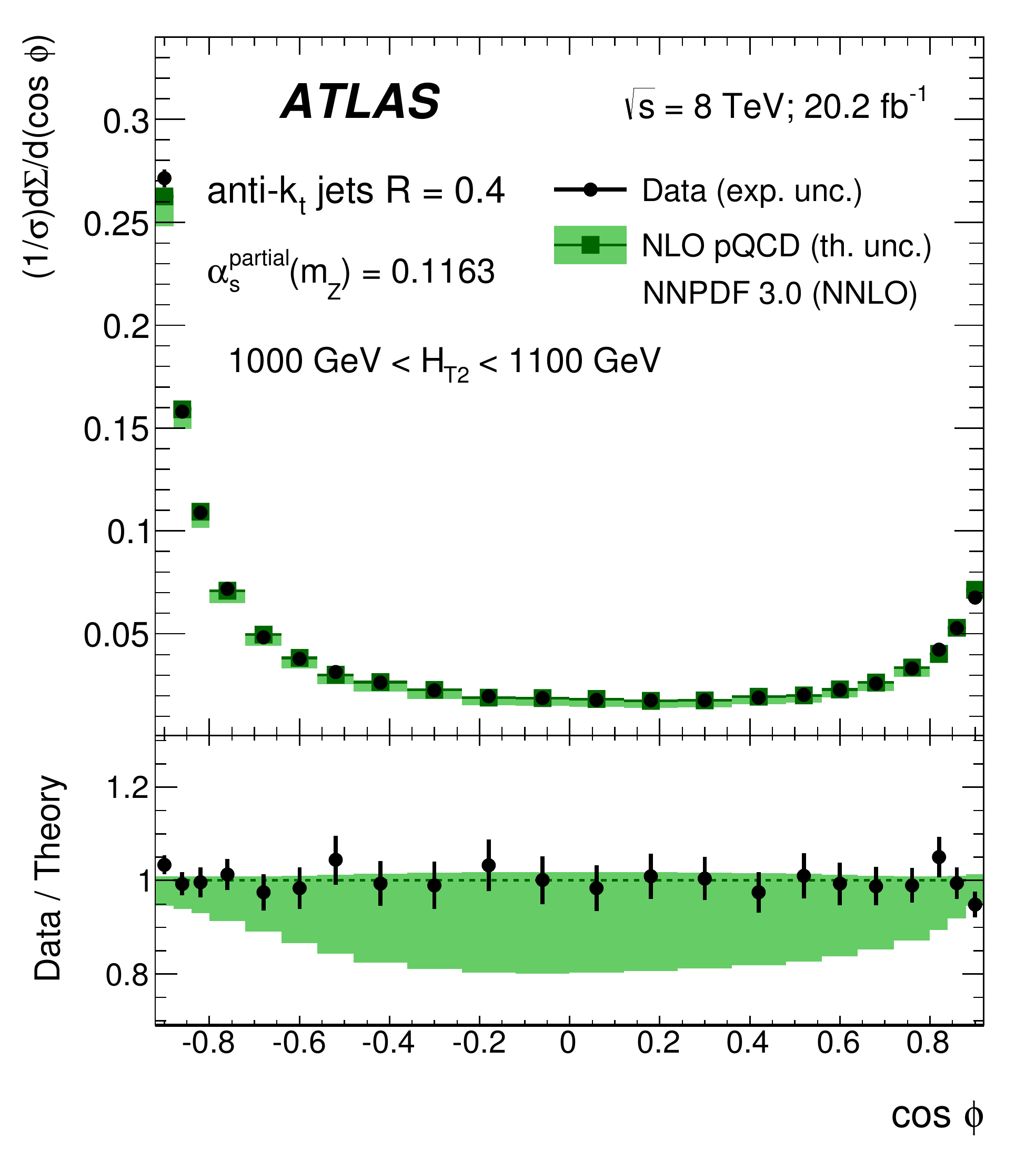}

\includegraphics[width=7.5cm,height=7.3cm]{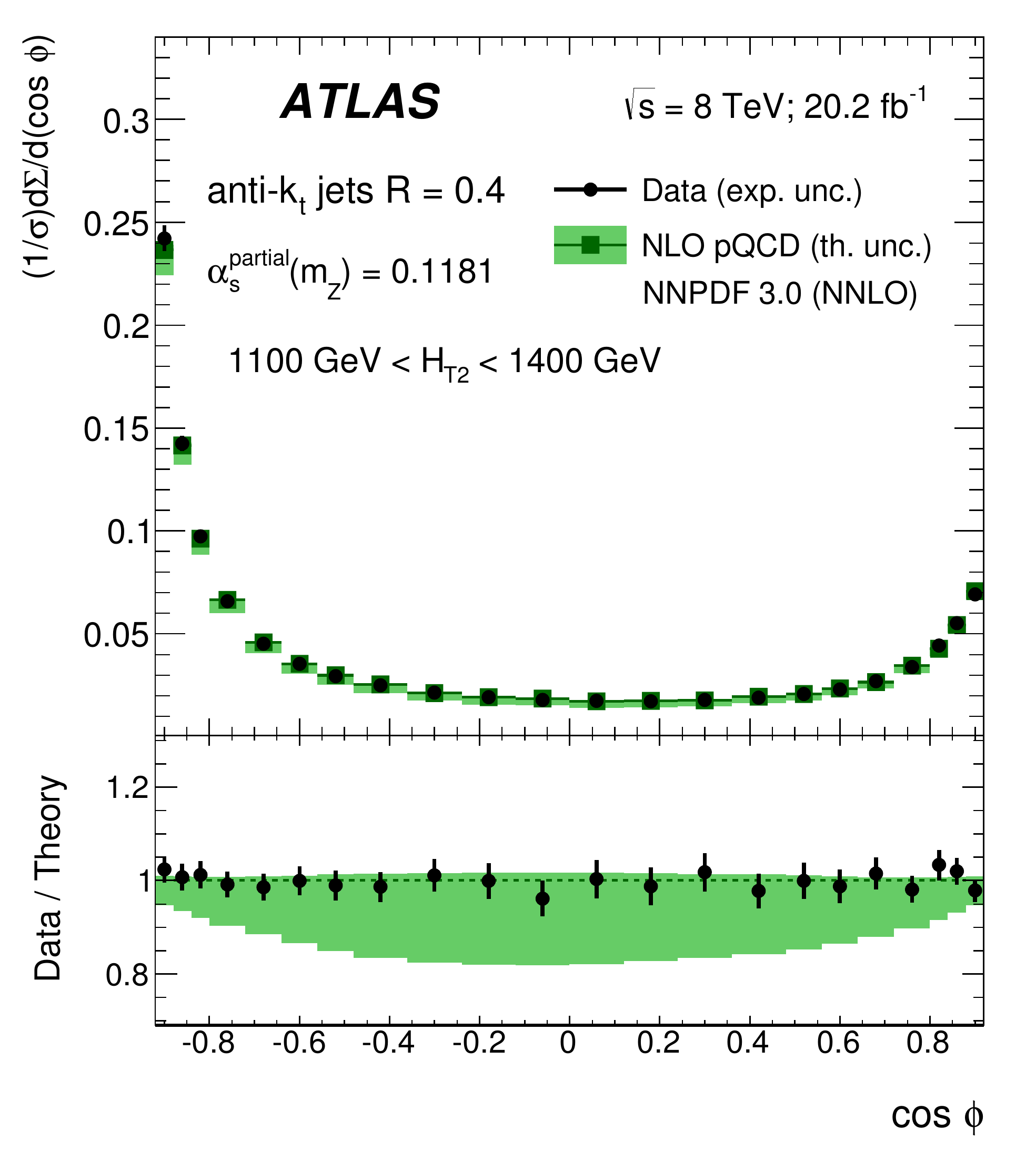}
\includegraphics[width=7.5cm,height=7.3cm]{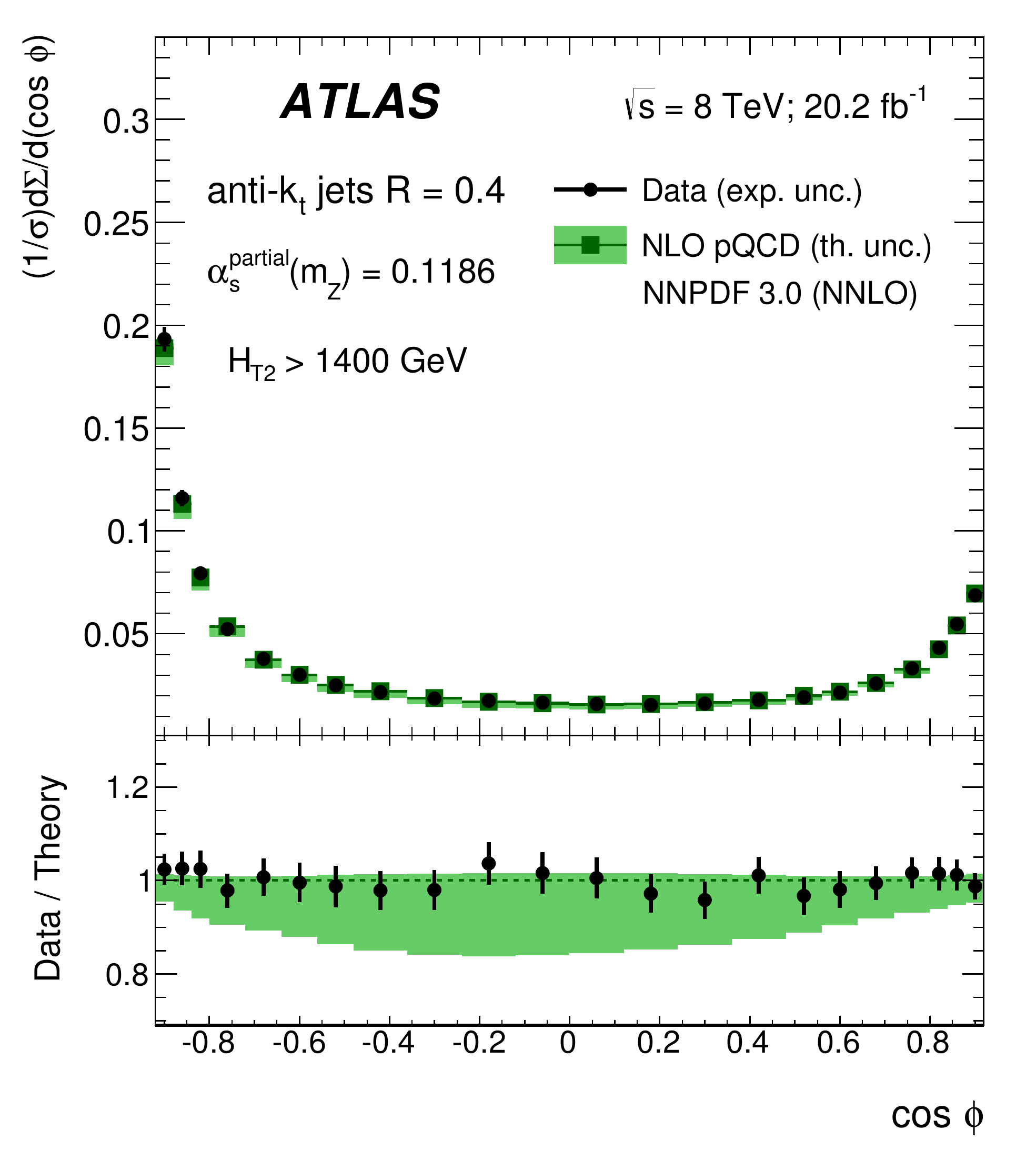}
\caption{Comparison of the TEEC data and the theoretical predictions after the fit. The value of $\alpha_{\mathrm{s}}(m_\Zboson)$ used in this comparison is fitted independently for each energy bin.}
\label{fig:teecAfterFit}
\end{figure}

\subsection{Global TEEC fit}
The combination of the previous results is done by considering all the $H_{\mathrm{T}2}$ bins into a single, global fit. The result obtained using the \textsc{NNPDF 3.0} PDF set has the largest PDF uncertainty and thus, in order to be conservative, it is the one quoted as the final value of $\alpha_s(m_\Zboson)$.

The impact of the correlations of the JES uncertainties on the result is studied by considering two additional correlation scenarios, one with stronger and one with weaker correlation assumptions~\cite{JES}. From the envelope of these results, an additional uncertainty of 0.0007 is assigned in order to cover this difference.

The results for $\alpha_{\mathrm{s}}(m_\Zboson)$ are summarised in Table~\ref{tab:teecGlobal} for each of the four PDF sets investigated in this analysis
\begin{table}[H]
\footnotesize
\begin{center}
\begin{tabular}{llc}
\hline
\T\B PDF & $\alpha_{\mathrm{s}}(m_\Zboson)$ value & $\chi^2/N_{\mathrm{dof}}$\\
\hline
\T \B MMHT 2014 & 0.1151 $\pm$ 0.0008 (exp.) $^{+ 0.0064 }_{- 0.0047 }$ (scale) $\pm$ 0.0012 (PDF) $\pm$ 0.0002 (NP) & 173 / 131 \\
\T \B CT14 & 0.1165 $\pm$ 0.0010 (exp.) $^{+ 0.0067 }_{- 0.0061 }$ (scale) $\pm$ 0.0016 (PDF) $\pm$ 0.0003 (NP) & 161 / 131 \\
\T \B NNPDF 3.0 & 0.1162 $\pm$ 0.0011 (exp.) $^{+ 0.0076 }_{- 0.0061 }$ (scale) $\pm$ 0.0018 (PDF) $\pm$ 0.0003 (NP) & 174 / 131 \\
\T \B HERAPDF 2.0 & 0.1177 $\pm$ 0.0008 (exp.) $^{+ 0.0064 }_{- 0.0040 }$ (scale) $\pm$ 0.0005 (PDF) $\pm$ 0.0002 (NP) $^{+ 0.0008 }_{- 0.0007 }$ (mod) & 169 / 131 \\
\hline
\end{tabular}
\end{center}
\caption{The results for $\alpha_{\mathrm{s}}$ from fits to the TEEC using different PDFs. The uncertainty referred to as NP is the one related to the non-perturbative corrections. The uncertainty labelled as `mod' corresponds to the \textsc{HERAPDF} modelling and parameterisation uncertainty.}
\label{tab:teecGlobal}
\end{table}
As a result of considering all the data, the experimental uncertainties are reduced with respect to the partial fits. Also, it should be noted that the values of $\alpha_{\mathrm{s}}$ extracted with different PDF sets show good agreement with each other within the PDF uncertainties, and are compatible with the latest world average value $\alpha_{\mathrm{s}}(m_\Zboson) = 0.1181 \pm 0.0011$~\cite{pdg}.

The final result for the TEEC fit is
\[
\alpha_{\mathrm{s}}(m_\Zboson) = 0.1162 \pm 0.0011 \mbox{ (exp.) } ^{+ 0.0076 }_{- 0.0061 } \mbox{ (scale)} \pm 0.0018 \mbox{ (PDF)} \pm 0.0003 \mbox{ (NP)}.\\
\]
A comparison of the results for $\alpha_{\mathrm{s}}$ from the global and partial fits is shown in Figure~\ref{fig:asRunTEEC}. In this figure, the results from previous experiments~\cite{atlasEEC, cmsR32, cms3jet, cmsIncJet, cmsIncJet2, cmsTTbar, d0AngCorr, d0IncJet} are also shown, together with the world average band~\cite{pdg}. Agreement between this result and the ones from other experiments is very good, even though the experimental uncertainties in this analysis are smaller than in previous measurements in hadron colliders.
\begin{figure}[H]
\vspace{4.cm}
\centering
\includegraphics[width=16.5cm,height=11.5cm]{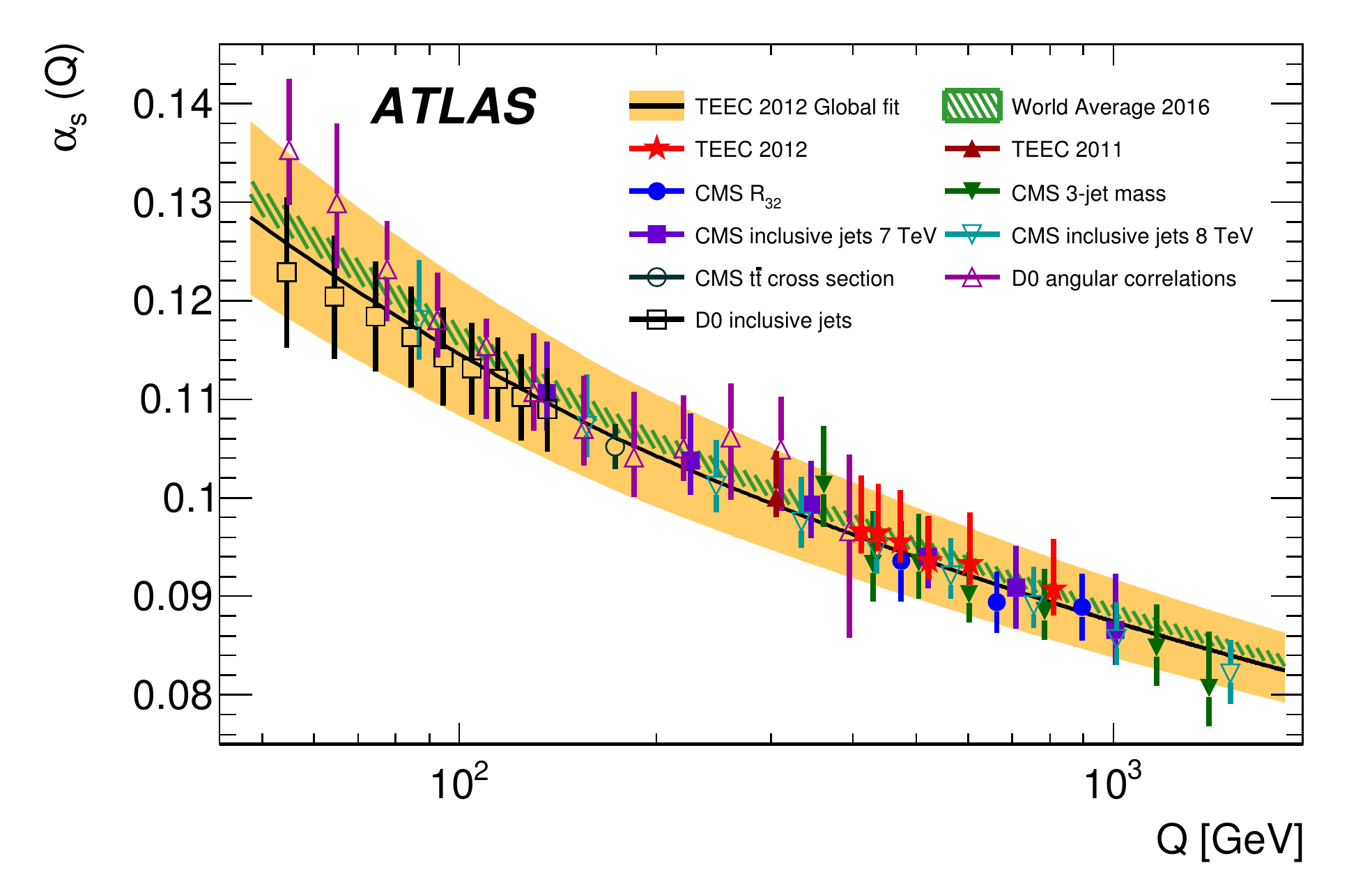}
\caption{Comparison of the values of $\alpha_{\mathrm{s}}(Q)$ obtained from fits to the TEEC functions at the energy scales given by $\langle H_{\mathrm{T}2} \rangle /2$ (red star points) with the uncertainty band from the global fit (orange full band) and the 2016 world average (green hatched band). Determinations from other experiments are also shown as data points. The error bars, as well as the orange full band, include all experimental and theoretical sources of uncertainty. The strong coupling constant is assumed to run according to the two-loop solution of the RGE.}
\label{fig:asRunTEEC}
\end{figure}

\newpage

\subsection{Fits to individual ATEEC functions}
\label{sec:teecFits}
The values of $\alpha_{\mathrm{s}}$ extracted from the fits to the measured ATEEC functions are summarised in Table~\ref{tab:asATEEC1}, together with the values of the $\chi^2$ functions at the minima.
\begin{table}[H]
\small
\begin{center}
\begin{tabular}{ccc}
\hline
\T\B $\langle Q \rangle$ (GeV) & $\alpha_{\mathrm{s}}(m_Z)$ value (\textsc{NNPDF 3.0}) & $\chi^2/N_{\mathrm{dof}}$\\
\hline
\T \B 412 & 0.1209 $\pm$ 0.0036 (exp.) $^{+ 0.0085 }_{- 0.0031 }$ (scale) $\pm$ 0.0013 (PDF) $\pm$ 0.0004 (NP) & 10.6 / 10 \\
\T \B 437 & 0.1211 $\pm$ 0.0026 (exp.) $^{+ 0.0064 }_{- 0.0014 }$ (scale) $\pm$ 0.0015 (PDF) $\pm$ 0.0010 (NP) &  6.8 / 10 \\
\T \B 472 & 0.1203 $\pm$ 0.0028 (exp.) $^{+ 0.0060 }_{- 0.0013 }$ (scale) $\pm$ 0.0016 (PDF) $\pm$ 0.0002 (NP) &  8.8 / 10 \\
\T \B 522 & 0.1196 $\pm$ 0.0025 (exp.) $^{+ 0.0054 }_{- 0.0010 }$ (scale) $\pm$ 0.0017 (PDF) $\pm$ 0.0004 (NP) & 10.9 / 10 \\
\T \B 604 & 0.1176 $\pm$ 0.0031 (exp.) $^{+ 0.0058 }_{- 0.0008 }$ (scale) $\pm$ 0.0020 (PDF) $\pm$ 0.0005 (NP) &  6.4 / 10 \\
\T \B 810 & 0.1172 $\pm$ 0.0037 (exp.) $^{+ 0.0053 }_{- 0.0009 }$ (scale) $\pm$ 0.0022 (PDF) $\pm$ 0.0001 (NP) &  9.8 / 10 \\
\hline
\end{tabular}
\end{center}
\caption{Values of the strong coupling constant at the $\Zboson$ boson mass scale, $\alpha_{\mathrm{s}}(m_\Zboson)$ obtained from fits to the ATEEC function for each $H_{\mathrm{T}2}$ interval using the \textsc{NNPDF 3.0} parton distribution functions. The values of the average scale $\langle Q \rangle$ for each energy bin are shown in the first column, while the values of the $\chi^2$ function at the minimum are shown in the third column. The uncertainty referred to as NP is the one related to the non-perturbative corrections.}
\label{tab:asATEEC1}

\end{table}
The values extracted from the ATEEC show smaller scale uncertainties than their counterpart values from TEEC. This is understood to be due to the fact that the scale dependence is mitigated for the ATEEC distributions because, for the TEEC, this dependence shows some azimuthal symmetry. Also, it is important to note that the values of the $\chi^2$ indicate excellent compatibility between the data and the theoretical predictions. Good agreement, within the scale uncertainty, is also observed between these values and the ones extracted from fits to the TEEC, as well as among themselves and with the current world average. The nuisance parameters are compatible with zero within one standard deviation.\\
As before, the values of $\alpha_{\mathrm{s}}(Q^2)$ at the scales of the measurement are obtained by evolving the values in Table~\ref{tab:asATEEC1} using Equation~(\ref{eq:asRunNLO}). The results are given in Table~\ref{tab:asATEEC2}. As in the TEEC case, Figure~\ref{fig:ateecAfterFit} compares the data with the theoretical predictions after the fit. The appendix includes tables in which the values of $\alpha_{\mathrm{s}}(m_\Zboson)$ obtained from the ATEEC fits are extrapolated to different values of $Q$, given by the averages of kinematic quantities other than $H_{\mathrm{T}2}/2$.
\begin{table}[H]
\begin{center}
\begin{tabular}{cc}
\hline
\T\B $\langle Q \rangle$ (GeV) & $\alpha_{\mathrm{s}}(Q^2)$ value (\textsc{NNPDF 3.0})\\
\hline
\T \B 412 & 0.0992 $\pm$ 0.0024 (exp.) $^{+ 0.0056 }_{- 0.0020 }$ (scale) $\pm$ 0.0009 (PDF) $\pm$ 0.0002 (NP) \\
\T \B 437 & 0.0986 $\pm$ 0.0017 (exp.) $^{+ 0.0041 }_{- 0.0009 }$ (scale) $\pm$ 0.0010 (PDF) $\pm$ 0.0007 (NP) \\
\T \B 472 & 0.0973 $\pm$ 0.0018 (exp.) $^{+ 0.0038 }_{- 0.0008 }$ (scale) $\pm$ 0.0010 (PDF) $\pm$ 0.0001 (NP) \\
\T \B 522 & 0.0957 $\pm$ 0.0016 (exp.) $^{+ 0.0034 }_{- 0.0006 }$ (scale) $\pm$ 0.0011 (PDF) $\pm$ 0.0003 (NP) \\
\T \B 604 & 0.0930 $\pm$ 0.0019 (exp.) $^{+ 0.0035 }_{- 0.0005 }$ (scale) $\pm$ 0.0012 (PDF) $\pm$ 0.0003 (NP) \\
\T \B 810 & 0.0899 $\pm$ 0.0021 (exp.) $^{+ 0.0031 }_{- 0.0005 }$ (scale) $\pm$ 0.0013 (PDF) $\pm$ 0.0001 (NP) \\
\hline
\end{tabular}
\end{center}
\caption{Values of the strong coupling constant at the measurement scales, $\alpha_{\mathrm{s}}(Q^2)$ obtained from fits to the ATEEC function for each $H_{\mathrm{T}2}$ interval using the \textsc{NNPDF 3.0} parton distribution functions. The uncertainty referred to as NP is the one related to the non-perturbative corrections.}
\label{tab:asATEEC2}
\end{table}

\begin{figure}[H]
\centering
\includegraphics[width=7.5cm,height=7.3cm]{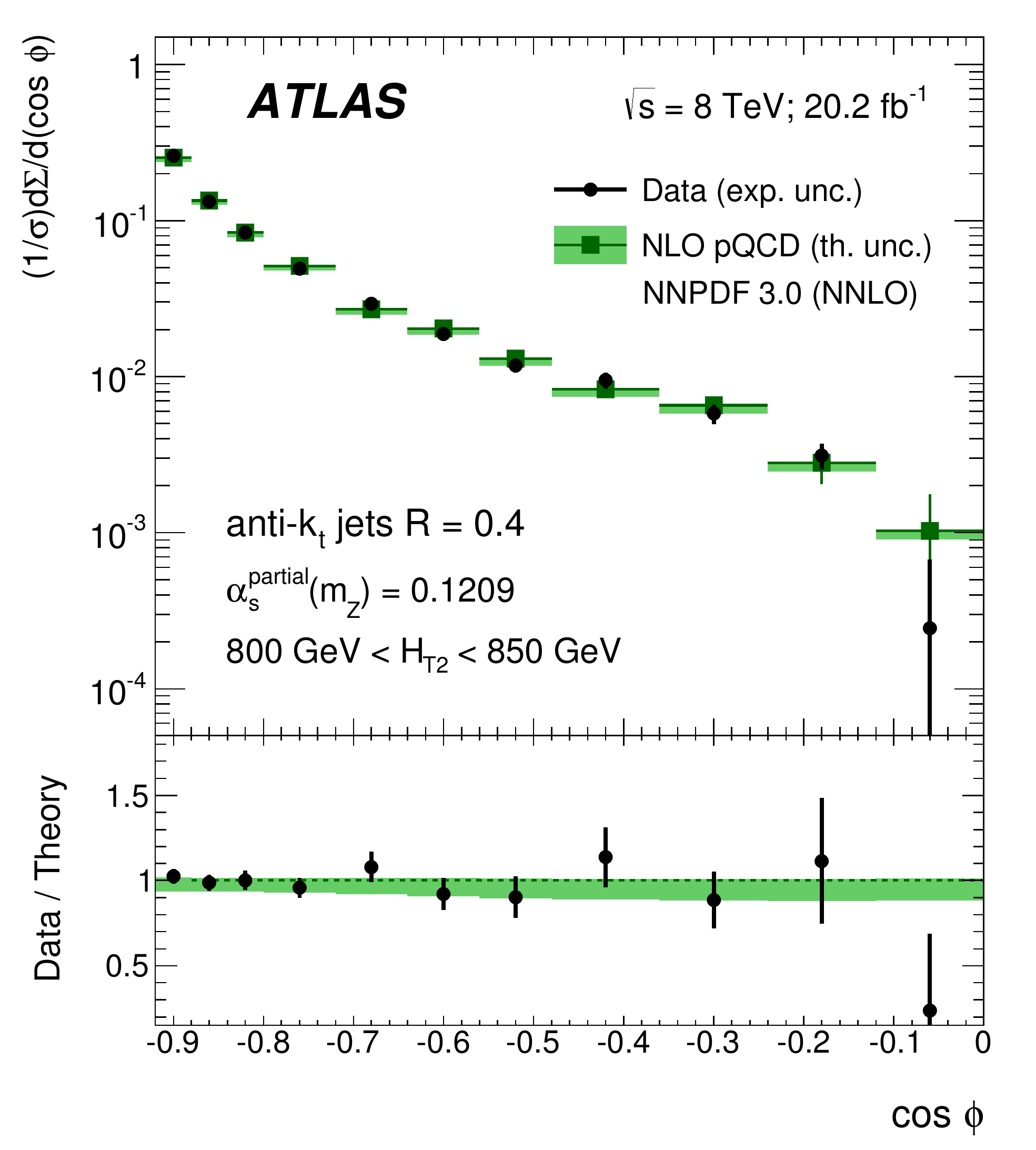}
\includegraphics[width=7.5cm,height=7.3cm]{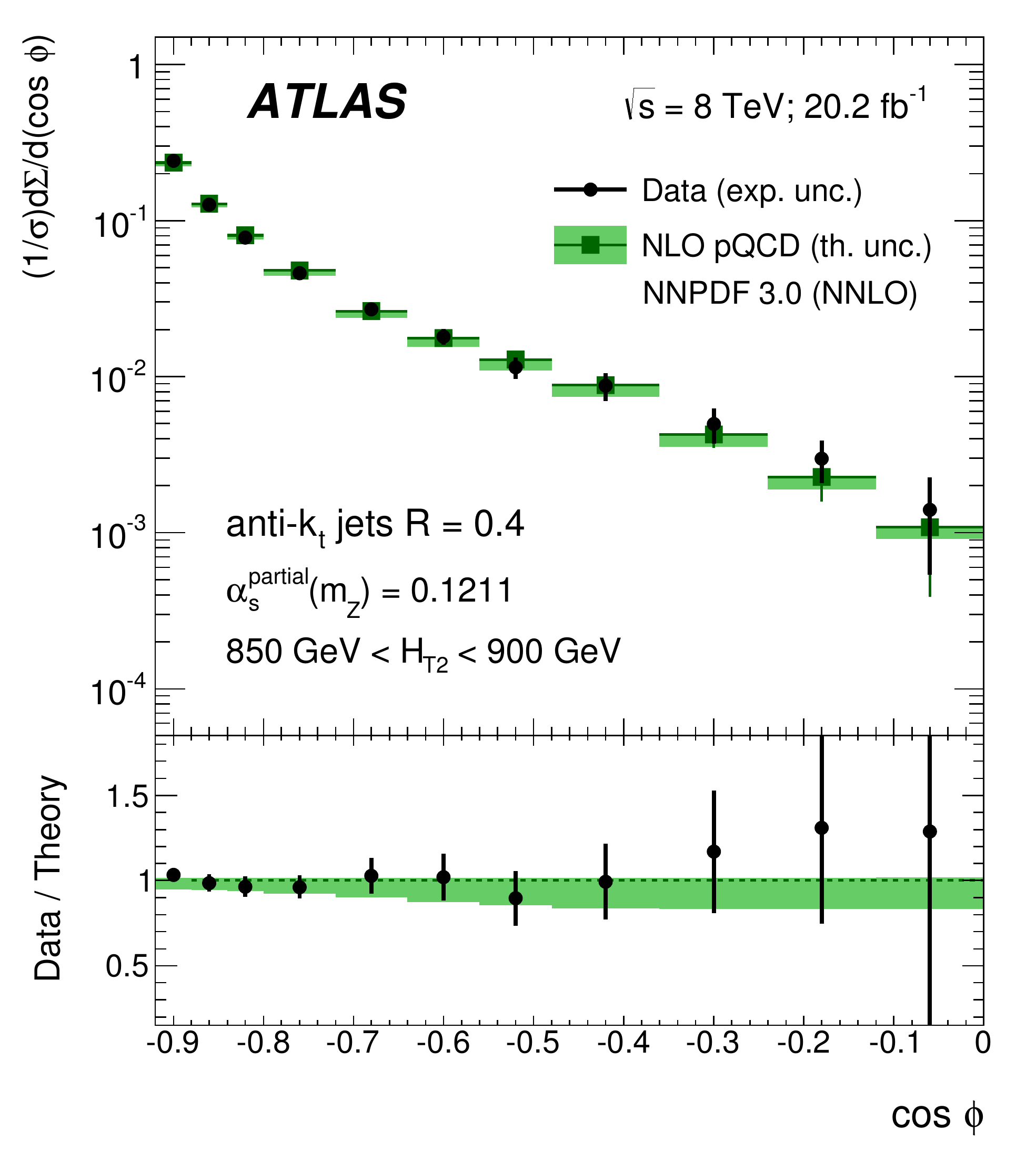}

\includegraphics[width=7.5cm,height=7.3cm]{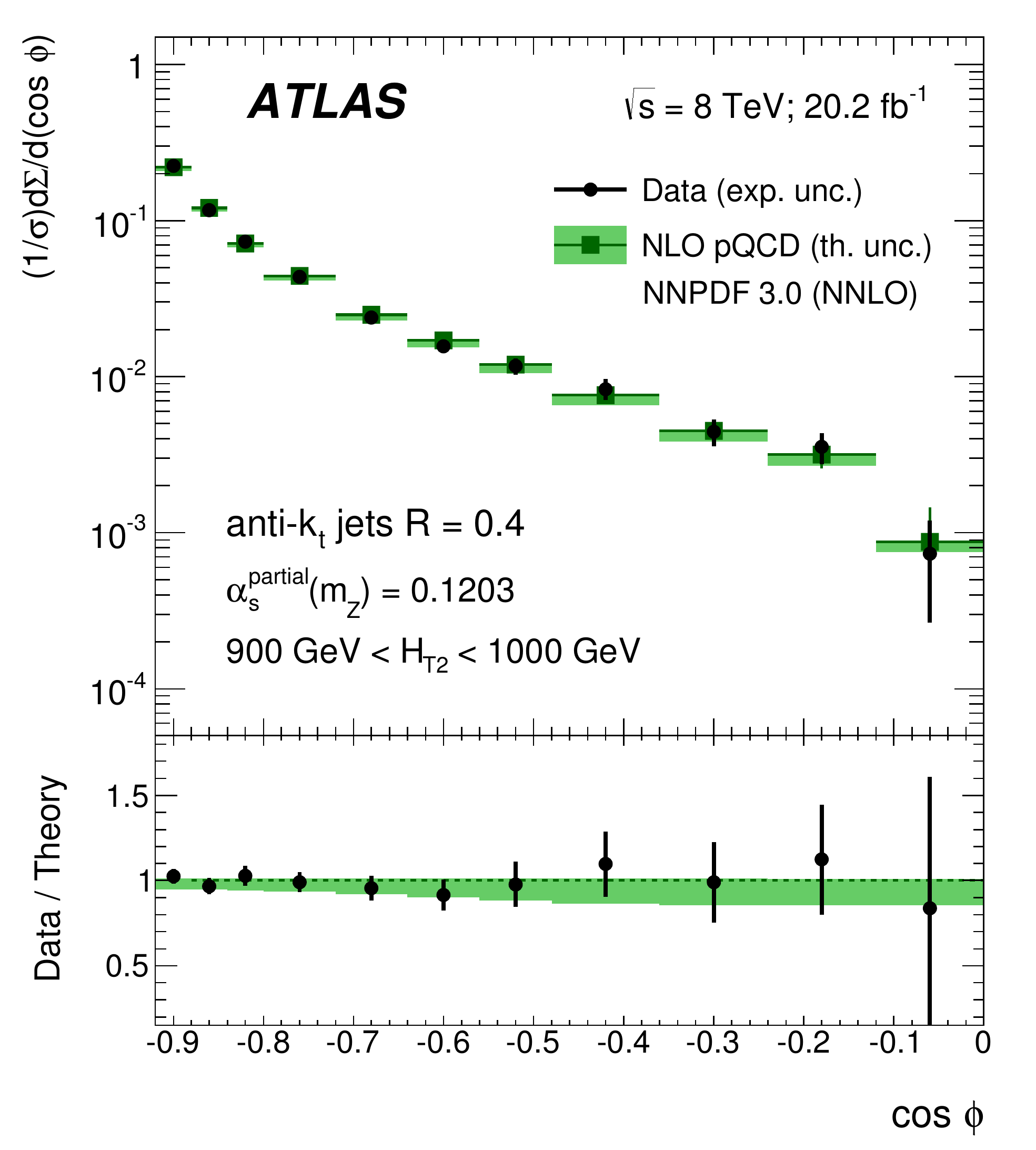}
\includegraphics[width=7.5cm,height=7.3cm]{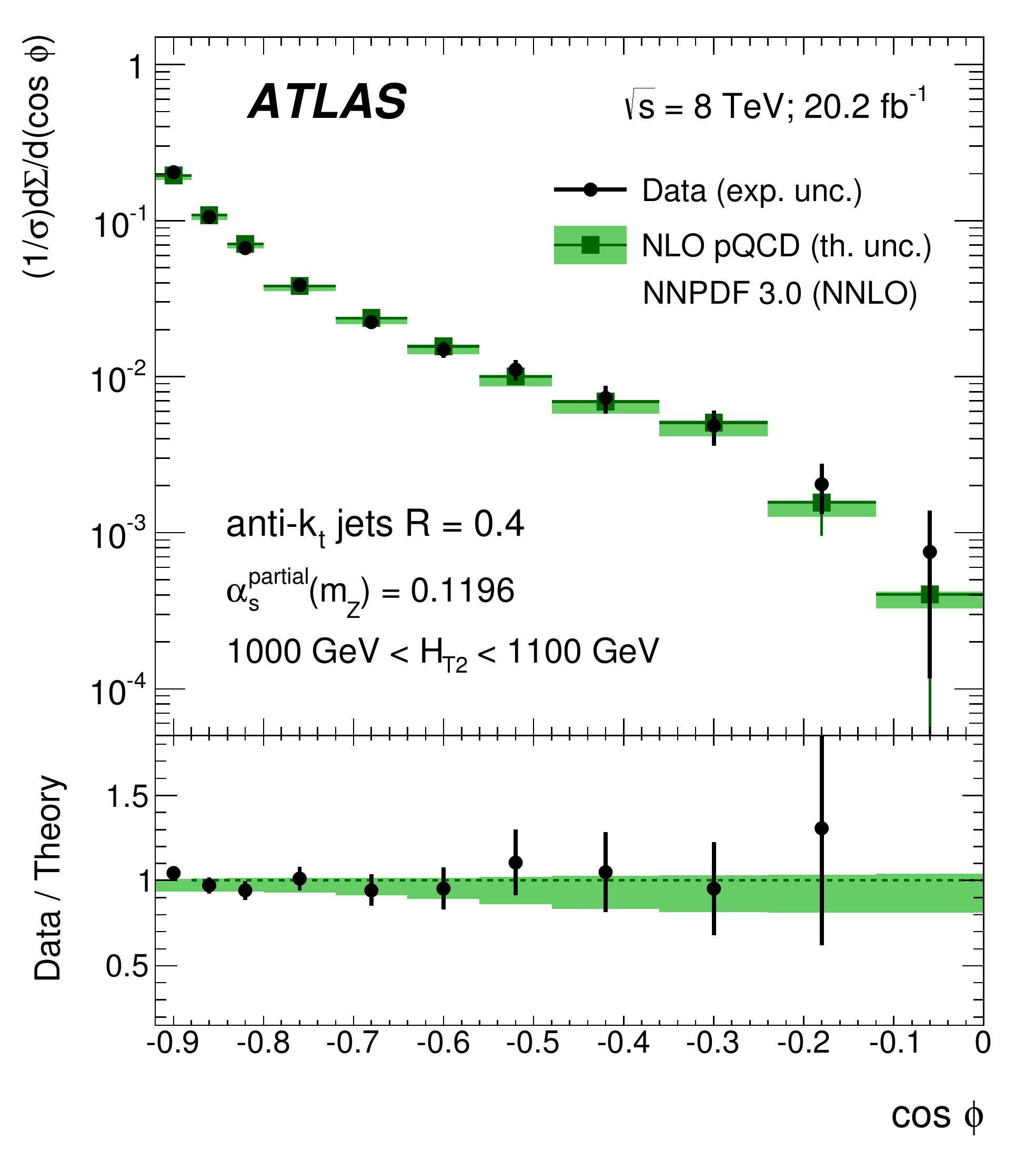}

\includegraphics[width=7.5cm,height=7.3cm]{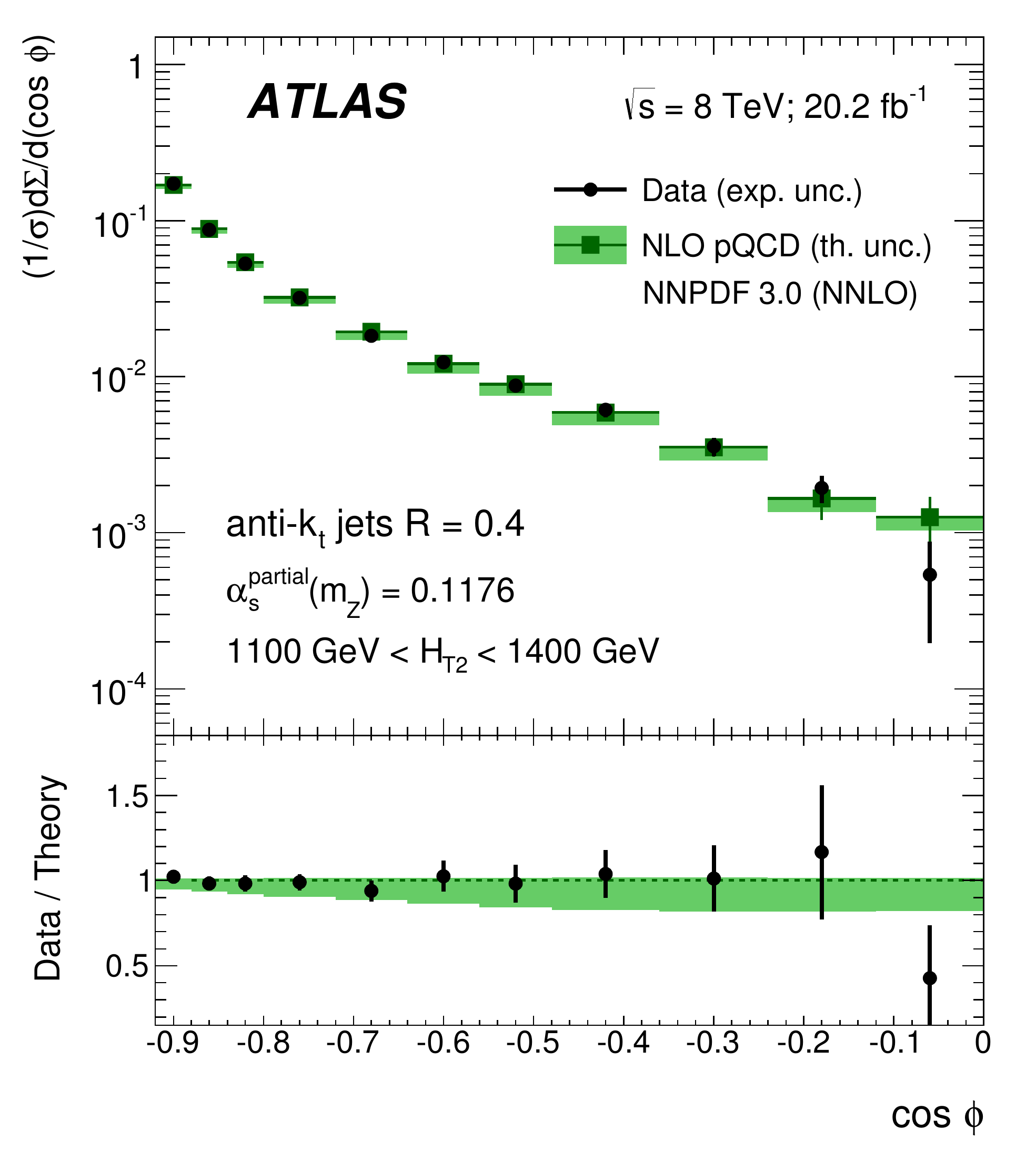}
\includegraphics[width=7.5cm,height=7.3cm]{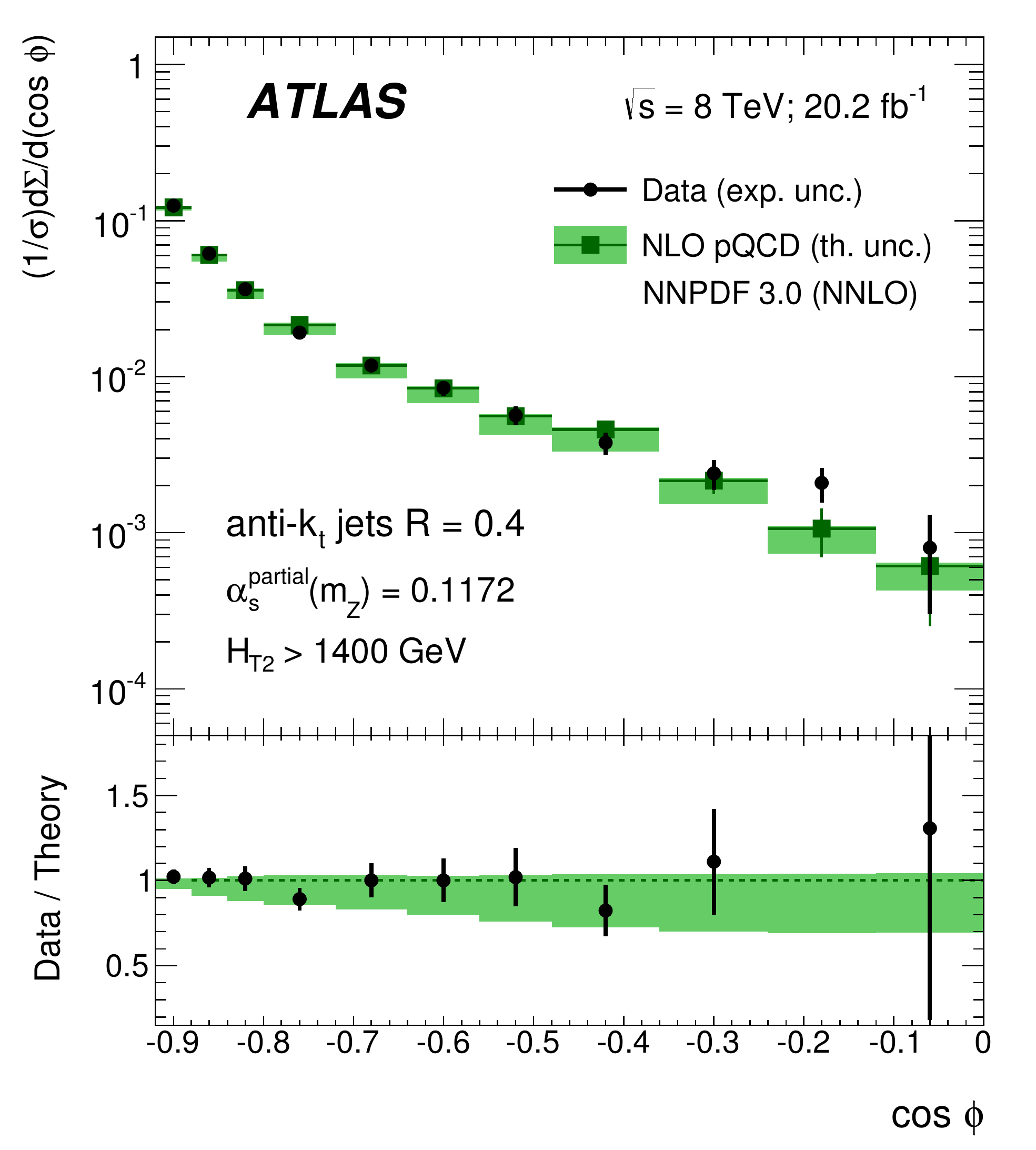}
\caption{Comparison of the ATEEC data and the theoretical predictions after the fit. The value of $\alpha_{\mathrm{s}}(m_\Zboson)$ used in this comparison is fitted independently for each energy bin.}
\label{fig:ateecAfterFit}
\end{figure}

\subsection{Global ATEEC fit}
As before, the global value of $\alpha_{\mathrm{s}}(m_\Zboson)$ is obtained from the combined fit of the ATEEC data in the six bins of $H_{\mathrm{T}2}$. Again, the \textsc{NNPDF 3.0} PDF set is used for the final result as it provides the most conservative choice. Also, as in the TEEC case, two additional correlation scenarios have been considered for the JES uncertainty. An additional uncertainty of 0.0003 is assigned in order to cover the differences.

The results are summarised in Table~\ref{tab:ateecGlobal} for the four sets of PDFs considered in the theoretical predictions.
\begin{table}[H]
\footnotesize
\begin{center}
\begin{tabular}{llc}
\hline
\T\B PDF & $\alpha_{\mathrm{s}}(m_\Zboson)$ value & $\chi^2/N_{\mathrm{dof}}$\\
\hline
\T \B MMHT 2014 & 0.1185 $\pm$ 0.0012 (exp.) $^{+ 0.0047 }_{- 0.0010 }$ (scale) $\pm$ 0.0010 (PDF) $\pm$ 0.0004 (NP) & 57.0 / 65 \\
\T \B CT14 & 0.1203 $\pm$ 0.0013 (exp.) $^{+ 0.0053 }_{- 0.0014 }$ (scale) $\pm$ 0.0015 (PDF) $\pm$ 0.0004 (NP) & 55.4 / 65 \\
\T \B NNPDF 3.0 & 0.1196 $\pm$ 0.0013 (exp.) $^{+ 0.0061 }_{- 0.0013 }$ (scale) $\pm$ 0.0017 (PDF) $\pm$ 0.0004 (NP) & 60.3 / 65 \\
\T \B HERAPDF 2.0 & 0.1206 $\pm$ 0.0012 (exp.) $^{+ 0.0050 }_{- 0.0014 }$ (scale) $\pm$ 0.0005 (PDF) $\pm$ 0.0002 (NP) $\pm$ 0.0007 (mod) & 54.2 / 65 \\
\hline
\end{tabular}
\end{center}
\caption{The results for $\alpha_{\mathrm{s}}$ from fits to the ATEEC using different PDFs. The uncertainty referred to as NP is the one related to the non-perturbative corrections. The uncertainty labelled as `mod' corresponds to the \textsc{HERAPDF} modelling and parameterisation uncertainty.}
\label{tab:ateecGlobal}
\end{table}
The values shown in Table~\ref{tab:ateecGlobal} are in good agreement with the values in Table~\ref{tab:teecGlobal}, obtained from fits to the TEEC functions. Also, it is important to note that the scale uncertainty is smaller in ATEEC fits than in TEEC fits. The values of the $\chi^2$ function at the minima show excellent agreement between the data and the pQCD predictions.

The final result for the ATEEC fit is
\[
\alpha_{\mathrm{s}}(m_\Zboson) = 0.1196 \pm 0.0013 \mbox{ (exp.) } ^{+ 0.0061 }_{- 0.0013 } \mbox{ (scale)} \pm 0.0017 \mbox{ (PDF)} \pm 0.0004 \mbox{ (NP)}.\\
\]
The values from Table~\ref{tab:asATEEC2} are compared with previous experimental results from Refs.~\cite{atlasEEC, cmsR32, cms3jet, cmsIncJet, cmsIncJet2, cmsTTbar, d0AngCorr, d0IncJet} in Figure~\ref{fig:asRunATEEC}, showing good compatibility, as well as with the value from the current world average~\cite{pdg}.

\begin{figure}[H]
\vspace{4.cm}
\centering
\includegraphics[width=16.5cm,height=11.5cm]{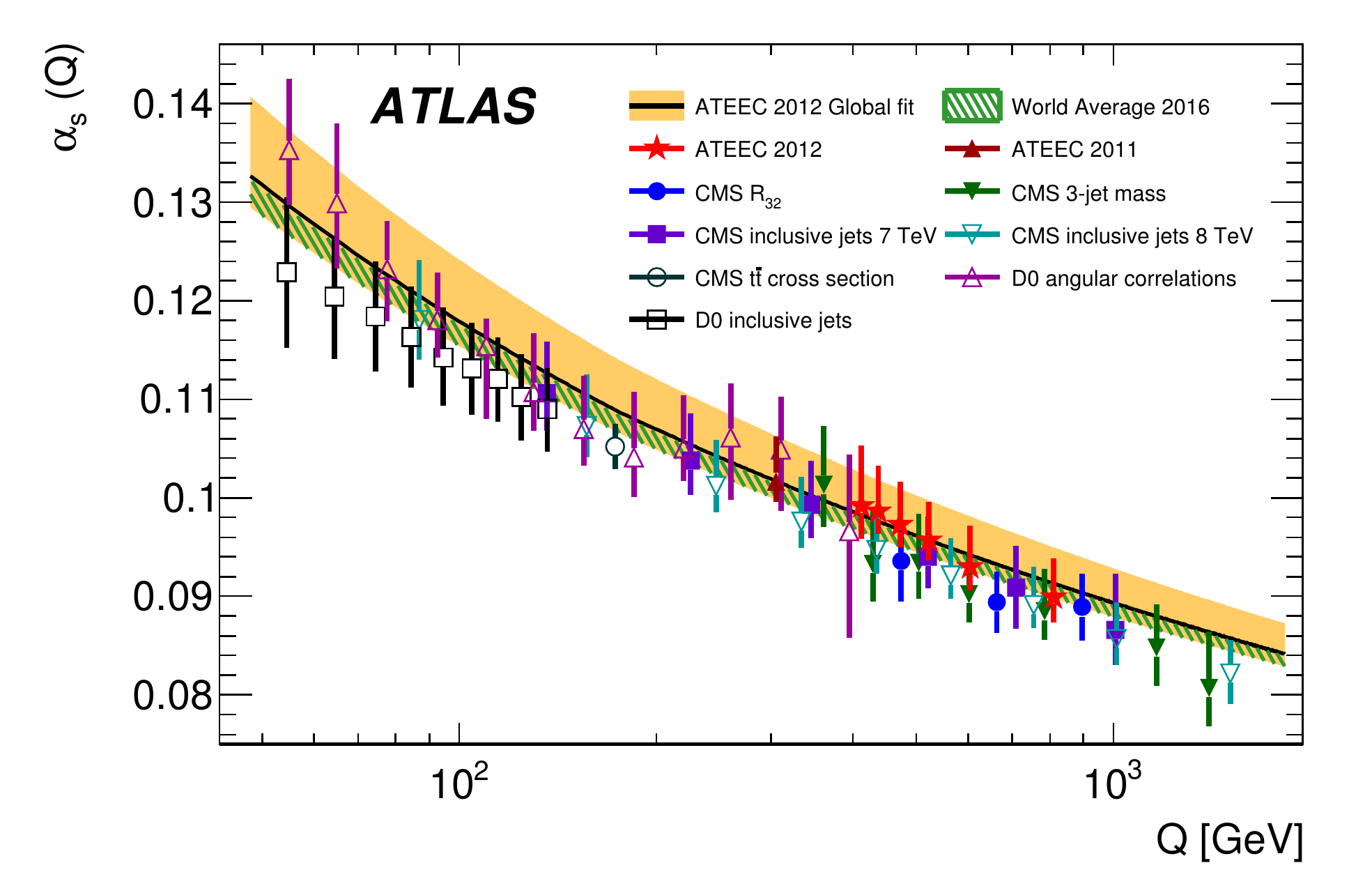}
\caption{Comparison of the values of $\alpha_{\mathrm{s}}(Q)$ obtained from fits to the ATEEC functions at the energy scales given by $\langle H_{\mathrm{T}2} \rangle /2$ (red star points) with the uncertainty band from the global fit (orange full band) and the 2016 world average (green hatched band). Determinations from other experiments are also shown as data points. The error bars, as well as the orange full band, include all experimental and theoretical sources of uncertainty. The strong coupling constant is assumed to run according to the two-loop solution of the RGE.}
\label{fig:asRunATEEC}
\end{figure}

\clearpage

\section{Conclusion}
\label{sec:conclusion}

The TEEC and ATEEC functions are measured in $20.2~\ifb$ of $pp$ collisions at a centre-of-mass energy $\sqrt{s} =  8~\TeV$ using the ATLAS detector at the LHC. The data, binned in six intervals of the sum of transverse momenta of the two leading jets, $H_{\mathrm{T}2} = p_{\mathrm{T}1}+p_{\mathrm{T}2}$, are corrected for detector effects and compared to the predictions of perturbative QCD, corrected for hadronisation and multi-parton interaction effects. The results show that the data are compatible with the theoretical predictions, within the uncertainties.

The data are used to determine the strong coupling constant $\alpha_{\mathrm{s}}$ and its evolution with the interaction scale $Q = (p_{\mathrm{T}1} + p_{\mathrm{T}2})/2$ by means of a $\chi^2$ fit to the theoretical predictions for both TEEC and ATEEC in each energy bin. Additionally, global fits to the TEEC and ATEEC data are performed, leading to 
\begin{eqnarray*}
\alpha_{\mathrm{s}}(m_\Zboson) = 0.1162 \pm 0.0011 \mbox{ (exp.) } ^{+ 0.0076 }_{- 0.0061 } \mbox{ (scale)} \pm 0.0018 \mbox{ (PDF)} \pm 0.0003 \mbox{ (NP)},\\
\alpha_{\mathrm{s}}(m_\Zboson) = 0.1196 \pm 0.0013 \mbox{ (exp.) } ^{+ 0.0061 }_{- 0.0013 } \mbox{ (scale)} \pm 0.0017 \mbox{ (PDF)} \pm 0.0004 \mbox{ (NP)},
\end{eqnarray*}
respectively. Conservatively, the values obtained using the \textsc{NNPDF 3.0} PDF set are chosen, as they provide the largest PDF uncertainty among the four PDF sets investigated. These two values are in good agreement with the determinations in previous experiments and with the current world average $\alpha_{\mathrm{s}}(m_\Zboson) = 0.1181 \pm 0.0011$. The correlation coefficient between the two determinations is $\rho = 0.60$.

The present results are limited by the theoretical scale uncertainties, which amount to 6\% of the value of $\alpha_{\mathrm{s}}(m_\Zboson)$ in the case of the TEEC determination and to 4\% in the case of the ATEEC. This uncertainty is expected to decrease as higher orders are calculated for the perturbative expansion.\\

\section*{Acknowledgements}

% Acknowledgements for papers with collision data
% Version 16-Nov-2017

% Standard acknowledgements start here
%----------------------------------------------
We thank CERN for the very successful operation of the LHC, as well as the
support staff from our institutions without whom ATLAS could not be
operated efficiently.

We acknowledge the support of ANPCyT, Argentina; YerPhI, Armenia; ARC, Australia; BMWFW and FWF, Austria; ANAS, Azerbaijan; SSTC, Belarus; CNPq and FAPESP, Brazil; NSERC, NRC and CFI, Canada; CERN; CONICYT, Chile; CAS, MOST and NSFC, China; COLCIENCIAS, Colombia; MSMT CR, MPO CR and VSC CR, Czech Republic; DNRF and DNSRC, Denmark; IN2P3-CNRS, CEA-DRF/IRFU, France; SRNSF, Georgia; BMBF, HGF, and MPG, Germany; GSRT, Greece; RGC, Hong Kong SAR, China; ISF, I-CORE and Benoziyo Center, Israel; INFN, Italy; MEXT and JSPS, Japan; CNRST, Morocco; NWO, Netherlands; RCN, Norway; MNiSW and NCN, Poland; FCT, Portugal; MNE/IFA, Romania; MES of Russia and NRC KI, Russian Federation; JINR; MESTD, Serbia; MSSR, Slovakia; ARRS and MIZ\v{S}, Slovenia; DST/NRF, South Africa; MINECO, Spain; SRC and Wallenberg Foundation, Sweden; SERI, SNSF and Cantons of Bern and Geneva, Switzerland; MOST, Taiwan; TAEK, Turkey; STFC, United Kingdom; DOE and NSF, United States of America. In addition, individual groups and members have received support from BCKDF, the Canada Council, CANARIE, CRC, Compute Canada, FQRNT, and the Ontario Innovation Trust, Canada; EPLANET, ERC, ERDF, FP7, Horizon 2020 and Marie Sk{\l}odowska-Curie Actions, European Union; Investissements d'Avenir Labex and Idex, ANR, R{\'e}gion Auvergne and Fondation Partager le Savoir, France; DFG and AvH Foundation, Germany; Herakleitos, Thales and Aristeia programmes co-financed by EU-ESF and the Greek NSRF; BSF, GIF and Minerva, Israel; BRF, Norway; CERCA Programme Generalitat de Catalunya, Generalitat Valenciana, Spain; the Royal Society and Leverhulme Trust, United Kingdom.

The crucial computing support from all WLCG partners is acknowledged gratefully, in particular from CERN, the ATLAS Tier-1 facilities at TRIUMF (Canada), NDGF (Denmark, Norway, Sweden), CC-IN2P3 (France), KIT/GridKA (Germany), INFN-CNAF (Italy), NL-T1 (Netherlands), PIC (Spain), ASGC (Taiwan), RAL (UK) and BNL (USA), the Tier-2 facilities worldwide and large non-WLCG resource providers. Major contributors of computing resources are listed in Ref.~\cite{ATL-GEN-PUB-2016-002}.
%----------------------------------------------

\printbibliography

\clearpage

\appendix

\part*{Appendix}
\addcontentsline{toc}{part}{Appendix}
This appendix contains tables in which the measured values of $\alpha_{\mathrm{s}}(m_\Zboson)$ are extrapolated to different values of $Q$ from the central results, given by the average $\pt$ of the third jet, $\langle p_{\mathrm{T}3} \rangle$, the average value of the three leading jets, $\langle (p_{\mathrm{T}1} + p_{\mathrm{T}2} + p_{\mathrm{T}3})\rangle / 3$ and the average value of the transverse momentum for each pair of jets $(i,j)$, $\langle (p_{\mathrm{T}1} + p_{\mathrm{T}2})\rangle / 2$ (Tables \ref{tab:asPt3TEEC}, \ref{tab:asHt3TEEC}, \ref{tab:asHtijTEEC}, \ref{tab:asPt3ATEEC}, \ref{tab:asHt3ATEEC}, \ref{tab:asHtijATEEC}).

\begin{table}[H]
\vspace{0.5cm}
\small
\begin{center}
\begin{tabular}{cc}
\hline
\T\B $\langle p_{\mathrm{T}3} \rangle$ (GeV) & $\alpha_{\mathrm{s}}(\langle p_{\mathrm{T}3}\rangle)$ value (TEEC, \textsc{NNPDF 3.0})\\
\hline
\T \B 169 & 0.1072 $\pm$ 0.0017 (exp.) $^{+ 0.0067 }_{- 0.0019 }$ (scale) $\pm$ 0.0011 (PDF) $\pm$ 0.0001 (NP) \\ 
\T \B 174 & 0.1074 $\pm$ 0.0014 (exp.) $^{+ 0.0060 }_{- 0.0014 }$ (scale) $\pm$ 0.0012 (PDF) $\pm$ 0.0002 (NP) \\ 
\T \B 179 & 0.1068 $\pm$ 0.0014 (exp.) $^{+ 0.0064 }_{- 0.0019 }$ (scale) $\pm$ 0.0012 (PDF) $\pm$ 0.0001 (NP) \\ 
\T \B 186 & 0.1052 $\pm$ 0.0014 (exp.) $^{+ 0.0054 }_{- 0.0013 }$ (scale) $\pm$ 0.0013 (PDF) $\pm$ 0.0001 (NP) \\ 
\T \B 197 & 0.1060 $\pm$ 0.0014 (exp.) $^{+ 0.0065 }_{- 0.0018 }$ (scale) $\pm$ 0.0014 (PDF) $\pm$ 0.0004 (NP) \\ 
\T \B 215 & 0.1052 $\pm$ 0.0018 (exp.) $^{+ 0.0066 }_{- 0.0027 }$ (scale) $\pm$ 0.0015 (PDF) $\pm$ 0.0003 (NP) \\
\hline
\end{tabular}
\end{center}
\caption{Values of $\alpha_{\mathrm{s}}$, obtained from TEEC fits, evolved to the average value of the third-jet transverse momentum in each event, $\langle p_{\mathrm{T}3}\rangle$ for each bin in $H_{\mathrm{T}2}$.}
\label{tab:asPt3TEEC}
\end{table}

\begin{table}[H]
\small
\begin{center}
\begin{tabular}{cc}
\hline
\T\B $\langle H_{\mathrm{T}3}/3 \rangle$ (GeV) & $\alpha_{\mathrm{s}}(\langle H_{\mathrm{T}3}/3 \rangle)$ value (TEEC, \textsc{NNPDF 3.0})\\
\hline
\T \B 289 & 0.1005 $\pm$ 0.0015 (exp.) $^{+ 0.0059 }_{- 0.0016 }$ (scale) $\pm$ 0.0010 (PDF) $\pm$ 0.0001 (NP) \\ 
\T \B 307 & 0.1004 $\pm$ 0.0013 (exp.) $^{+ 0.0052 }_{- 0.0012 }$ (scale) $\pm$ 0.0010 (PDF) $\pm$ 0.0002 (NP) \\ 
\T \B 332 & 0.0994 $\pm$ 0.0012 (exp.) $^{+ 0.0055 }_{- 0.0016 }$ (scale) $\pm$ 0.0010 (PDF) $\pm$ 0.0001 (NP) \\ 
\T \B 366 & 0.0973 $\pm$ 0.0012 (exp.) $^{+ 0.0046 }_{- 0.0011 }$ (scale) $\pm$ 0.0011 (PDF) $\pm$ 0.0001 (NP) \\ 
\T \B 423 & 0.0970 $\pm$ 0.0012 (exp.) $^{+ 0.0054 }_{- 0.0015 }$ (scale) $\pm$ 0.0012 (PDF) $\pm$ 0.0003 (NP) \\ 
\T \B 564 & 0.0943 $\pm$ 0.0014 (exp.) $^{+ 0.0053 }_{- 0.0022 }$ (scale) $\pm$ 0.0012 (PDF) $\pm$ 0.0002 (NP) \\
\hline
\end{tabular}
\end{center}
\caption{Values of $\alpha_{\mathrm{s}}$, obtained from TEEC fits, evolved to the average value of the average transverse momentum of the three leading jets in each event, $\langle (p_{\mathrm{T}1} + p_{\mathrm{T}2} + p_{\mathrm{T}3})\rangle / 3$ for each bin in $H_{\mathrm{T}2}$.}
\label{tab:asHt3TEEC}
\end{table}

\begin{table}[H]
\small
\begin{center}
\begin{tabular}{cc}
\hline
\T\B $\langle H_{\mathrm{T}ij}/2 \rangle$ (GeV) & $\alpha_{\mathrm{s}}(\langle H_{\mathrm{T}ij}/2 \rangle)$ value (TEEC, \textsc{NNPDF 3.0})\\
\hline
\T \B 366 & 0.0979 $\pm$ 0.0014 (exp.) $^{+ 0.0055 }_{- 0.0015 }$ (scale) $\pm$ 0.0009 (PDF) $\pm$ 0.0001 (NP) \\ 
\T \B 386 & 0.0978 $\pm$ 0.0012 (exp.) $^{+ 0.0049 }_{- 0.0012 }$ (scale) $\pm$ 0.0010 (PDF) $\pm$ 0.0002 (NP) \\ 
\T \B 413 & 0.0969 $\pm$ 0.0011 (exp.) $^{+ 0.0052 }_{- 0.0016 }$ (scale) $\pm$ 0.0010 (PDF) $\pm$ 0.0001 (NP) \\ 
\T \B 452 & 0.0951 $\pm$ 0.0011 (exp.) $^{+ 0.0044 }_{- 0.0011 }$ (scale) $\pm$ 0.0011 (PDF) $\pm$ 0.0001 (NP) \\ 
\T \B 515 & 0.0949 $\pm$ 0.0011 (exp.) $^{+ 0.0052 }_{- 0.0014 }$ (scale) $\pm$ 0.0011 (PDF) $\pm$ 0.0003 (NP) \\ 
\T \B 672 & 0.0925 $\pm$ 0.0014 (exp.) $^{+ 0.0051 }_{- 0.0021 }$ (scale) $\pm$ 0.0012 (PDF) $\pm$ 0.0002 (NP) \\
\hline
\end{tabular}
\end{center}
\caption{Values of $\alpha_{\mathrm{s}}$, obtained from TEEC fits, evolved to the average value of transverse momentum for every pair of jets in each event, $\langle (p_{\mathrm{T}i} + p_{\mathrm{T}j} \rangle / 2$ for each bin in $H_{\mathrm{T}2}$.}
\label{tab:asHtijTEEC}
\end{table}

\clearpage

\begin{table}[H]
\vspace{1.cm}
\small
\begin{center}
\begin{tabular}{cc}
\hline
\T\B $\langle p_{\mathrm{T}3} \rangle$ (GeV) & $\alpha_{\mathrm{s}}(\langle p_{\mathrm{T}3}\rangle)$ value (ATEEC, \textsc{NNPDF 3.0})\\
\hline
\T \B  169 & 0.1104 $\pm$ 0.0030 (exp.) $^{+ 0.0070 }_{- 0.0025 }$ (scale) $\pm$ 0.0011 (PDF) $\pm$ 0.0003 (NP) \\ 
\T \B  174 & 0.1101 $\pm$ 0.0022 (exp.) $^{+ 0.0052 }_{- 0.0011 }$ (scale) $\pm$ 0.0012 (PDF) $\pm$ 0.0008 (NP) \\ 
\T \B  179 & 0.1090 $\pm$ 0.0023 (exp.) $^{+ 0.0049 }_{- 0.0011 }$ (scale) $\pm$ 0.0013 (PDF) $\pm$ 0.0002 (NP) \\ 
\T \B  186 & 0.1079 $\pm$ 0.0021 (exp.) $^{+ 0.0044 }_{- 0.0008 }$ (scale) $\pm$ 0.0014 (PDF) $\pm$ 0.0003 (NP) \\ 
\T \B  197 & 0.1056 $\pm$ 0.0025 (exp.) $^{+ 0.0046 }_{- 0.0006 }$ (scale) $\pm$ 0.0016 (PDF) $\pm$ 0.0004 (NP) \\ 
\T \B  215 & 0.1041 $\pm$ 0.0029 (exp.) $^{+ 0.0042 }_{- 0.0007 }$ (scale) $\pm$ 0.0017 (PDF) $\pm$ 0.0001 (NP) \\ 
\hline
\end{tabular}
\end{center}
\caption{Values of $\alpha_{\mathrm{s}}$, obtained from ATEEC fits, evolved to the average value of the third-jet transverse momentum in each event, $\langle p_{\mathrm{T}3}\rangle$ for each bin in $H_{\mathrm{T}2}$}
\label{tab:asPt3ATEEC}
\end{table}

\begin{table}[H]
\vspace{1.cm}
\small
\begin{center}
\begin{tabular}{cc}
\hline
\T\B $\langle H_{\mathrm{T}3}/3 \rangle$ (GeV) & $\alpha_{\mathrm{s}}(\langle H_{\mathrm{T}3}/3 \rangle)$ value (ATEEC, \textsc{NNPDF 3.0})\\
\hline
\T \B  289 & 0.1033 $\pm$ 0.0026 (exp.) $^{+ 0.0061 }_{- 0.0022 }$ (scale) $\pm$ 0.0009 (PDF) $\pm$ 0.0003 (NP) \\ 
\T \B  307 & 0.1027 $\pm$ 0.0019 (exp.) $^{+ 0.0045 }_{- 0.0010 }$ (scale) $\pm$ 0.0011 (PDF) $\pm$ 0.0007 (NP) \\ 
\T \B  332 & 0.1013 $\pm$ 0.0019 (exp.) $^{+ 0.0042 }_{- 0.0009 }$ (scale) $\pm$ 0.0011 (PDF) $\pm$ 0.0001 (NP) \\ 
\T \B  366 & 0.0996 $\pm$ 0.0017 (exp.) $^{+ 0.0037 }_{- 0.0007 }$ (scale) $\pm$ 0.0012 (PDF) $\pm$ 0.0003 (NP) \\ 
\T \B  423 & 0.0966 $\pm$ 0.0021 (exp.) $^{+ 0.0038 }_{- 0.0005 }$ (scale) $\pm$ 0.0013 (PDF) $\pm$ 0.0003 (NP) \\ 
\T \B  564 & 0.0934 $\pm$ 0.0023 (exp.) $^{+ 0.0033 }_{- 0.0006 }$ (scale) $\pm$ 0.0014 (PDF) $\pm$ 0.0001 (NP) \\ 
\hline
\end{tabular}
\end{center}
\caption{Values of $\alpha_{\mathrm{s}}$, obtained from ATEEC fits, evolved to the average value of the average transverse momentum of the three leading jets in each event, $\langle (p_{\mathrm{T}1} + p_{\mathrm{T}2} + p_{\mathrm{T}3})\rangle / 3$ for each bin in $H_{\mathrm{T}2}$}
\label{tab:asHt3ATEEC}
\end{table}

\begin{table}[H]
\vspace{1.cm}
\small
\begin{center}
\begin{tabular}{cc}
\hline
\T\B $\langle H_{\mathrm{T}ij}/2 \rangle$ (GeV) & $\alpha_{\mathrm{s}}(\langle H_{\mathrm{T}ij}/2 \rangle)$ value (ATEEC, \textsc{NNPDF 3.0})\\
\hline
\T \B  366 & 0.1005 $\pm$ 0.0025 (exp.) $^{+ 0.0058 }_{- 0.0021 }$ (scale) $\pm$ 0.0009 (PDF) $\pm$ 0.0002 (NP) \\ 
\T \B  386 & 0.1000 $\pm$ 0.0018 (exp.) $^{+ 0.0043 }_{- 0.0009 }$ (scale) $\pm$ 0.0010 (PDF) $\pm$ 0.0007 (NP) \\ 
\T \B  413 & 0.0987 $\pm$ 0.0018 (exp.) $^{+ 0.0040 }_{- 0.0009 }$ (scale) $\pm$ 0.0010 (PDF) $\pm$ 0.0001 (NP) \\ 
\T \B  452 & 0.0973 $\pm$ 0.0017 (exp.) $^{+ 0.0035 }_{- 0.0007 }$ (scale) $\pm$ 0.0011 (PDF) $\pm$ 0.0003 (NP) \\ 
\T \B  515 & 0.0946 $\pm$ 0.0020 (exp.) $^{+ 0.0037 }_{- 0.0005 }$ (scale) $\pm$ 0.0013 (PDF) $\pm$ 0.0003 (NP) \\ 
\T \B  672 & 0.0917 $\pm$ 0.0022 (exp.) $^{+ 0.0032 }_{- 0.0006 }$ (scale) $\pm$ 0.0013 (PDF) $\pm$ 0.0001 (NP) \\ 
\hline
\end{tabular}
\end{center}
\caption{Values of $\alpha_{\mathrm{s}}$, obtained from ATEEC fits, evolved to the average value of transverse momentum for every pair of jets in each event, $\langle (p_{\mathrm{T}i} + p_{\mathrm{T}j} \rangle / 2$ for each bin in $H_{\mathrm{T}2}$.}
\label{tab:asHtijATEEC}
\end{table}

\newpage 
% ATLAS Collaboration author list
% Data extracted on 01-May-2017 for paper reference STDM-2016-10
\begin{flushleft}
{\Large The ATLAS Collaboration}

\bigskip

M.~Aaboud$^\textrm{\scriptsize 137d}$,
G.~Aad$^\textrm{\scriptsize 88}$,
B.~Abbott$^\textrm{\scriptsize 115}$,
J.~Abdallah$^\textrm{\scriptsize 8}$,
O.~Abdinov$^\textrm{\scriptsize 12}$$^{,*}$,
B.~Abeloos$^\textrm{\scriptsize 119}$,
S.H.~Abidi$^\textrm{\scriptsize 161}$,
O.S.~AbouZeid$^\textrm{\scriptsize 139}$,
N.L.~Abraham$^\textrm{\scriptsize 151}$,
H.~Abramowicz$^\textrm{\scriptsize 155}$
% ...
\bigskip
\\
% ...
$^{8}$ Department of Physics, The University of Texas at Arlington, Arlington TX, United States of America\\
% ...
$^{12}$ Institute of Physics, Azerbaijan Academy of Sciences, Baku, Azerbaijan\\
% ...
$^{88}$ CPPM, Aix-Marseille Universit{\'e} and CNRS/IN2P3, Marseille, France\\
% ...
$^{115}$ Homer L. Dodge Department of Physics and Astronomy, University of Oklahoma, Norman OK, United States of America\\
% ...
$^{119}$ LAL, Univ. Paris-Sud, CNRS/IN2P3, Universit{\'e} Paris-Saclay, Orsay, France\\
% ...
$^{137}$ $^{(a)}$ Facult{\'e} des Sciences Ain Chock, R{\'e}seau Universitaire de Physique des Hautes Energies - Universit{\'e} Hassan II, Casablanca; $^{(b)}$ Centre National de l'Energie des Sciences Techniques Nucleaires, Rabat; $^{(c)}$ Facult{\'e} des Sciences Semlalia, Universit{\'e} Cadi Ayyad, LPHEA-Marrakech; $^{(d)}$ Facult{\'e} des Sciences, Universit{\'e} Mohamed Premier and LPTPM, Oujda; $^{(e)}$ Facult{\'e} des sciences, Universit{\'e} Mohammed V, Rabat, Morocco\\
% ...
$^{139}$ Santa Cruz Institute for Particle Physics, University of California Santa Cruz, Santa Cruz CA, United States of America\\
% ...
$^{151}$ Department of Physics and Astronomy, University of Sussex, Brighton, United Kingdom\\
% ...
$^{155}$ Raymond and Beverly Sackler School of Physics and Astronomy, Tel Aviv University, Tel Aviv, Israel\\
% ...
$^{161}$ Department of Physics, University of Toronto, Toronto ON, Canada\\
% ...
$^{a}$ Also at Department of Physics, King's College London, London, United Kingdom\\
$^{b}$ Also at Institute of Physics, Azerbaijan Academy of Sciences, Baku, Azerbaijan\\
$^{c}$ Also at Novosibirsk State University, Novosibirsk, Russia\\
$^{d}$ Also at TRIUMF, Vancouver BC, Canada\\
$^{e}$ Also at Department of Physics {\&} Astronomy, University of Louisville, Louisville, KY, United States of America\\
% ...
$^{*}$ Deceased
\end{flushleft}

\end{document}